\documentclass[a4paper,11pt]{article}
\usepackage{jheppub}
\usepackage{lineno}
\usepackage{multirow}
\usepackage{booktabs}
\usepackage{slashed}
\usepackage{enumitem}
\usepackage{bbm}

\allowdisplaybreaks[2]
\newcommand{\sss}{\scriptscriptstyle}
\newcommand{\nn}{\nonumber \\}
\newcommand{\as}{\alpha_s}
\newcommand{\aew}{\alpha_e}
\newcommand{\eps}{\epsilon}
\newcommand{\eQ}{e_Q}
\newcommand{\eq}{e_q}
\newcommand{\nq}{n_q}
\newcommand{\qA}{q^{\prime}}
\newcommand{\eqA}{e_{q^{\prime}}}
\newcommand{\nqA}{n_{q^{\prime}}}
\newcommand{\ps}{\slashed{p}}
\newcommand{\Ps}{\slashed{P}}
\def\MSbar{\overline{\mathrm{MS}}}
\def\SigmaMass{m_{\sigma}}
\def\mMS{\overline{m}}
\def\mOS{m_{\mathrm{os}}}

\def\ZpOm{\mathrm{C}_{\overline{m}}}

\newcommand{\OS}{\text{os}}
\newcommand{\logu}{L_{\mu}}
\newcommand{\logums}{L_{\bar{\mu}}}
\newcommand{\ZmOS}{Z_{m}^{\OS}}
\newcommand{\ZmMSb}{Z_{\mMS}}
\newcommand{\ZQOS}{Z_{Q}^{\OS}}
\newcommand{\ZqOS}{Z_{q}^{\OS}}
\newcommand{\Zff}{Z_{m}^{f}}

\newcommand{\GeV}{\text{GeV}}
\newcommand{\MeV}{\text{MeV}}

\DeclareMathOperator{\Li}{Li}

\def\form{\texttt{FORM}}

\def\diagen{\texttt{DiaGen}}
\def\idsolver{\texttt{IdSolver}}

\title{\boldmath 
Three-loop QCD+QED corrections to on-shell quark renormalization
}

\author[a]{Long Chen,}
\emailAdd{longchen@sdu.edu.cn}

\author[a]{Hong-Yang Han,}
\emailAdd{hhyang7@mail.sdu.edu.cn}

\author[a]{Zhe Li,}
\emailAdd{lizhep@sdu.edu.cn}

\author[b]{Marco Niggetiedt}
\emailAdd{marco.niggetiedt@physik.uzh.ch}

\affiliation[a]{School of Physics, Shandong University, Jinan, Shandong 250100, China}
\affiliation[b]{Physik-Institut, Universit\"at Z\"urich, Winterthurerstrasse 190, CH-8057 Z\"urich, Switzerland}

\abstract{
We present the three-loop QCD+QED mixed corrections to the on-shell quark mass and wave-function renormalization constants through orders $\mathcal{O}(\alpha_s^m\alpha^n)$ with $m+n=3$. 
We further derive the three-loop relation between the pole mass and the $\overline{\text{MS}}$ mass of a heavy quark, including the complete mixed QCD+QED contributions. The corresponding quark-mass anomalous dimension in the presence of both interactions is also extracted. 
Moreover, we provide the explicit conversion formulae, up to the same perturbative order, between the pole mass and the trace-anomaly subtracted $\sigma$-mass of heavy quark.
}

\keywords{}

\begin{document}
\preprint{CPTNP-2026-005, ZU-TH 05/26}
\maketitle
\flushbottom

\section{Introduction}
\label{sec:Intro}

Quark masses are fundamental parameters of the Standard Model of particle physics, and their precise determination is crucial for various high-precision tests of the model.
However, the color confinement inherent in Quantum Chromodynamics (QCD) makes the experimental extraction of quark masses highly non-trivial, as quarks are not observed as isolated free particles but are typically confined within hadrons (with the exception of the top quark).
Moreover, the quark-mass parameters in the QCD Lagrangian depend on the particular renormalization procedure required to subtract the intermediate ultraviolet divergences in the scattering amplitudes.
Consequently, there is no theoretically unique way to define the ``physical'' mass of an isolated (heavy) quark.
Nevertheless, there are two most commonly used process-independent perturbative mass definitions for massive quarks in the perturbation theory, the pole mass and $\MSbar$ mass definition. 
The pole mass for a particle in perturbation theory is defined through a special ``physical'' subtraction condition where the zero-point of the full inverse propagator in its momentum, i.e.~the pole position of the propagator, coincides with the renormalized on-shell mass, while the residue at the pole fixes the corresponding wave-function renormalization constant. 
Throughout this paper, we therefore use the terms ``on-shell mass'' and ``perturbative pole mass'' interchangeably. 
It is well-known that the notion of pole mass for an elementary fermion in a fundamental gauge theory is not only renormalization-scheme independent, but also well-defined to \textit{any} but \textit{finite} order within perturbation theory~\cite{Tarrach:1980up,Bigi:1994em,Beneke:1994sw,Breckenridge:1994gs,Smith:1996xz,Kronfeld:1998di,Gambino:1999ai}. 
However, due to infrared renormalon effects, it suffers from an intrinsic non-perturbative ambiguity of order $\Lambda_{\text{QCD}}$ and therefore is not a strictly well-defined physical observable~\cite{Bigi:1994em,Beneke:1994sw} and thus could not be used to arbitrarily high precision; 
the complete quark propagator in the full QCD does not have a simple pole and thus the concept of quark pole mass cannot be defined outside perturbation theory. 
Another commonly used definition is based on the modified minimal subtraction ($\overline{\text{MS}}$) scheme~\cite{tHooft:1973mfk,Bardeen:1978yd}, leading to a short-distance mass that is free from the leading infrared renormalon ambiguities~\cite{Bigi:1994em,Beneke:1998ui}. 
The perturbative results reformulated in terms of $\overline{\text{MS}}$ mass usually exhibit improved perturbative convergence compared to those using the on-shell pole mass, which makes it particularly well suited for higher-order perturbative computations, especially in high-energy scattering processes.

Motivated by different theoretical or practical considerations, in addition to the pole and $\MSbar$ masses, several useful alternative short-distance mass definitions of heavy quark have been proposed in the literature.
Prominent examples include the kinetic mass~\cite{Bigi:1994ga,Bigi:1996si,Czarnecki:1997sz,Fael:2020iea}, the potential-subtracted mass~\cite{Beneke:1998rk}, the 1S mass~\cite{Hoang:1998ng}, the MSR mass~\cite{Hoang:2008yj,Hoang:2017suc}, and the (minimal) renormalon-subtracted mass~\cite{Pineda:2001zq,Komijani:2017vep}. These mass schemes are explicitly designed to remove the infrared renormalon contained in the pole mass, thereby providing short-distance quantities with improved perturbative behavior. 
In addition, regularization-independent renormalization schemes such as the RI/MOM mass~\cite{Martinelli:1994ty,Franco:1998bm,Chetyrkin:1999pq} and the RI/(m)SMOM mass~\cite{Aoki:2007xm,Sturm:2009kb,Almeida:2010ns,Kniehl:2020sgo,Bednyakov:2020ugu,Boyle:2016wis,Chen:2025seb} play an important role in non-perturbative determinations of heavy-quark masses using Lattice QCD~\cite{Wilson:1974sk,Creutz:1980zw,Hamber:1981zn}.
These mass definitions can be converted into each other within perturbation theory, which is frequently needed in practical calculations as well as in the extraction of mass values from experiments (as the mass definitions with the best practice vary). 
To this end, the pole or on-shell mass typically serves as conceptually a convenient \textit{regularization-and-renormalization-independent} bridging pivot. 
However, due to the well-known leading IR-renormalon issue in the perturbative definition of the pole mass, the short-distance $\MSbar$ mass is, instead, more often employed as the reference mass when comparing determinations for quark masses (especially at low-energy scales) using different definitions and approaches. 
The relation between $\MSbar$ mass and the pole mass is thus among the theoretical ingredients needed for precise determination of heavy-quark masses from high-energy physics. 
In order to achieve high precision, it is mandatory to know the conversion relations between the different mass schemes as precisely as possible, e.g.~to high perturbative orders and incorporate as much existing interactions as possible.
Recently, a remarkably accurate result for the $\MSbar$ $b$-quark mass was reported in ref.~\cite{Colquhoun:2025pnh} (extending the earlier study~\cite{Hatton:2021syc}) using Lattice methods which has an uncertainty less than $0.2\%$. 
The incorporation of QED effects becomes necessary for quark-mass determination at this precision.

Apart from the precision quark-mass determination, the on-shell renormalization of heavy quark mass and wave-function are among the essential ingredients for high-order perturbative calculations of on-shell S-matrix elements for scattering processes involving external heavy quarks. 
These corrections are among the crucial ingredients for reliably connecting the measured physical observables at high-energy colliders to the Standard Model parameters and for reducing theoretical uncertainties to the levels commensurate with experimental precision.
With the forthcoming accumulation of high-precision experimental data at the High-Luminosity LHC~\cite{Azzi:2017iwa,ATLAS:2018kci,CMS:2013wfa,ATLAS:2025rva} and the construction and operation of future high-luminosity $e^+e^-$ colliders \cite{ILC:2013jhg, FCC:2018evy, CEPCStudyGroup:2018ghi,CLICdp:2018cto}, unprecedented accuracy in experimental measurements of heavy-quark physics will impose increasingly stringent requirements on the corresponding theoretical predictions, particularly regarding the accuracy of perturbative calculations. 
For example, high-precision QCD corrections to the inclusive and differential results for $H/Z\to b\bar b$, $t\to bW$, $e^+e^-\to Q\bar Q$ have been derived at the level of a few per-mille~\cite{Baikov:2005rw,Baikov:2008jh,Chen:2022wit, Wang:2024ilc,Chen:2023osm,Chen:2026gin,Chen:2026jwl,Chen:2022vzo}, in order to match the projected experimental precision at future high-energy, high-luminosity colliders~\cite{ILC:2013jhg,FCC:2018evy,CEPCStudyGroup:2018ghi, CLICdp:2018cto,Amjad:2015mma,deBlas:2019rxi,Schwienhorst:2022yqu, Cepeda:2019klc,Azzi:2019yne}.
The mixed QCD+QED (and electroweak) corrections to these processes are expected to contribute at a comparable level and hence are needed for reliable theoretical predictions at the sub-percent or even per-mille level (ideally with their associated errors kept subdominant to the experimental ones), especially for non-inclusive high-precision observables such as the forward-backward asymmetries~\cite{Amjad:2013tlv,CLICdp:2018esa,Cobal:2025shq,Rohrig:2025bea} (where their actual numerical impact might be larger than a simple estimate based on the size of couplings).

The on-shell renormalization constant of heavy quark mass and wave-function, as well as the corresponding relation between the $\MSbar$ mass and the pole mass in QCD have been computed up to two-loop order \cite{Tarrach:1980up,Gray:1990yh,Avdeev:1997sz,ferguson1999analysis}, three-loop order \cite{Chetyrkin:1999ys,Chetyrkin:1999qi,Melnikov:2000qh,Melnikov:2000zc,Marquard:2007uj}, and even four-loop order \cite{Marquard:2015qpa,Marquard:2016dcn,Marquard:2018rwx} (albeit with a few four-loop non-logarithmic terms known only numerically) respectively, and the estimation of the five and six-loop contributions are also provided in ref.~\cite{Kataev:2019zfx}.
The electroweak and QCD-electroweak mixed corrections up to two-loops 
have been studied in refs.~\cite{Hempfling:1994ar,Jegerlehner:2003py,Jegerlehner:2003sp,Faisst:2004gn,Martin:2005ch,Eiras:2005yt,Jegerlehner:2012kn,Kniehl:2015nwa,Martin:2016xsp}, with the pure QED corrections known up to three loops~\cite{Melnikov:2000zc}.
However, the three-loop QCD+QED mixed contributions to the on-shell mass and wave-function renormalization constants, i.e. corrections of~$\mathcal{O}(\alpha_s^m\alpha^n)$ with $m+n=3$ and $m\, n>0$, have not yet been available in the literature. 
In this work, we close this gap by computing the complete QCD+QED mixed corrections to these renormalization constants up to three-loop order. 
Based on these results, we further determine the corresponding three-loop relation between the $\MSbar$ mass and the pole mass, thereby extending the known perturbative conversion formula to include the full set of both QCD and QED effects at this order.

Recently, a new mass definition known as the trace-anomaly-subtracted $\sigma$-mass has been proposed \cite{Chen:2025iul,Chen:2025zfa}, which can be interpreted as a residual Higgs-generated mass for an on-shell heavy quark obtained by subtracting away the trace-anomaly contribution from its perturbative pole mass.
The $\sigma$-mass definition for heavy quark is not only gauge-invariant, regularization/renormalization scheme- and scale-independent, but also proved~\cite{Chen:2025zfa} to be free from the leading IR-renormalon ambiguity. 
As such, it can serve as a unique \textit{common reference mass scheme}, particularly suitable for comparing and combining quark-mass determinations derived from various physical observables using different methods.
The explicit three-loop QCD perturbative results for the mass ratio between $\sigma$-mass and pole mass, as well as the ratio between $\sigma$-mass and $\MSbar$ mass have been presented in ref.~\cite{Chen:2025iul}.
In this paper, we extend the formula for the $m_\sigma$ of a given heavy quark to take into account the mixed QCD+QED effects at three-loop order.
~\\

The remainder of this paper is structured as follows. 
In section \ref{sec:method}, the formalism for the on-shell renormalization constants of the heavy-quark mass and wave-function is recapitulated, followed by an outline of the computational strategy employed in this work.
Section \ref{sec:analytic_expr} presents the closed-form results of the three-loop mixed QCD+QED heavy quark-mass and wave-function renormalization constants, together with the complete mass relation between the pole mass and $\MSbar$ mass at the same order.
In addition, we provide the explicit results for the relations with the trace-anomaly subtracted $\sigma$-mass of heavy quark.
Furthermore, we also quantify these QED and mixed QCD+QED effects on the heavy-quark mass conversions.
We conclude in section \ref{sec:Conclusion}.

\section{Preliminaries and technicalities}
\label{sec:method}
To be self-contained, we provide in this section a pedagogical review of the on-shell renormalization conditions for quark fields and the derivation of the master formula for the on-shell heavy quark mass and wave-function renormalization constants employed in our calculation, mainly based on the  refs.~\cite{Chetyrkin:1999qi,Melnikov:2000zc,Marquard:2007uj}.
Subsequently, the technical details of our computational setup are exposed, including the generation of the contributing Feynman diagrams, the reduction of corresponding amplitudes and eventually the extraction of the renormalization constants in question.

\subsection{On-shell quark renormalization}
\label{sec:OS_renorm}
The heavy $Q$-quark mass and wave-function renormalization constants, $\ZmOS$ and $\ZQOS$, are introduced formally by 
\begin{align}\label{eq:def_mos}
m_{B}=\ZmOS\, \mOS\,,
\qquad
\psi_B = \sqrt{\ZQOS}\, \psi_R\,,
\end{align}
where $m_B$ and $\psi_B$ denote the bare quark mass and quark field, while $\mOS$ and $\psi_R$ stand for the quark on-shell/pole mass and renormalized quark field, respectively.
The on-shell renormalization constants $\ZmOS$ and $\ZQOS$ are determined by the renormalization condition: 
\begin{enumerate}[label=(\roman*)]
    \item The location of the pole of the heavy quark propagator coincides with its renormalized mass.
    \item The residue of the renormalized propagator at this pole is equal to one.
\end{enumerate}

In the following, we briefly sketch the derivation of the on-shell renormalization constants for the heavy quark mass $\ZmOS$ and the wave-function $\ZQOS$. 
Let us begin with definition of the complete (connected) two-point bare Green's function $G_B$ for a massive quark:
\begin{align}
i\, G_{B}(\ps) 
&= \int \mathrm{d}^4\,x \, e^{+i p \cdot x} \,
 \langle 0 \big|\, \hat{\mathrm{T}} \{\psi_B(x)\, \bar{\psi}_B(0) \} \,\big| 0 \rangle_{conn.} \nonumber\\
  &= \frac{i}{\ps - m_B} \sum_{n=0}^{\infty} \Big(i\Sigma_B(\ps,\, m_B)\, \frac{i}{\ps - m_B} \Big)^n \nn
 &= \frac{i}{\ps - m_B + \Sigma_B(\ps,\, m_B)}\,,
\end{align} 
where $\Sigma_B$ is the bare one-particle-irreducible (1PI) quark self-energy.
The dependence of $\Sigma_B$ on the bare couplings $\as^B, \aew^B$, and bare mass $m_B$ is understood by default.
The corresponding renormalized Green's function $G_R$ reads
\begin{align}
i\, G_{R}(\ps) 
&= \int \mathrm{d}^4\,x \, e^{+i p \cdot x} \,
 \langle 0 \big|\, \hat{\mathrm{T}} \{\psi_R(x)\, \bar{\psi}_R(0) \} \,\big| 0 \rangle_{conn.} \nn
  &= \frac{i}{\ps - m_R + \Sigma_R(\ps,\,  m_R)} 
\end{align}
with $\Sigma_R$ being the renormalized 1PI self-energy.
By definition \eqref{eq:def_mos}, the bare and renormalized Green's functions are related as $G_B(\ps) = \ZQOS\, G_R(\ps)$,
which immediately implies the following relation between their inverses:
\begin{align}
\ZQOS \big(\ps - m_B + \Sigma_B(\ps,\, m_B)\big)
=
\ps - m_R + \Sigma_R(\ps,\,  m_R)\,.
\end{align}
Here, the wave-function renormalization constant $\ZQOS$ absorbs all overall UV-divergences of $G_B(\ps)$ left after all sub-divergences are subtracted by the renormalization of the mass and coupling, rendering $G_R(\ps)$ UV-finite.

Within the on-shell renormalization, the renormalization conditions require that, the renormalized Green's function $G_{R}(\ps)$ has a simple pole at the on-shell mass $\ps = \mOS$ with unit residue, 
\begin{align}
iG_R(\slashed p) \xrightarrow{\ps\to \mOS} \frac{i}{\ps - \mOS}\,.
\end{align}
This condition can be equivalently expressed in terms of the renormalized self-energy function as
\begin{equation}
\Sigma_R(\ps = \mOS,\, m_R=\mOS)
 = m_R - \mOS
=0\,,
\qquad
\frac{\partial\, \Sigma_R(\ps,\, \mOS)}{\partial\, \ps} \Big|_{\ps = \mOS} =0\,.
\end{equation}
Correspondingly, in terms of the bare self-energy function, the on-shell renormalization conditions can be reformulated as
\begin{align}\label{eq:osRC_SigB}
&
\Sigma_B(\ps = \mOS,\, m_B=\ZmOS\, \mOS)
=m_B- \mOS
=\big(  \ZmOS -1 \big)\, \mOS \,, \nn
&
\ZQOS
\Big(1 + \frac{\partial\,}{\partial\, \ps} 
\Sigma_B(\ps,\, m_B=\ZmOS\, \mOS) \big|_{\ps = \mOS}\Big)
=1\,.   
\end{align}
 Therefore, the UV divergence in $\Sigma_B(\slashed{p} = m_p,\, m_B=Z_m\, \mOS)$ right at on-shell $\slashed{p} = m_p$ point can be renormalized just via an \textit{additive} renormalization by simply subtracting away $m_B$, which does not need any multiplicative wave-function renormalization; namely, it does not involve eventually or in essence $\ZQOS$ at all (which actually contains intermediate IR-soft divergences due to masslessness of gauge bosons).
It is important to know that $\Sigma_B$ in eq.~\eqref{eq:osRC_SigB} denotes the full bare self-energies, defined originally with bare mass $m_B$ and bare coupling $\alpha_s^B, \aew^B$ (which can be expressed in terms of the renormalized Lagrangian parameters).

In general, by virtue of the Lorentz invariance, the bare 1PI quark self-energy $\Sigma_B(\ps,m_B)$ can be conveniently decomposed in terms of two independent form factors:
\begin{align}\label{eq:SE_Dec}
\Sigma_B(\ps, \, m_B=\ZmOS\, \mOS) =
\mOS \Sigma_{m}(p^{2}, \mOS)+(\ps - \mOS) \Sigma_{d}(p^{2}, \mOS )
\end{align}
where $\Sigma_{m,d}$ are scalar functions of $p^2$ and the quark on-shell mass $\mOS$.
In terms of the form factors introduced in above equation, we have  $\frac{\partial}{\partial\, \slashed{p}}\Sigma_B$, appearing in eq.~\eqref{eq:osRC_SigB}, evaluated at the on-shell momentum configuration reading
\begin{align}\label{eq:dev_SigB}
\frac{\partial\,}{\partial\, \ps} 
\Sigma_B\big|_{\ps = \mOS} &=  
\mOS\, \frac{\partial}{\partial \ps} \Sigma_{m}(p^{2}, \mOS)
+\frac{\partial}{\partial \ps} \big((\ps - \mOS) \Sigma_{d}(p^{2}, \mOS ) \big) \nn 
&=
\mOS\, 2\,\ps\, \frac{\partial}{\partial p^2} \Sigma_{m}(p^{2}, \mOS)
+\Sigma_{d}(p^{2}, \mOS )
+(\ps - \mOS)\,2\,\ps\, \frac{\partial}{\partial p^{2}} \Sigma_{d}(p^{2}, \mOS ) \nn
&=
2\, \mOS^{2} \frac{\partial}{\partial p^{2}} \Sigma_{m}\left(p^{2}, \mOS\right)\big|_{p^{2}=\mOS^{2}}
+ \Sigma_{d}\left(p^{2}, \mOS\right)\big|_{p^{2}=\mOS^{2}}\,.
\end{align}
Note that the derivative in $\frac{\partial}{\partial\, \slashed{p}} \Sigma_B(\slashed{p},\, m_B)\big|_{\slashed{p} = \mOS}$ applies only to the $p$-dependence in the bare 1PI-self-energy function $\Sigma_B(\slashed{p},\, m_B)$ (with this derivative evaluated eventually at on-shell momentum configuration $\slashed{p} \, u(p) = \mOS\, u(p)$).
The operation of taking derivative in the external momentum in the derivative \textit{commutes} with loop integration in dimensional regularization.
Consequently, the operation of taking derivatives in the external momenta can be done at the integrand level of the dimensionally-regularized loop amplitudes or loop integrals.

Substituting the decomposition ~\eqref{eq:SE_Dec} of the bare quark self-energy $\Sigma_B(\slashed{p},\, m_B)$ and the derivative $\frac{\partial\,}{\partial\, \ps} \Sigma_B\big|_{\ps = \mOS} $ in eq.~\eqref{eq:dev_SigB} into the on-shell renormalization conditions in eq.~\eqref{eq:osRC_SigB}, we derive the following formula for the on-shell quark mass and wave-function renormalization constants, $\ZmOS$ and $\ZQOS$, in terms of the bare self-energy functions:
\begin{align}\label{eq:expr_ZmOS_ZqOS}
\ZmOS &= 1 + 
\Sigma_m(p^2, \mOS)\big|_{p^2=\mOS^2}\,, \nn
(\ZQOS)^{-1} &= 1 + 
2\, \mOS^{2} \frac{\partial}{\partial p^{2}} \Sigma_{m}\left(p^{2}, \mOS\right)\big|_{p^{2}=\mOS^{2}}
+ \Sigma_{d}\left(p^{2}, \mOS\right)\big|_{p^{2}=\mOS^{2}}\,.
\end{align}
Here, the $\Sigma_{m,d}(p^2, \mOS)$ represent the full bare self-energies (which may be defined originally with bare mass $m_B$ and bare couplings) rewritten in terms of the on-shell mass $\mOS$ and mass renormalization constant $\ZmOS$ by $m_B=\ZmOS\mOS$.
The derivatives with respect to $p^{2}$ are evaluated at the on-shell point $p^{2}=\mOS^{2}$.  
If the external quark is massless, then $\ZmOS = 1$ due to the chiral symmetry in pure gauge theories, as $\Sigma_m = 0$ and  accordingly $(\ZqOS)^{-1} = 1 + \Sigma_{d}\left(p^{2}=0\right)$ which would reduce 1 unless there are massive particles running inside the loop corrections to $\Sigma_{d}$.

To facilitate the extraction of the quantities in r.h.s.~of~\eqref{eq:expr_ZmOS_ZqOS} in practical calculations, especially the derivative in momentum, one may introduce the auxiliary power expansion parameter $t$ in the following parameterization of the external momentum~\cite{Melnikov:2000zc,Marquard:2007uj,Marquard:2018rwx},
\begin{align}
p^\mu = P^\mu (1+t)\,, \qquad P^2 = \mOS^2\,.
\end{align}
One considers then the following projection of the bare self-energy~\eqref{eq:SE_Dec}, 
\begin{align}\label{eq:OSproj}
T_{1}\equiv\operatorname{Tr}\Big[\frac{(\Ps+\mOS)}{4\, P^{2}} \Sigma_B(\ps, \mOS)\Big]
=\Sigma_{m}(p^{2}, \mOS ) + t\, \Sigma_{d}(p^{2}, \mOS )\,,
\end{align}
where the momentum and mass in the external projector $(\Ps+\mOS)/ (4\, P^{2})$ are exactly the on-shell $P$ and $\mOS$ of the external quark, rather than $t$-dependent off-shell momentum or bare mass. 
In the case of the external quark being massless, then $\mOS = 0$, $\Sigma_m = 0$ and in the limit $p^2 \rightarrow 0$ the projection~\eqref{eq:OSproj} effectively reduces to 
\begin{align}
\frac{1}{4\,p^2}\,
\mathrm{Tr}
\big[\slashed p\, 
\big( \Sigma_B(\slashed p) =\slashed p\,\Sigma_d(p^2) \big)
\big] \big|_{p^2 \rightarrow 0} = \Sigma_d(p^2 = 0)\,,
\end{align}
which would be scaleless and vanish in dimensional regularization unless there are massive particles running inside the loop corrections. 
Expanding the bare self-energy projection, defined in eq.~\eqref{eq:OSproj}, in the small parameter $t$, and retaining terms up to the first power, one ends up with  
\begin{align}
T_{1}=
\Sigma_{m}\left(p^{2}, \mOS\right)\Big|_{p^{2}=\mOS^{2}}
+t\,\Big[
2\, \mOS^{2} \frac{\partial}{\partial p^{2}} \Sigma_{m}\left(p^{2}, \mOS\right)\big|_{p^{2}=\mOS^{2}}
+\Sigma_{d}\left(p^{2}, \mOS\right)\big|_{p^{2}=\mOS^{2}}
\Big]
+\mathcal{O}\big(t^{2}\big)\, .
\end{align}
Comparing this expansion with eq.~\eqref{eq:expr_ZmOS_ZqOS}, one sees that the leading term and the coefficient of the linear term in $t$ of this projection expansion can be used to extract the on-shell quark mass renormalization constant $\ZmOS$ and the wave-function renormalization constant $\ZQOS$, respectively.
We note again that the $\Sigma_{m}$ and $\Sigma_{d}$ in the above equation are the full bare self-energies, which may be defined and determined originally with bare mass Lagrangian $m_B$ and bare couplings $\as^B, \aew^B$, but one is free to rewrite the bare $m_B$, $\as^B, \aew^B$ systematically and iteratively in terms of the renormalized counter-parts (which we usually do for the final expressions presented in the renormalized perturbation theory).

\subsection{Computational setup}
The perturbative QCD+QED corrections to the matrix elements involved for extracting the on-shell heavy quark renormalizations are computed in terms of Feynman diagrams, which are subsequently manipulated in a similar fashion to our previous calculations~\cite{Ahmed:2021spj,Chen:2021gxv,Chen:2023lus,Chen:2022lun}.
More specifically, symbolic expressions of the contributing Feynman diagrams to three-loop order are generated by the diagram generator~\diagen~\cite{diagen}.\footnote{The C++ library \diagen~provides, besides diagram generation for arbitrary Feynman rules, topological analysis tools and an interface to the C++ library \idsolver~that allows to directly apply integration-by-parts identities to the integrals occurring in the generated diagrams. \idsolver~has been originally written for the calculation of ref.~\cite{Czakon:2004bu}, while \diagen~predates this software.} 
For cross-checks, we have prepared in parallel an alternative computational setup based on \texttt{FeynArts} package \cite{Hahn:2000kx} to generate Feynman diagrams and the corresponding amplitudes. 
The computations are done in a general Lorentz-covariant gauge for the gluon field, with the general-covariant-gauge fixing parameter $\xi$ defined through the gluon propagator $\frac{i}{k^2} \left(-g^{\mu\nu} + \xi \,\frac{k^{\mu} k^{\mu}}{k^2} \right)$, $k$ being the momentum of the gluon.
For simplicity, the gauge-fixing parameter in the photon propagator is set the same as the $\xi$ introduced in the gluon propagator.
Retaining the $\xi$-dependence up to three-loop order not only validates the $\xi$-independence of the mass renormalization constants, but is actually needed for the full result for the on-shell field renormalization constants which do exhibit $\xi$-dependence in DR from three-loop onward~\cite{Melnikov:2000zc,Chen:2025iul}.

\begin{figure}[htbp]
\centering
\includegraphics[width = 1.0\textwidth]{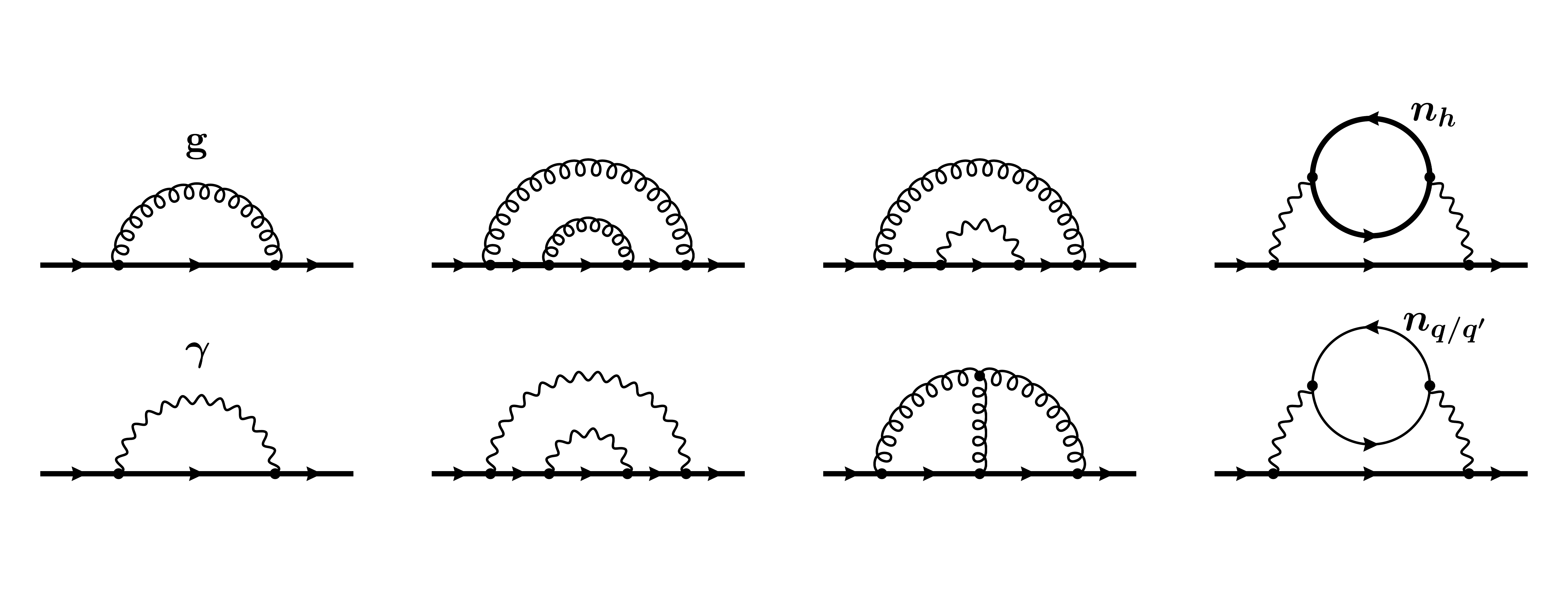} 
\caption{Representative Feynman diagrams contributing to the mixed QCD+QED corrections to the heavy quark self-energy up to two-loop order.
The first column displays the diagrams at one-loop order, while the remaining columns show illustrative examples of the two-loop contribution. 
The last row displays diagrams with a closed heavy-quark ($n_h$) and light-quark ($n_q$ or $n_{\qA}$) loops. (The thick solid line denotes the massive quark, whereas the thin solid line represents massless quarks.)
}
\label{fig:FeynDia_2-loops}
\end{figure}

\begin{figure}[htbp]
\centering
\includegraphics[width = 1.0\textwidth]{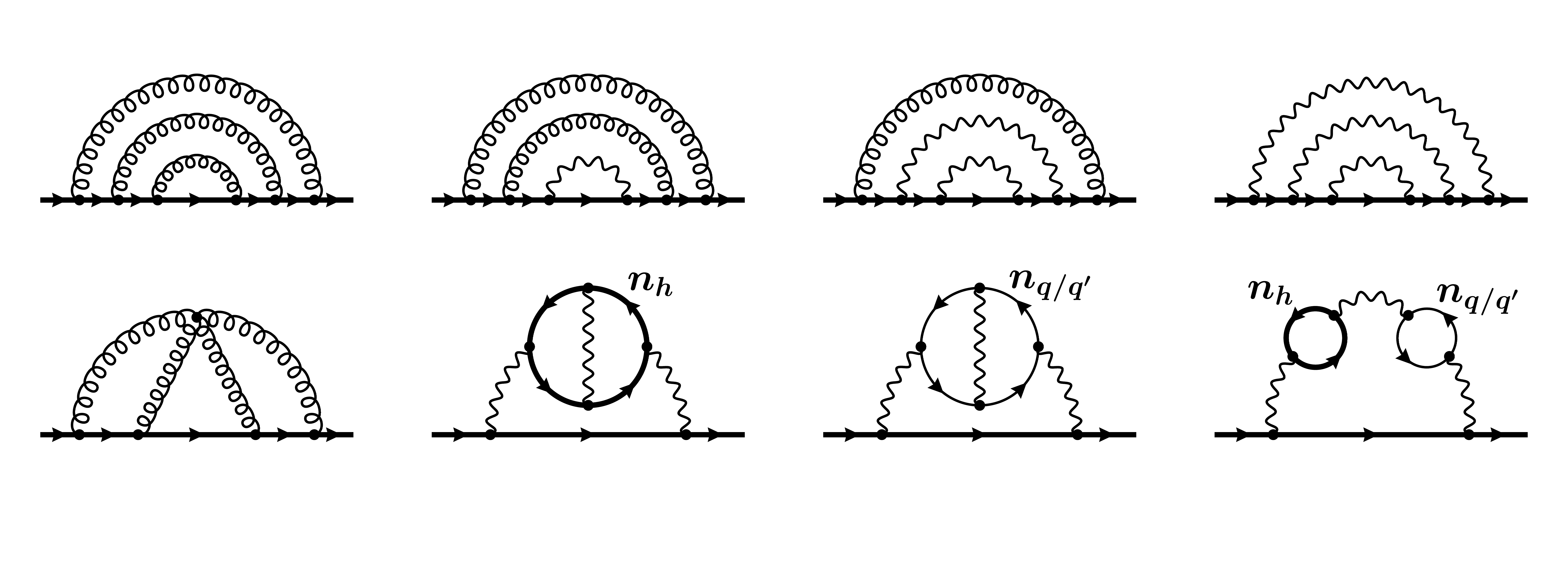} 
\caption{Representative Feynman diagrams for the mixed QCD+QED corrections to the heavy quark self-energy at three-loop order.
The first row shows typical contributions at orders $\as^3$, $\as^2\aew$, $\as\aew^2$, and $\aew^3$, respectively.
The second row displays diagrams with a four-gluon vertex, as well as closed heavy-quark ($n_h$) and light-quark ($n_q$ or $n_{\qA}$) loops, including a diagram containing both.
}
\label{fig:FeynDia_3-loops}
\end{figure}

At one-loop order, two Feynman diagrams contribute to the heavy-quark self-energy in the presence of QCD and QED interactions, which are shown in the first column of figure~\ref{fig:FeynDia_2-loops}, representing the one-loop pure QCD and pure QED corrections, respectively. 
At two-loop order, in addition to the pure QCD $\as^2$ and pure QED $\aew^2$ corrections, mixed QCD+QED corrections of order $\as \aew$ arise. Furthermore, contributions involving gluon self-interactions and closed heavy/light quark loops are present in the remaining columns of figure~\ref{fig:FeynDia_2-loops}.
Representative Feynman diagrams contributing to the mixed QCD+QED corrections to the heavy quark self-energy at three-loop order are depicted in figure~\ref{fig:FeynDia_3-loops}.
The first row illustrates typical diagrams contributing at orders $\as^3, \as^2\aew, \as\aew^2$, and $\aew^3$, respectively. The second row shows representative contributions involving a four-gluon vertex, a closed heavy-quark loop, a closed light-quark loop, as well as a diagram containing both heavy- and light-quark loops.
In our present calculational setup, we consider a single massive quark flavor carrying electric charge $\eQ$, together with two distinct massless quark flavors with electric charges $e_q$ and $e_{\qA}$, and multiplicities $n_q$ and $n_{\qA}$, respectively. 
Accordingly, the total number of massless quarks included thus reads $n_l = n_q + n_{\qA}$.
In addition, we denote by $n_h$ the number of massive quark flavors with mass equal to the on-shell mass $\mOS$; throughout this paper we set $n_h = 1$, while keeping $\nq$ and $\nqA$ arbitrary.


The Feynman-rules substitution, the color algebra with a generic SU($N_c$) group and $D$-dimensional Lorentz as well as the Dirac algebra are performed using \form~\cite{Vermaseren:2000nd} and \texttt{CalcLoop} \cite{calcloop}. 
After tensor reduction, the amplitudes are expressed as linear combinations of a significant number of scalar Feynman integrals, which can be categorized into several integral families. 
Scalar Feynman integrals belonging to the same family are interrelated and can be systematically reduced to a finite set of irreducible master integrals (MIs) by means of integration-by-parts (IBP) identities~\cite{Tkachov:1981wb,Chetyrkin:1981qh}.
In this work, we employ \texttt{Kira} \cite{Maierhofer:2017gsa,Klappert:2020nbg} and \texttt{Blade} \cite{Guan:2024byi}, as well as \idsolver~\cite{diagen}, to perform the IBP reduction, wherein the Laporta algorithm \cite{Laporta:2000dsw} and block triangular form \cite{Guan:2019bcx,Guan:2024byi} are implemented for solving IBP identities.

Subsequently, we numerically compute all MIs at the on-shell kinematic point utilizing the publicly available package \texttt{AMFlow}~\cite{Liu:2017jxz,Liu:2020kpc,Liu:2021wks,Liu:2022mfb,Liu:2022chg} with high-precision numerical accuracy.
Additionally, to ensure the accuracy and reliability of our results, all MIs have been independently cross-verified by approaching the on-shell limit from generic off-shell kinematics.
This is implemented through the generalized power-logarithmic series expansion~\cite{Czakon:2008zk,Lee:2017qql,Liu:2017jxz,Moriello:2019yhu,Hidding:2020ytt} of the differential equations~\cite{Kotikov:1990kg,Kotikov:1991pm,Remiddi:1997ny,Gehrmann:1999as,Argeri:2007up}, where the package \texttt{AMFlow} is employed to evaluate the integrals at the boundary points.
As expected, a perfect agreement was found between our two computational approaches~\cite{Fleischer:1998dw}.
Finally, the analytic expressions are reconstructed from the high-precision numerical results using the PSLQ algorithm~\cite{ferguson1999analysis}.
We note that the required three-loop on-shell propagator-type MIs have previously been obtained in terms of expansions to sufficiently high powers in the dimensional regulator~\cite{Melnikov:2000zc,Lee:2010ik}.

\section{Results up to three loops in mixed QCD+QED }
\label{sec:analytic_expr}
In this section, we present the explicit analytic expressions for the on-shell heavy-quark mass and wave-function renormalization constants, the relation between the $\MSbar$ and on-shell masses, as well as the trace-anomaly-subtracted $\sigma$-mass, up to three-loop order in QCD+QED. 
Throughout this paper, all renormalization constants as well as the mass relations and the trace-anomaly subtracted $\sigma$-mass, are expressed in terms of renormalized couplings $\as$ and $\aew$, which are defined as
\begin{align}\label{eq:aR_def}
\as^B
= \mu^{2\eps}\Big(\frac{e^{\gamma_E}}{4\pi}\Big)^{\eps} Z_{\as} \as \,,
\qquad
\aew^B
= \mu^{2\eps}\Big(\frac{e^{\gamma_E}}{4\pi}\Big)^{\eps} Z_{\aew} \aew \,.
\end{align}
Here, $Z_{\as}$ and $Z_{\aew}$ denote the $\MSbar$ renormalization constants of the strong and electromagnetic couplings, respectively.
We have derived the mixed QCD+QED contributions to the $\MSbar$ coupling renormalization constants $Z_{\as}$ and $Z_{\aew}$, which are needed in the calculation of the on-shell quark renormalization constants at three-loop order and of which the explicit expressions are presented in appendix~\ref{appendix:Alpha}.
For the sake of readers' convenience, a supplemental file containing the electronic form for all analytic expressions presented in this section is provided.

\subsection{On-shell quark mass renormalization constant $\ZmOS$}
\label{sec:ZmOS}

The on-shell quark mass renormalization constant $\ZmOS$ up to three loops in mixed QCD+QED can be conveniently cast into the following form:
\begin{align}\label{eq:ZmOS_exp}
\ZmOS = 1 +
\sum_{l=1}^{3}
\sum^{i+j=l}_{i=0}
\as^i\, \aew^j\, Z_{m,l}^{(i,j)}\,,
\end{align}
where $l$ denotes the loop order, while $i$ and $j$ stand for the power of the QCD and QED couplings, respectively. Here and in the following, $\as$ and $\aew$ are understood as the rescaled couplings $\as/(4\pi)$ and $\aew/(4\pi)$.

In the following, we provide the explicit analytic expressions for the coefficients in the perturbative expansion~\eqref{eq:ZmOS_exp} of the on-shell quark mass renormalization constant $\ZmOS$.
The one-loop QCD and QED on-shell quark mass renormalization constants read
\begin{align}
Z_{m,1}^{(1,0)} = {}&
-C_F \Bigg[\,
 \frac{3}{\eps}
+ 4
+ 3 \logu
+ \eps \Big(
     8
    + \frac{1}{4} \pi^2
    + 4 \logu
    + \frac{3}{2} \logu^2
\Big)
\nn
& + \eps^2 \Big(
      \frac{1}{3} \pi^2
    + 16
    - \zeta_3
    + \Big(
         8
        + \frac{1}{4} \pi^2
      \Big) \logu
    + 2 \logu^2
    + \frac{1}{2} \logu^3
\Big)
\Bigg]\,, 
\nn
Z_{m,1}^{(0,1)} = {} &
- \eQ^2\Bigg[\, \frac{3 }{\eps}
+  4
    + 3 \logu
+ \eps\, \Big(
     8
    + \frac{1}{4} \pi^2
    + 4 \logu
    + \frac{3}{2} \logu^2
\Big)
\nn
& 
+ \eps^2 \Big(
    \frac{1}{3} \pi^2
    + 16
    - \zeta_3
    + \Big(
         8
        + \frac{1}{4} \pi^2
      \Big) \logu
    + 2 \logu^2
    + \frac{1}{2} \logu^3
\Big)
\Bigg]
\,,
\end{align}
where $\logu \equiv \log (\mu^2/\mOS^2)$ and $\eQ$ denotes the electric charge of the massive quark. 
The two-loop QCD+QED on-shell quark mass renormalization constants are given by
\begin{align}
Z_{m,2}^{(2,0)} = {}&
C_F^2 \Bigg[
    \frac{9}{2 \eps^2}
    + \frac{1}{\eps} \Big(
        \frac{45}{4}
        + 9 \logu
    \Big)
    + \Big(
        \frac{199}{8}
        - \frac{17}{4} \pi^2
        + 8 \pi^2 \log2
        - 12 \zeta_3
        + \frac{45}{2} \logu
        + 9 \logu^2
    \Big)
\nn
& 
    + \eps \Big(
        \frac{677}{16}
        - \frac{205}{8} \pi^2
        + \frac{14}{5} \pi^4
        + 48 \pi^2 \log2
        - 16 \pi^2 \log^2 2
        - 8 \log^4 2
        - 135 \zeta_3
        - 192 \Li_4\!\big(\tfrac{1}{2}\big)
\nn
& 
        + \Big(
            \frac{199}{4}
            - \frac{17}{2} \pi^2
            + 16 \pi^2 \log2
            - 24 \zeta_3
        \Big)\logu 
        + \frac{45}{2} \logu^2
        + 6 \logu^3
    \Big)
\Bigg]
\nn
& + C_A C_F \Bigg[\,
    \frac{11}{2 \eps^2}
    - \frac{97}{12 \eps}
    + \Big(
        - \frac{1111}{24}
        + \frac{4}{3} \pi^2
        - 4 \pi^2 \log2
        + 6 \zeta_3
        - \frac{185}{6} \logu
        - \frac{11}{2} \logu^2
    \Big)
\nn
& 
    + \eps \Big(
        - \frac{8581}{48}
        + \frac{271}{72} \pi^2
        - \frac{7}{5} \pi^4
        - 24 \pi^2 \log2
        + 8 \pi^2 \log^2 2
        + 4 \log^4 2
        + 52 \zeta_3
        + 96 \Li_4\!\big(\tfrac{1}{2}\big)
\nn
& 
        + \Big(
            - \frac{1463}{12}
            + \frac{7}{4} \pi^2
            - 8 \pi^2 \log2
            + 12 \zeta_3
        \Big)\logu 
        - \frac{229}{6} \logu^2
        - \frac{11}{2} \logu^3
    \Big)
\Bigg]
\nn
& + C_F n_h \Bigg[
        - \frac{1}{\eps^2}
        + \frac{5}{6 \eps}
        + \Big(
            \frac{143}{12}
            - \frac{4}{3} \pi^2
            + \frac{13}{3} \logu
            + \logu^2
        \Big)
        + \eps \Big(
            \frac{1133}{24}
            - \frac{227}{36} \pi^2
            + 8 \pi^2 \log2
            \nn & 
            - 28 \zeta_3
            + \Big(
                \frac{175}{6}
                - \frac{5}{2} \pi^2
            \Big) \logu
            + \frac{17}{3} \logu^2
            + \logu^3
        \Big)
    \Bigg]
\nn
& 
    + C_F n_l \Bigg[
        - \frac{1}{\eps^2}
        + \frac{5}{6 \eps}
        + \Big(
            \frac{71}{12}
            + \frac{2}{3} \pi^2
            + \frac{13}{3} \logu
            + \logu^2
        \Big)
        + \eps \Big(
            \frac{581}{24}
            + \frac{97}{36} \pi^2
            + 8 \zeta_3
            + \Big(
                \frac{103}{6}
                \nn & 
                + \frac{3}{2} \pi^2
            \Big) \logu
            + \frac{17}{3} \logu^2
            + \logu^3
        \Big)
\Bigg]\,,
\nn
Z_{m,2}^{(1,1)} ={} &
 \eQ^2 C_F \Bigg[\,
    \frac{9}{\eps^2}
    + \frac{1}{\eps} \Big(
        \frac{45}{2}
        + 18 \logu
    \Big)
    + \Big(
        \frac{199}{4}
        - \frac{17}{2} \pi^2
        + 16 \pi^2 \log2
        - 24 \zeta_3
        + 45 \logu
        + 18 \logu^2
    \Big)
\nn
 &   + \eps \Big(
        \frac{677}{8}
        - \frac{205}{4} \pi^2
        + \frac{28}{5} \pi^4
        + 96 \pi^2 \log2
        - 32 \pi^2 \log^2 2
        - 16 \log^4 2
        - 270 \zeta_3
        - 384 \Li_4\!\big(\tfrac{1}{2}\big)
\nn
&        +  \Big(
            \frac{199}{2}
            - 17 \pi^2
            + 32 \pi^2 \log2
            - 48 \zeta_3
        \Big)\logu
        + 45 \logu^2
        + 12 \logu^3
    \Big)
\Bigg]\,,
\nn
 Z_{m,2}^{(0,2)} ={} &
\eQ^4  n_h N_c \Bigg[
    - \frac{2}{\eps^2}
    + \frac{5}{3\eps}
    + \Big(
        \frac{143}{6}
        - \frac{8}{3} \pi^2
        + \frac{26}{3}\logu
        + 2 \logu^2
    \Big)
   + \eps \Big(
        \frac{1133}{12}
        - \frac{227}{18} \pi^2
        + 16\pi^2 \log2
         \nn & 
        - 56 \zeta_3
        + \Big(
            \frac{175}{3}
            - 5\pi^2
        \Big)\logu
        + \frac{34}{3}\logu^2
        + 2 \logu^3
    \Big)
\Bigg]
\nn &
+ \eQ^2\big(
    \eq^2 \nq
    + \eqA^2 \nqA
\big)  N_c \Bigg[
    - \frac{2}{\eps^2}
    + \frac{5}{3\eps}
    + \Big(
        \frac{71}{6}
        + \frac{4}{3} \pi^2
        + \frac{26}{3}\logu
        + 2 \logu^2
    \Big) 
+ \eps \Big(
        \frac{581}{12}
        + \frac{97}{18} \pi^2
        \nn &   
        + 16 \zeta_3
        + \Big(
            \frac{103}{3}
            + 3\pi^2
        \Big)\logu
        + \frac{34}{3}\logu^2
        + 2 \logu^3
    \Big)
\Bigg]
\nn &
+ \eQ^4 \Bigg[\,
    \frac{9}{2\eps^2}
    + \frac{1}{\eps} \Big(
        \frac{45}{4}
        + 9 \logu
    \Big)
    + \Big(
        \frac{199}{8}
        - \frac{17}{4} \pi^2
        + 8\pi^2 \log2
        - 12 \zeta_3
        + \frac{45}{2}\logu
        + 9 \logu^2
    \Big)
\nn
&    + \eps \Big(
        \frac{677}{16}
        - \frac{205}{8} \pi^2
        + \frac{14}{5} \pi^4
        + 48\pi^2 \log2
        - 16\pi^2 \log^2 2
        - 8 \log^4 2
        - 135 \zeta_3
        - 192 \Li_4\!\big(\tfrac{1}{2}\big)
\nn
&        + \Big(
            \frac{199}{4}
            - \frac{17}{2}\pi^2
            + 16\pi^2 \log2
            - 24 \zeta_3
        \Big)\logu 
        + \frac{45}{2}\logu^2
        + 6 \logu^3
    \Big)
\Bigg]
\,,
\end{align}
where $n_h$ denotes the number of massive quark flavors, while $q$ and $\qA$ label two distinct massless quark flavors with electric charges $e_q$ and $e_{\qA}$, and multiplicities $n_q$ and $n_{\qA}$, respectively (hence $n_l = n_q + n_{\qA}$).
For the sake of readability, the lengthy three-loop-order QCD+QED expressions for $\ZmOS$ are collected in Appendix~\ref{appendix:ZmOS}.

Our pure QCD results for the on-shell quark mass renormalization constant $\ZmOS$ up to three loops are in exact agreement with the expressions presented in refs.~\cite{Melnikov:2000zc,Marquard:2007uj,Barnreuther:2013qvf}. 
Our QED results are formulated for two distinct massless quark flavors with electric charges $e_q$ and $e_{\qA}$, and multiplicities $n_q$ and $n_{\qA}$, respectively, thereby covering all phenomenologically relevant configurations in the SM.
The pure QED corrections at two-loop order and the mixed QCD–QED contributions at three-loop order contain only terms linear in the flavor multiplicities $n_h$, $\nq$, and $\nqA$. 
In contrast, at three-loop order in pure QED, quartic structures in the flavor multiplicities emerge, namely $n_h^2, n_h\nq, n_h\nqA, \nq^2, \nqA^2$, and $\nq\nqA$.

\subsection{On-shell heavy-quark wave-function renormalization constant $\ZQOS$ }
\label{sec:ZQOS}

The on-shell heavy-quark wave-function renormalization constant $\ZQOS$ up to three loops in mixed QCD+QED can be conveniently cast into the following form:
\begin{align}\label{eq:ZQOS_exp}
\ZQOS = 1 +
\sum_{l=1}^{3}
\sum^{i+j=l}_{i=0}
\as^i\, \aew^j\, Z_{Q,l}^{(i,j)}\,.
\end{align}
The one-loop QCD and QED on-shell quark wave-function renormalization constants read
\begin{align}
Z_{Q,1}^{(1,0)} = {}&
-C_F \Bigg[
\frac{3}{\eps}
+4
+3 \logu
+ \eps\Big(
8
+\frac{1}{4} \pi^2
+4 \logu
+\frac{3}{2} \logu^2
\Big)
\nn &
+ \eps^2\Big(
16
+\frac{1}{3} \pi^2
-\zeta_3
+\big(8+\frac{1}{4} \pi^2 \big)\logu
+2 \logu^2
+\frac{1}{2} \logu^3
\Big)
\Bigg]\,,
\nn
Z_{Q,1}^{(0,1)} = {} &
-\eQ^2\Bigg[
\frac{3}{\eps}
+4
+3 \logu
+ \eps\Big(
8
+\frac{1}{4} \pi^2
+4 \logu
+\frac{3}{2} \logu^2
\Big)
\nn &
+ \eps^2\Big(
16
+\frac{1}{3} \pi^2
-\zeta_3
+\big(8+\frac{1}{4} \pi^2 \big)\logu
+2 \logu^2
+\frac{1}{2} \logu^3
\Big)
\Bigg]\,.
\end{align}
The two-loop QCD+QED on-shell wave-function renormalization constants are given by
\begin{align}
Z_{Q,2}^{(2,0)} = {}&
C_F^2 \Bigg[\,
  \frac{9}{2 \eps^2}
  + \frac{1}{\eps} \Big(\frac{51}{4}+9 \logu\Big)
  + \frac{433}{8}
  - \frac{49}{4}\pi^2
  + 16\pi^2 \log2
  - 24 \zeta_3
  + \frac{51}{2}\logu
  + 9 \logu^2
\nn &
  + \eps\Big(
    \frac{211}{16}
    - \frac{339}{8}\pi^2
    + \frac{28}{5}\pi^4
    + 92\pi^2 \log2
    - 32\pi^2 \log^2 2
    - 16\log^4 2
    - 297 \zeta_3
    \nn &
    - 384\Li_4\!\big(\tfrac12\big)
    + \Big(
      \frac{433}{4}
      - \frac{49}{2}\pi^2
      + 32\pi^2 \log2
      - 48 \zeta_3
    \Big)\logu
    + \frac{51}{2}\logu^2
    + 6 \logu^3
  \Big)
\Bigg]
\nn &
+ C_A C_F \Bigg[
  \frac{11}{2 \eps^2}
  - \frac{127}{12 \eps}
  - \frac{1705}{24}
  + 5\pi^2
  - 8\pi^2 \log2
  + 12 \zeta_3
  - \frac{215}{6}\logu
  - \frac{11}{2}\logu^2
\nn &
  + \eps\Big(
    -\frac{9907}{48}
    + \frac{769}{72}\pi^2
    - \frac{14}{5}\pi^4
    - 46\pi^2 \log2
    + 16\pi^2 \log^2 2
    + 8\log^4 2
    + 129 \zeta_3
    \nn &
    + 192\Li_4\!\big(\tfrac12\big)
    + \Big(
      -\frac{2057}{12}
      + \frac{109}{12}\pi^2
      - 16\pi^2 \log2
      + 24 \zeta_3
    \Big)\logu
    - \frac{259}{6}\logu^2
    - \frac{11}{2}\logu^3
  \Big)
\Bigg]
\nn &
+ C_F n_h  \Bigg[\,
  \frac{1}{\eps}\Big(\frac12+2 \logu\Big)
  + \frac{947}{36}
  - \frac{5}{2}\pi^2
  + \frac{11}{3}\logu
  + 3 \logu^2
  + \eps\Big(
    \frac{17971}{216}
    - \frac{445}{36}\pi^2
    \nn &
    + 16\pi^2 \log2
    - \frac{170}{3}\zeta_3
    + \Big(
        \frac{1043}{18}
        - \frac{29}{6}\pi^2
      \Big)\logu
    + 5 \logu^2
    + \frac{7}{3}\logu^3
  \Big)
\,\Bigg]
\nn &
+ C_F n_l \Bigg[
  -\frac{1}{\eps^2}
  + \frac{11}{6 \eps}
  + \frac{113}{12}
  + \frac{2}{3}\pi^2
  + \frac{19}{3}\logu
  + \logu^2
  + \eps\Big(
    \frac{851}{24}
    + \frac{127}{36}\pi^2
    + 8 \zeta_3
    \nn &
    + \Big(
        \frac{145}{6}
        + \frac{3}{2}\pi^2
      \Big)\logu
    + \frac{23}{3}\logu^2
    + \logu^3
  \Big)
\Bigg]\,,
\nn
Z_{Q,2}^{(1,1)} ={} &
\eQ^2 C_F \Bigg[\, 
  \frac{9}{\eps^2}
  + \frac{1}{\eps} \Big(\frac{51}{2}+18 \logu\Big)
  + \frac{433}{4}
  - \frac{49}{2}\pi^2
  + 32\pi^2 \log2
  - 48 \zeta_3
  + 51 \logu
  + 18 \logu^2
\nn &\qquad
  + \eps \Big(
    \frac{211}{8}
    - \frac{339}{4}\pi^2
    + \frac{56}{5}\pi^4
    + 184\pi^2 \log2
    - 64\pi^2 \log^2 2
    - 32\log^4 2
    - 594 \zeta_3
    \nn &
    - 768\Li_4\!\big(\tfrac{1}{2}\big)
    + \Big(
        \frac{433}{2}
        - 49\pi^2
        + 64\pi^2 \log2
        - 96 \zeta_3
      \Big)\logu
    + 51 \logu^2
    + 12 \logu^3
  \Big)
\Bigg]\,,
\nn
Z_{Q,2}^{(0,2)} ={} &
\eQ^4n_h N_c\Bigg[\,
  \frac{1}{\eps}(1+4 \logu)
  + \frac{947}{18}
  - 5\pi^2
  + \frac{22}{3}\logu
  + 6 \logu^2
  + \eps \Big(
    \frac{17971}{108}
    - \frac{445}{18}\pi^2
    + 32\pi^2 \log2
    \nn &
    - \frac{340}{3}\zeta_3
    + \Big(
        \frac{1043}{9}
        - \frac{29}{3}\pi^2
      \Big)\logu
    + 10 \logu^2
    + \frac{14}{3}\logu^3
  \Big)
\Bigg]
\nn &
+ \eQ^2\,(\eq^2 \nq+\eqA^2 \nqA) N_c \Bigg[
  -\frac{2}{\eps^2}
  + \frac{11}{3\eps}
  + \frac{113}{6}
  + \frac{4}{3}\pi^2
  + \frac{38}{3}\logu
  + 2 \logu^2
  + \eps \Big(
    \frac{851}{12}
    + \frac{127}{18}\pi^2
    \nn &
    + 16 \zeta_3
    + \Big(
        \frac{145}{3}
        + 3\pi^2
      \Big)\logu
    + \frac{46}{3}\logu^2
    + 2 \logu^3
  \Big)
\Bigg]
\nn &
+ \eQ^4 \Bigg[\,
  \frac{9}{2\eps^2}
  + \frac{1}{\eps}\Big(\frac{51}{4}+9 \logu\Big)
  + \frac{433}{8}
  - \frac{49}{4}\pi^2
  + 16\pi^2 \log2
  - 24 \zeta_3
  + \frac{51}{2}\logu
  + 9 \logu^2
\nn &
  + \eps \Big(
    \frac{211}{16}
    - \frac{339}{8}\pi^2
    + \frac{28}{5}\pi^4
    + 92\pi^2 \log2
    - 32\pi^2 \log^2 2
    - 16\log^4 2
    - 297 \zeta_3
    \nn &
    - 384\Li_4\!\big(\tfrac{1}{2}\big)
    + \Big(
        \frac{433}{4}
        - \frac{49}{2}\pi^2
        + 32\pi^2 \log2
        - 48 \zeta_3
      \Big)\logu
    + \frac{51}{2}\logu^2
    + 6 \logu^3
  \Big)
\Bigg]
\, .
\end{align}
For the sake of readability, the lengthy three-loop-order QCD+QED expressions for the on-shell wave-function renormalization constant $\ZQOS$ of a heavy-quark are collected in Appendix~\ref{appendix:ZQOS}.
For completeness, the explicit expressions for the on-shell wave-function renormalization constant $\ZqOS$ of a massless quark with charge $\eq$, computed up to three-loop order in mixed QCD+QED, are presented in Appendix~\ref{appendix:ZqOS}.

The pure QCD part of our result for the on-shell heavy quark wave-function renormalization constant $\ZQOS$ up to three-loops is in exact agreement with the results provided in ref.~\cite{Melnikov:2000zc,Marquard:2007uj,Barnreuther:2013qvf}.
Note that our pure and mixed QED contributions are formulated for two distinct massless quark flavors with charges $e_q$ and $e_{\qA}$ and multiplicities $n_q$ and $n_{\qA}$, thereby covering all phenomenologically relevant SM configurations.
While the two-loop pure QED and three-loop mixed QCD–QED corrections are linear in the flavor multiplicities, quartic flavor structures arise at three loops in pure QED.

\subsection{Mass relation between the $\MSbar$ and the on-shell masses }
\label{sec:Zmff}

The quark masses in the on-shell and $\MSbar$ renormalization schemes are renormalized multiplicatively. The connection between renormalized and bare quark masses is defined as
\begin{align}\label{eq:mMS_def}
    m_B=\ZmOS \mOS = \ZmMSb \mMS\,,
\end{align}
where $m_B$ is the bare quark mass, and $\ZmOS$ and $\ZmMSb$ are the corresponding quark mass renormalization constants in the on-shell and $\MSbar$ scheme, respectively.
Owing to the infrared finiteness of the on-shell quark mass~\cite{Tarrach:1980up,Breckenridge:1994gs,Smith:1996xz,Kronfeld:1998di}, the finite mass relation between the $\MSbar$ and the on-shell quark mass is defined as the following ratio: 
\begin{align}\label{eq:Zff_def}
\Zff \equiv  
\Zff(\alpha_s,\, \alpha_e\,,\mu/\mOS) =
\frac{\mMS}{\mOS}
 =
\frac{\ZmOS(\alpha_s,\, \alpha_e\,,\mu/\mOS)}{ \ZmMSb(\alpha_s,\, \alpha_e)}\,,
\end{align}
which admits the following perturbative expansion up to three loops in mixed QCD+QED:
\begin{align}\label{eq:Zff_exp}
\Zff = 1 +
\sum_{l=1}^{3}
\sum^{i+j=l}_{i=0}
\as^i\, \aew^j\, z_{m,l}^{(i,j)}\,.
\end{align}

To calculate the mass relation $\Zff$ in mixed QCD+QED corrections at the three-loop order, we need to calculate the mass renormalization factors in both the on-shell and $\MSbar$ schemes at three-loop orders, respectively. 
In the previous subsection, we present the explicit analytic expression for the three-loop mixed QCD+QED quark mass renormalization constant in the on-shell scheme.
The $\MSbar$ quark mass renormalization constant has been computed up to four \cite{Chetyrkin:1997dh,Vermaseren:1997fq,Chetyrkin:2004mf} and five loops~\cite{Baikov:2014qja,Luthe:2016xec,Baikov:2017ujl}. 
However, the three-loop mixed QCD+QED quark mass renormalization constant in the $\MSbar$ scheme is still unknown.
Exploiting the defining form of $\MSbar$ renormalization constant, $\ZmMSb$ can be extracted from the explicit expression for $\ZmOS$ by constructing the following ansatz,
\begin{align}\label{eq:ZmMS_ansatz}
\ZmMSb
=
1
+
\sum_{l=1}^{3}
\sum_{k=1}^{l}
\frac{1}{\epsilon^k}
\sum^{i+j=l}_{i=0}
\as^i\, \aew^j\, Z_{\mMS,l,k}^{(i,j)}\,.
\end{align}
The unknown coefficients are fixed by requiring that the ratio
$\Zff$ is free of UV poles in dimensional regularization,
\begin{align}
\Zff \big|_{\epsilon^{-k}} = 0,
\qquad
\forall\, \quad
k \ge 1 \,.
\end{align}
This condition uniquely determines the $\MSbar$ mass renormalization constant $\ZmMSb$ and simultaneously yields a finite conversion factor $\Zff$ between the $\MSbar$ and on-shell masses.\\

In the following, we present the explicit analytic results for the perturbative expansion~\eqref{eq:Zff_exp} of the mass conversion factor $\Zff$.
The one-loop QCD and QED contributions read 
\begin{align}
z_{m,1}^{(1,0)} = 
-C_F \big( 4 + 3 \logu \big)\,, 
\qquad
z_{m,1}^{(0,1)} = 
-\eQ^{2} \big( 4 + 3 \logu \big)\,.
\end{align}
The two-loop  mixed QCD+QED contributions to the mass relation are given by
\begin{align}
z_{m,2}^{(2,0)} = {}
&
 C_F^2 \Big(
    \frac{7}{8}
    - 5\pi^2
    + 8\pi^2 \log2
    - 12 \zeta_3
    + \frac{21}{2} \logu
    + \frac{9}{2} \logu^2
\Big)
\nn
&
+ C_A C_F \Big(
    -\frac{1111}{24}
    + \frac{4}{3} \pi^2
    - 4\pi^2 \log2
    + 6 \zeta_3
    - \frac{185}{6} \logu
    - \frac{11}{2} \logu^2
\Big) 
\nn
&
+ C_F n_h \Big(
    \frac{143}{12}
    - \frac{4}{3} \pi^2
    + \frac{13}{3} \logu
    + \logu^2
  \Big)
  + C_F n_l \Big(
    \frac{71}{12}
    + \frac{2}{3} \pi^2
    + \frac{13}{3} \logu
    + \logu^2
  \Big)
\,,
\nn
z_{m,2}^{(1,1)} = {}&
\eQ^2 C_F \Big(
    \frac{7}{4}
    - 10\pi^2
    + 16\pi^2 \log2
    - 24 \zeta_3
    + 21 \logu
    + 9 \logu^2
\Big)\,,
\nn
z_{m,2}^{(0,2)} = {}&
\eQ^4 n_h N_c \Big(
    \frac{143}{6}
    - \frac{8}{3} \pi^2
    + \frac{26}{3} \logu
    + 2 \logu^2
\Big)
    \nn &
+\eQ^2 \big( \eq^2 \nq + \eqA^2 \nqA \big) N_c
\Big(
    \frac{71}{6}
    + \frac{4}{3} \pi^2
    + \frac{26}{3} \logu
    + 2 \logu^2
\Big)
    \nn &
+\eQ^4 \Big(
    \frac{7}{8}
    - 5\pi^2
    + 8\pi^2 \log2
    - 12 \zeta_3
    + \frac{21}{2} \logu
    + \frac{9}{2} \logu^2
\Big) \,.
\end{align}
At three-loop order, it is convenient to further decompose the pure QCD contribution $z_{m,3}^{(3,0)}$ defined in eq.~\eqref{eq:Zff_exp} into components corresponding to the different color structures:
\begin{align}
z_{m,3}^{(3,0)} = {}&
C_F^3\, z_{m,\sss{F^3}}^{(3,0)}
+ C_A C_F^2\, z_{m,\sss{AF^2}}^{(3,0)}
+ C_A^2 C_F\, z_{m,\sss{A^2F}}^{(3,0)}
+ C_F^2 n_h\, z_{m,\sss{F^2h}}^{(3,0)}
\nn 
&
+ C_F^2 n_l\, z_{m,\sss{F^2l}}^{(3,0)}
+ C_A C_F n_h\, z_{m,\sss{AFh}}^{(3,0)}
+ C_A C_F n_l\, z_{m,\sss{AFl}}^{(3,0)}
\nn 
&
+ C_F n_h^2\, z_{m,\sss{Fh^2}}^{(3,0)}
+ C_F n_l^2\, z_{m,\sss{Fl^2}}^{(3,0)}
+ C_F n_h n_l \, z_{m,\sss{Fhl}}^{(3,0)}\,,
\end{align}
where,
\begin{align}
z_{m,\sss{F^3}}^{(3,0)} = {}& 
- \frac{2969}{12}
  - \frac{613}{3} \pi^2
  - \frac{4}{3} \pi^4
  + 464 \pi^2 \log2
  + 32 \pi^2 \log^2 2
  - 32 \log^4 2
  - 324 \zeta_3
  - 4 \pi^2 \zeta_3
  \nn &
  + 40 \zeta_5
  - 768\Li_4\!\big(\tfrac{1}{2}\big)
  + \Big(
      - \frac{489}{8}
      + 15 \pi^2
      - 24 \pi^2 \log2
      + 36 \zeta_3
  \Big)\logu 
  - \frac{27}{2} \logu^2
  - \frac{9}{2} \logu^3
  \,,
  \nn 
 z_{m,\sss{AF^2}}^{(3,0)} = {}
 &
 \frac{13189}{72}
  + \frac{509}{9} \pi^2
  + \frac{260}{27} \pi^4
  - \frac{248}{9} \pi^2 \log2
  - \frac{496}{9} \pi^2 \log^2 2
  - \frac{32}{9} \log^4 2
  - \frac{1546}{3} \zeta_3
  \nn 
  &
  - 76 \pi^2 \zeta_3
  + 180 \zeta_5
  - \frac{256}{3}\Li_4\!\big(\tfrac{1}{2}\big)
  +\Big(
      \frac{5813}{24}
      - \frac{122}{3} \pi^2
      + \frac{212}{3} \pi^2 \log2
      - 106 \zeta_3
  \Big) \logu 
  \nn 
  &
  + 109 \logu^2
  + \frac{33}{2} \logu^3
  \,, \nn
  z_{m,\sss{A^2F}}^{(3,0)} = {} &
-\frac{1322545}{1944}
  - \frac{1955}{54} \pi^2
  - \frac{179}{54} \pi^4
  - \frac{920}{9} \pi^2 \log2
  + \frac{176}{9} \pi^2 \log^2 2
  + \frac{88}{9} \log^4 2
  \nn
  &
  + \frac{2686}{9} \zeta_3
  + 51 \pi^2 \zeta_3
  - 130 \zeta_5
  + \frac{704}{3}\Li_4\!\big(\tfrac{1}{2}\big)
  + \Big(
      -\frac{13243}{27}
      + \frac{88}{9} \pi^2
      - \frac{88}{3} \pi^2 \log2
      \nn &
      + 44 \zeta_3
  \Big)\logu 
  - \frac{2341}{18} \logu^2
  - \frac{121}{9} \logu^3 
  \,, \nn
 z_{m,\sss{F^2h}}^{(3,0)} = {} &
  \frac{1067}{18}
  - \frac{680}{27} \pi^2
  + \frac{182}{135} \pi^4
  + \frac{256}{9} \pi^2 \log2
  - \frac{32}{9} \pi^2 \log^2 2
  + \frac{32}{9} \log^4 2
  - \frac{212}{3} \zeta_3
  \nn 
  &
  + \frac{256}{3}\Li_4\!\big(\tfrac{1}{2}\big)
  + \Big(
      -\frac{151}{12}
      + \frac{32}{3} \pi^2
      - \frac{32}{3} \pi^2 \log2
      - 8 \zeta_3
  \Big)\logu 
  - 13 \logu^2
  - 3 \logu^3
  \,, \nn
  z_{m,\sss{F^2l}}^{(3,0)} = {} &
  \frac{1283}{18}
  + \frac{208}{9} \pi^2
  - \frac{238}{135} \pi^4
  - \frac{352}{9} \pi^2 \log2
  + \frac{64}{9} \pi^2 \log^2 2
  + \frac{32}{9} \log^4 2
  + \frac{220}{3} \zeta_3
  \nn &
  + \frac{256}{3}\Li_4\!\big(\tfrac{1}{2}\big)
  + \Big(
      \frac{65}{12}
      + \frac{14}{3} \pi^2
      - \frac{32}{3} \pi^2 \log2
      - 8 \zeta_3
  \Big)\logu 
  - 13 \logu^2
  - 3 \logu^3
  \,, \nn
z_{m,\sss{AFh}}^{(3,0)} = {}&
\frac{144959}{486}
  - \frac{898}{9} \pi^2
  - \frac{172}{135} \pi^4
  + \frac{1024}{9} \pi^2 \log2
  + \frac{16}{9} \pi^2 \log^2 2
  - \frac{16}{9} \log^4 2
  - \frac{218}{9} \zeta_3
  \nn &
  + 4 \pi^2 \zeta_3
  - 20 \zeta_5
  - \frac{128}{3}\Li_4\!\big(\tfrac{1}{2}\big)
  + \Big(
      \frac{4664}{27}
      - \frac{104}{9} \pi^2
      + \frac{16}{3} \pi^2 \log2
      + 16 \zeta_3
  \Big)\logu 
  \nn &
  + \frac{373}{9} \logu^2
  + \frac{44}{9} \logu^3
  \,, \nn
z_{m,\sss{AFl}}^{(3,0)} = {} &
\frac{70763}{486}
  + \frac{350}{27} \pi^2
  + \frac{38}{135} \pi^4
  + \frac{176}{9} \pi^2 \log2
  - \frac{32}{9} \pi^2 \log^2 2
  - \frac{16}{9} \log^4 2
  + \frac{178}{9} \zeta_3
  \nn &
  - \frac{128}{3}\Li_4\!\big(\tfrac{1}{2}\big)
  + \Big(
      \frac{3476}{27}
      + \frac{28}{9} \pi^2
      + \frac{16}{3} \pi^2 \log2
      + 16 \zeta_3
  \Big)\logu 
  + \frac{373}{9} \logu^2
  + \frac{44}{9} \logu^3
  \,, \nn
z_{m,\sss{Fh^2}}^{(3,0)} = {} &
-\frac{9481}{486}
  + \frac{64}{135} \pi^2
  + \frac{88}{9} \zeta_3
  + \Big(
      -\frac{394}{27}
      + \frac{16}{9} \pi^2
  \Big)\logu 
  - \frac{26}{9} \logu^2
  - \frac{4}{9} \logu^3
  \,, \nn
z_{m,\sss{Fl^2}}^{(3,0)} = {} &
 -\frac{2353}{486}
  - \frac{52}{27} \pi^2
  - \frac{56}{9} \zeta_3
  + \Big(
      -\frac{178}{27}
      - \frac{8}{9} \pi^2
  \Big)\logu 
  - \frac{26}{9} \logu^2
  - \frac{4}{9} \logu^3 
\,, \nn
z_{m,\sss{Fhl}}^{(3,0)} = {} &
-\frac{5917}{243}
  + \frac{52}{27} \pi^2
  + \frac{32}{9} \zeta_3
  + \Big(
      -\frac{572}{27}
      + \frac{8}{9} \pi^2
  \Big)\logu 
  - \frac{52}{9} \logu^2
  - \frac{8}{9} \logu^3
  \,.
\end{align}
The mixed QCD+QED corrections at three-loops are given by
\begin{align}
z_{m,3}^{(2,1)} = {}& 
\eQ^2 C_F^2 \Bigg[
    -\frac{2969}{4}
    - 613 \pi^2
    - 4 \pi^4
    + 1392 \pi^2 \log2
    + 96 \pi^2 \log^2 2
    - 96 \log^4 2
    - 972 \zeta_3
    - 12 \pi^2 \zeta_3
  \nn
    &
    + 120 \zeta_5
    - 2304\Li_4\!\big(\tfrac{1}{2}\big)
    - \Big(
        \frac{1467}{8}
        - 45 \pi^2
        + 72 \pi^2 \log2
        - 108 \zeta_3
    \Big)\logu 
    - \frac{81}{2} \logu^2
    - \frac{27}{2} \logu^3
\, \Bigg]
\nn &
+
\eQ^2 C_A C_F\Bigg[\,
    \frac{13189}{72}
    + \frac{509}{9} \pi^2
    + \frac{260}{27} \pi^4
    - \frac{248}{9} \pi^2 \log2
    - \frac{496}{9} \pi^2 \log^2 2
    - \frac{32}{9} \log^4 2
        \nn &
    - \frac{1546}{3} \zeta_3
    - 76 \pi^2 \zeta_3
    + 180 \zeta_5
    - \frac{256}{3}\Li_4\!\big(\tfrac{1}{2}\big)
+ \Big(
        \frac{5813}{24}
        - \frac{122}{3} \pi^2
        + \frac{212}{3} \pi^2 \log2
        - 106 \zeta_3
    \Big)\logu 
     \nn &
    + 109 \logu^2
    + \frac{33}{2} \logu^3
\Bigg] 
\nn &
+
\eQ^2 n_h  C_F \Bigg[\,
    \frac{1067}{18}
    - \frac{680}{27} \pi^2
    + \frac{182}{135} \pi^4
    + \frac{256}{9} \pi^2 \log2
    - \frac{32}{9} \pi^2 \log^2 2
    + \frac{32}{9} \log^4 2
\nn &
    - \frac{212}{3} \zeta_3
    + \frac{256}{3}\Li_4\!\big(\tfrac{1}{2}\big)
    + \Big(
        -\frac{151}{12}
        + \frac{32}{3} \pi^2
        - \frac{32}{3} \pi^2 \log2
        - 8 \zeta_3
    \Big)\logu 
    - 13 \logu^2
    - 3 \logu^3
\,\Bigg]
\nn &
+
\eQ^2 n_l C_F \Bigg[\,
    \frac{85}{36}
    + \frac{190}{9} \pi^2
    - \frac{44}{27} \pi^4
    - \frac{352}{9} \pi^2 \log2
    + \frac{64}{9} \pi^2 \log^2 2
    + \frac{32}{9} \log^4 2
    + \frac{352}{3} \zeta_3
 \nn &   
    + \frac{256}{3}\Li_4\!\big(\tfrac{1}{2}\big)
    + \Big(
        -\frac{301}{12}
        + \frac{14}{3} \pi^2
        - \frac{32}{3} \pi^2 \log2
        + 16 \zeta_3
    \Big)\logu 
    - 16 \logu^2
    - 3 \logu^3
\, \Bigg]
\nn &
+
(\eq^2\nq +\eqA^2\nqA)C_F \Bigg[\,
    \frac{827}{12}
    + 2 \pi^2
    - \frac{2}{15} \pi^4
    - 44 \zeta_3
    + \Big(
        \frac{61}{2}
        - 24 \zeta_3
    \Big)\logu 
    + 3 \logu^2
\,\Bigg]\,,
\end{align}
and
\begin{align}
 z_{m,3}^{(1,2)} = {}&   
 \eQ^4 n_h N_c C_F \Bigg[\,
    \frac{1067}{9}
    - \frac{1360}{27} \pi^2
    + \frac{364}{135} \pi^4
    + \frac{512}{9} \pi^2 \log2
    - \frac{64}{9} \pi^2 \log^2 2
    + \frac{64}{9} \log^4 2
\nn &
    - \frac{424}{3} \zeta_3
    + \frac{512}{3}\Li_4\!\big(\tfrac{1}{2}\big)
    + \Big(
        -\frac{151}{6}
        + \frac{64}{3} \pi^2
        - \frac{64}{3} \pi^2 \log2
        - 16 \zeta_3
    \Big)\logu 
    - 26 \logu^2
    - 6 \logu^3
\,\Bigg]
\nn &
+
\eQ^2(\eq^2\nq+\eqA^2\nqA) N_c C_F \Bigg[\,
    \frac{1283}{9}
    + \frac{416}{9} \pi^2
    - \frac{476}{135} \pi^4
    + \frac{440}{3} \zeta_3
    + \frac{512}{3}\Li_4\!\big(\tfrac{1}{2}\big)
    - \frac{704}{9} \pi^2 \log2
        \nn &
    + \frac{128}{9} \pi^2 \log^2 2
    + \frac{64}{9} \log^4 2
    + \Big(
        \frac{65}{6}
        + \frac{28}{3} \pi^2
        - \frac{64}{3} \pi^2 \log2
        - 16 \zeta_3
    \Big)\logu 
    - 26 \logu^2
    - 6 \logu^3
\,\Bigg]
    \nn &
+
\eQ^4 C_F\Bigg[
    - \frac{2969}{4}
    - 613 \pi^2
    - 4 \pi^4
    + 1392 \pi^2 \log2
    + 96 \pi^2 \log^2 2
    - 96 \log^4 2
    - 972 \zeta_3
    \nn &
    - 12 \pi^2 \zeta_3
    + 120 \zeta_5
    - 2304\Li_4\!\big(\tfrac{1}{2}\big)
    + \Big(
        - \frac{1467}{8}
        + 45 \pi^2
        - 72 \pi^2 \log2
        + 108 \zeta_3
    \Big)\logu 
        \nn &
    - \frac{81}{2} \logu^2
    - \frac{27}{2} \logu^3
\,\Bigg]\,.
\end{align}
At three-loop order in pure QED, quartic structures in the flavor multiplicities arise, specifically $n_h^2$, $n_h\nq$, $n_h\nqA$, $\nq^2$, $\nqA^2$, and $\nq\nqA$.
In view of the corresponding electric-charge factors, it is thus natural to organize $z_{m,3}^{(0,3)}$, defined in eq.~\eqref{eq:Zff_exp}, into separate contributions classified by the underlying quark-flavor structures as follows:
\begin{align}
z_{m,3}^{(0,3)} = {}&
\eQ^6  n_h^2 N_c^2 z^{(0,3)}_{h^2}
+ \eQ^4 \eq^2 n_h \nq N_c^2  z^{(0,3)}_{hq}
+ \eQ^4 \eqA^2 n_h\nqA N_c^2 z^{(0,3)}_{h\qA}
+ \eQ^2 \eq^4 \nq^2 N_c^2 z^{(0,3)}_{q^2}
\nn
&
+ \eQ^2 \eqA^4 \nqA^2 N_c^2  z^{(0,3)}_{{\qA}^2}
+ \eQ^2 \eq^2  \eqA^2 \nq\nqA N_c^2 z^{(0,3)}_{q\qA}
+ \eQ^2 \eq^4 \nq N_c  z^{(0,3)}_{\eq^4}
+ \eQ^2 \eqA^4 \nqA N_c z^{(0,3)}_{\eqA^4}
\nn 
&
+ \eQ^4 \eq^2 \nq N_c  z^{(0,3)}_{\eq^2}
+ \eQ^4 \eqA^2 \nqA N_c z^{(0,3)}_{\eqA^2}
+ \eQ^6 n_h N_c z^{(0,3)}_{h} 
+ \eQ^6 z^{(0,3)}_{Q}\,,
\end{align}
where,
\begin{align}
 z^{(0,3)}_{h^2} = {}&
 -\frac{18962}{243}
  + \frac{256}{135} \pi^2
  + \frac{352}{9} \zeta_3
  +\Big(
      -\frac{1576}{27}
      + \frac{64}{9} \pi^2
    \Big) \logu 
  - \frac{104}{9} \logu^2
  - \frac{16}{9} \logu^3
\,, \nn
z^{(0,3)}_{hq} = {} &
 -\frac{23668}{243}
  + \frac{208}{27} \pi^2
  + \frac{128}{9} \zeta_3
  + \Big(
      -\frac{2288}{27}
      + \frac{32}{9} \pi^2
    \Big)\logu 
  - \frac{208}{9} \logu^2
  - \frac{32}{9} \logu^3
\,, \nn
z^{(0,3)}_{h\qA} = {} &
z^{(0,3)}_{hq} 
\,, \nn 
z^{(0,3)}_{q^2} = {} &
-\frac{4706}{243}
  - \frac{208}{27} \pi^2
  - \frac{224}{9} \zeta_3
  +  \Big(
      -\frac{712}{27}
      - \frac{32}{9} \pi^2
    \Big)\logu
  - \frac{104}{9} \logu^2
  - \frac{16}{9} \logu^3
\,, \nn
z^{(0,3)}_{{\qA}^2} = {}
& z^{(0,3)}_{q^2} 
\,, \nn
z^{(0,3)}_{q\qA} = {}
& 
-\frac{9412}{243}
   - \frac{416}{27} \pi^2
   - \frac{448}{9} \zeta_3
   + \Big(
       -\frac{1424}{27}
       - \frac{64}{9} \pi^2
     \Big)\logu 
   - \frac{208}{9} \logu^2
   - \frac{32}{9} \logu^3
\,, \nn
z^{(0,3)}_{\eq^4}  = {}
& 
\frac{827}{6}
+ 4 \pi^2
- \frac{4}{15} \pi^4
- 88 \zeta_3
+ \logu \big( 61 - 48 \zeta_3 \big)
+ 6 \logu^2
\,, \nn
 z^{(0,3)}_{\eqA^4} = {}&
 z^{(0,3)}_{\eq^4} 
 \,, \nn
z^{(0,3)}_{\eq^2} = {}
& 
\frac{85}{18}
+ \frac{380}{9} \pi^2
- \frac{88}{27} \pi^4
- \frac{704}{9} \pi^2 \log2
+ \frac{128}{9} \pi^2 \log^2 2
+ \frac{64}{9} \log^4 2
+ \frac{704}{3} \zeta_3
+ \frac{512}{3}\Li_4\!\big(\tfrac{1}{2}\big)
\nn &
+ \Big(
-\frac{301}{6}
+ \frac{28}{3} \pi^2
- \frac{64}{3} \pi^2 \log2
+ 32 \zeta_3
\Big)\logu 
- 32 \logu^2
- 6 \logu^3
\,, \nn
 z^{(0,3)}_{\eqA^2} = {}&
 z^{(0,3)}_{\eq^2} 
 \,, \nn
z^{(0,3)}_{h} = {}
& 
\frac{1067}{9}
- \frac{1360}{27} \pi^2
+ \frac{364}{135} \pi^4
+ \frac{512}{9} \pi^2 \log2
- \frac{64}{9} \pi^2 \log^2 2
+ \frac{64}{9} \log^4 2
- \frac{424}{3} \zeta_3
\nn &
+ \frac{512}{3}\Li_4\!\big(\tfrac{1}{2}\big)
+ \Big(
-\frac{151}{6}
+ \frac{64}{3} \pi^2
- \frac{64}{3} \pi^2 \log2
- 16 \zeta_3
\Big)\logu 
- 26 \logu^2
- 6 \logu^3
\,, \nn
z^{(0,3)}_{Q} = {}
& 
-\frac{2969}{12}
- \frac{613}{3} \pi^2
- \frac{4}{3} \pi^4
+ 464 \pi^2 \log2
+ 32 \pi^2 \log^2 2
- 32 \log^4 2
- 324 \zeta_3
- 4 \pi^2 \zeta_3
\nn 
&
+ 40 \zeta_5
- 768\Li_4\!\big(\tfrac{1}{2}\big)
+ \Big(
-\frac{489}{8}
+ 15 \pi^2
- 24 \pi^2 \log2
+ 36 \zeta_3
\Big)\logu 
- \frac{27}{2} \logu^2
- \frac{9}{2} \logu^3
\,.
\end{align}
The explicit expression for the $\MSbar$ quark mass renormalization constant $\ZmMSb$ is derived along the way and  presented in appendix~\ref{appendix:ZmMS}.
Our pure QCD results of $\Zff$ up to three-loops are in exact agreement with the expressions provided in ref.~\cite{Melnikov:2000qh} as implemented in \texttt{RunDec}~\cite{Chetyrkin:2000yt,Herren:2017osy}.

\subsection{$\sigma$-mass for heavy quark in mixed QCD+QED}
\label{sec:sigma_Mass}

Refs.~\cite{Chen:2025iul,Chen:2025zfa} proposed recently a new gauge-invariant, regularization/renormalization scheme- and scale-independent definition for quark mass, which is proven~\cite{Chen:2025zfa} to be free from the leading IR-renormalon ambiguity by showing that the leading IR-renormalon divergence in the perturbative pole mass of a massive quark~\cite{Bigi:1994em,Beneke:1994sw,Beneke:1994rs,Smith:1996xz} resides entirely in the contribution from the trace anomaly in QCD. 
In the context of Standard Model, this $\sigma$-mass may be interpreted as a residual Higgs-generated mass for an on-shell heavy quark obtained by subtracting away the trace-anomaly contribution from its perturbative pole mass. 
Therefore, this mass definition nicely combines the merits of both the perturbative pole-mass and $\MSbar$-mass definition, while elegantly circumventing their respective unappealing and undesirable features.
The explicit three-loop QCD perturbative results for the mass ratio between $\sigma$-mass and pole mass, as well as the ratio between $\sigma$-mass and $\MSbar$ mass are presented in ref.~\cite{Chen:2025iul}.
We are now ready to extend the results further to take into account the mixed QCD+QED effects at three-loop order.

In general, the on-shell wave-function renormalization constants (with mixed QCD and QED corrections) are needed to compute the on-shell matrix elements of the trace-anomaly operator, responsible for subtracting a particular class of divergences associated with external on-shell legs. 
However, as far as the determination of the relation between the $\SigmaMass$ and on-shell mass for a heavy quark is concerned, it is also feasible to take a short-cut approach by extending the formula for $\SigmaMass$ derived in ref.~\cite{Chen:2025zfa} to take into account the presence of QED interactions in addition to QCD interactions.
To this end, the anomalous dimensions of the QCD and QED couplings (up to two-loop), as well as the mass (up to three-loop), shall be involved as well; however, this information can be extracted from the coupling and mass renormalization constants which are needed anyway in this computation.
\\

If the QED interaction is turned on in addition to the QCD interaction considered in ref.~\cite{Chen:2025zfa}, it is not hard to see that the formula for the $\SigmaMass$ mass of a massive quark remains virtually the same, namely 
\begin{equation}\label{eq:sigmass_formula}
\SigmaMass = 
 \mOS \, \Big( 1 + \frac{\overline{m}}{\ZpOm} \frac{\partial\, \ZpOm\big(\alpha_s,\, \alpha_e\,, \frac{\mu}{\overline{m}} \big) }{ \partial\, \overline{m} } \Big)\,,    
\end{equation}
where the new finite pole-to-$\MSbar$ mass conversion factor $\ZpOm$ is defined as the inverse of $\Zff$ introduced in eq.~\eqref{eq:Zff_def}, 
\begin{equation}\label{eq:Zpm_def}
\ZpOm\equiv  \ZpOm(\alpha_s,\,\alpha_e\,, \mu/\overline{m}) = \frac{\mOS}{\overline{m}} = \frac{Z_{\overline{m}} (\alpha_s,\, \alpha_e\,)}{ \ZmOS(\alpha_s,\, \alpha_e\,,\mu/\mOS)} \Big|_{\epsilon \rightarrow 0}\,.    
\end{equation}
We note that the explicit logarithmic mass-dependence in $\ZpOm$ used in \eqref{eq:sigmass_formula} must be expressed using $\overline{m}$, rather than $\mOS$ (which would result in a different explicit $\mu$-dependence in the same ratio).
Repeating the derivation as in ref.~\cite{Chen:2025zfa}, we eventually end up with the following more explicit form for the ratio of $\SigmaMass$ to $\mOS$, 
\begin{eqnarray}\label{eq:TASmass_formula}
Z_{\sigma} \, \equiv \, \frac{\SigmaMass}{\mOS} 
\, = \,
\frac{1}{1 - 2 \gamma_m }
\,+\, 2\, 
\frac{
\beta_s\, \frac{\partial \ln\big( \ZpOm\big(\alpha_s\,,\alpha_e\,, \frac{\mu}{\overline{m}}\big) \big)}{\partial \ln\big(\alpha_s\big)}
\,+\,
\beta_e\, \frac{\partial \ln\big( \ZpOm\big(\alpha_s\,,\alpha_e\,, \frac{\mu}{\overline{m}}\big) \big)}{\partial \ln\big(\alpha_e\big)}\,
}{1 - 2 \gamma_m}
\end{eqnarray} 
in terms of the anomalous dimensions of the $\MSbar$-renormalized couplings and mass in the presence of the QCD+QED interactions (but with just one quark kept massive). 
An appealing feature of \eqref{eq:TASmass_formula} is that, owing to the leading perturbative term of $\beta_s$ and $\beta_e$ being either $\mathcal{O}(\alpha_s)$ or $\mathcal{O}(\alpha_e)$, the perturbative result for $Z_{\sigma}$ at $N$-loop involves the perturbative expression of $\ZpOm$ only up to one loop-order less; and similarly for $\beta_s$ and $\beta_e$ since the logarithmic derivatives of $\ZpOm$ in couplings start from one-loop order too.
As is well-known, the explicit expressions for the anomalous dimensions $\beta_s$ and $\beta_e$ can be directly read-off from the $1/\epsilon$ poles of the respective $\MSbar$ renormalization constants $Z_{\as}$ and $Z_{\aew}$ presented in eq.~\eqref{eq:aR_expr}, and similarly the quark-mass anomalous dimension $\gamma_m$ from the $\ZmMSb$ given in appendix~\ref{appendix:ZmMS}.

With all ingredients ready at our disposal, we obtain the following explicit expression for the ratio between $m_{\sigma}$ and $\mOS$ up to three-loop orders in QCD+QED,
\begin{align}\label{eq:Zsigma_exp}
Z_{\sigma} = {}&
1
- \as (6C_F) - \aew (6\eQ^{2})
\nn &
+ 
\as^2 \Bigg[\,
33 C_F^{2}
+ C_A C_F
\Big(
- \frac{185}{3}
- 22 \logums
\Big)
+ C_F (n_h + n_l)
\Big(
\frac{26}{3}
+ 4 \logums
\Big)
\,\Bigg]
\nn &
+ \as\aew \Bigg[
66 \eQ^{2}C_F
 \Bigg]
+ 
\aew^2\Bigg[\,
33 \eQ^{4}
+ \eQ^{2}\,
\big( \eQ^{2} n_h + \eq^{2} \nq + \eqA^{2} \nqA \big)N_c\,
\Big( \frac{52}{3} + 8 \logums \Big)
\, \Bigg]
\nn &
+
\as^3\Bigg[
-309 C_F^3
+
C_A C_F^2 \Big(
  876
  - \frac{220}{3}\pi^2
  + \frac{352}{3}\pi^2 \log2
  - 176 \zeta_3
  + 374 \logums
\Big)
\nn &
+ C_A^2 C_F \Big(
  -\frac{26486}{27}
  + \frac{176}{9}\pi^2
  - \frac{176}{3}\pi^2 \log2
  + 88 \zeta_3
  - \frac{4682}{9} \logums
  - \frac{242}{3} \logums^2
\Big)
\nn &
+ C_F^2 (n_h+n_l) \Big(
  -55
  + \frac{40}{3}\pi^2
  - \frac{64}{3}\pi^2 \log2
  - 16 \zeta_3
  - 56 \logums
\Big)
\nn &
+ C_A C_F n_h \Big(
  \frac{9328}{27}
  - \frac{208}{9}\pi^2
  + \frac{32}{3}\pi^2 \log2
  + 32 \zeta_3
  + \frac{1492}{9} \logums
  + \frac{88}{3} \logums^2
\Big)
\nn &
+ C_A C_F n_l \Big(
  \frac{6952}{27}
  + \frac{56}{9}\pi^2
  + \frac{32}{3}\pi^2 \log2
  + 32 \zeta_3
  + \frac{1492}{9} \logums
  + \frac{88}{3} \logums^2
\Big)
\nn &
- C_F n_h^2 \Big(
  \frac{788}{27}
  - \frac{32}{9}\pi^2
  + \frac{104}{9} \logums
  + \frac{8}{3} \logums^2
\Big)
- C_F n_l^2 \Big(
  \frac{356}{27}
  + \frac{16}{9}\pi^2
  + \frac{104}{9} \logums
  + \frac{8}{3} \logums^2
\Big)
\nn &
- C_F n_h n_l \Big(
  \frac{1144}{27}
  - \frac{16}{9}\pi^2
  + \frac{208}{9} \logums
  + \frac{16}{3} \logums^2
\Big)
\Bigg]
\nn &
+ \as^2\aew \Bigg[
  -927 \eQ^{2}C_F^2
+ \eQ^{2}C_A C_F\Big(
  876
  - \frac{220}{3}\pi^{2}
  + \frac{352}{3}\pi^{2}\log 2
  - 176 \zeta_3
  + 374 \logums
\Big)
\nn &
+ \eQ^{2}n_h C_F \Big(
  -55
  + \frac{40}{3}\pi^{2}
  - \frac{64}{3}\pi^{2}\log2
  - 16 \zeta_3
  - 56 \logums
\Big)
\nn &
+ (\eq^{2}\nq + \eqA^{2}\nqA) C_F \Big(
  61
  - 48 \zeta_3
  + 12 \logums
\Big)
\nn &
+ \eQ^{2} n_l C_F  \Big(
  -116
  + \frac{40}{3}\pi^{2}
  - \frac{64}{3}\pi^{2}\log2
  + 32 \zeta_3
  - 68 \logums
\Big)
\Bigg]
\nn &
+ \as\aew^2\Bigg[
\eQ^{4} n_h N_c C_F\Big(
  -110
  + \frac{80}{3}\pi^{2}
  - \frac{128}{3}\pi^{2}\log 2
  - 32 \zeta_3
  - 112 \logums
\Big)
\nn &
+ \eQ^{2}(\eq^{2}\nq+ \eqA^{2}\nqA )N_c C_F \Big(
  -110
  + \frac{80}{3}\pi^{2}
   - \frac{128}{3}\pi^{2}\log 2
  - 32 \zeta_3
  - 112 \logums
\Big)
\nn &
-927 \eQ^{4} C_F
 \Bigg]
 \nn &
 + \aew^3\Bigg[\,
 \eQ^{6}n_h^{2} N_c^{2}\Big(
  -\frac{3152}{27}
  + \frac{128}{9}\pi^{2}
  - \frac{416}{9}\logums
  - \frac{32}{3}\logums^{2}
\Big)
\nn &
+ \eQ^{4}n_h(\eq^{2} \nq + \eqA^{2} \nqA) N_c^{2}\Big(
  -\frac{4576}{27}
  + \frac{64}{9}\pi^{2}
  - \frac{832}{9}\logums
  - \frac{64}{3}\logums^{2}
\Big)
\nn &
+ \eQ^{2}(\eq^{4}\nq^{2}+\eqA^{4}\nqA^{2})N_c^{2}\Big(
  -\frac{1424}{27}
  - \frac{64}{9}\pi^{2}
  - \frac{416}{9}\logums
  - \frac{32}{3}\logums^{2}
\Big)
\nn &
+ \eQ^{2}\eq^{2}\eqA^{2}\nq \nqA N_c^{2}\Big(
  -\frac{2848}{27}
  - \frac{128}{9}\pi^{2}
  - \frac{832}{9}\logums
  - \frac{64}{3}\logums^{2}
\Big)
\nn &
+ 
  \eQ^{2}(\eq^{4}\nq + \eqA^{4}\nqA )N_c\Big(
    122
    - 96 \zeta_3
    + 24 \logums
  \Big)
  \nn &
  + \eQ^{4}(\eq^{2}\nq + \eqA^{2}\nqA )N_c\Big(
    -232
    + \frac{80}{3}\pi^{2}
    - \frac{128}{3}\pi^{2}\log 2
    + 64 \zeta_3
    - 136 \logums
  \Big)
\nn &
+ \eQ^{6}n_h N_c\Big(
  -110
  + \frac{80}{3}\pi^{2}
  - \frac{128}{3}\pi^{2}\log 2
  - 32 \zeta_3
  - 112 \logums
\Big)
\nn &
- 309 \eQ^{6}
 \,\Bigg]\,,
\end{align}
where $\logums \equiv \log (\mu^2/\mMS^2)$.
The pure QCD part of $Z_{\sigma}$ is in full agreement with the result provided in ref.~\cite{Chen:2025iul}.

With the relationship~\eqref{eq:TASmass_formula}, one can also readily obtain the conversion factor from the $\mMS$ to $\SigmaMass$:
\begin{eqnarray}\label{eq:TASmass_over_MSbar}
Z_{\bar{\sigma}}
\equiv
\frac{\SigmaMass}{\mMS} =Z_{\sigma}\, \ZpOm 
\, = \,
\frac{\ZpOm \,+\, 2\,
\beta_s\, \alpha_s\, \frac{\partial\,  \ZpOm\big(\alpha_s\,, \alpha_e\,,\mu/\overline{m}\big)}{\partial\, \alpha_s} 
+
2\beta_e\, \alpha_e\, \frac{\partial\,  \ZpOm\big(\alpha_s\,,\alpha_e\,, \mu/\overline{m}\big)}{\partial\, \alpha_e}
}{1 - 2 \gamma_m }\,.
\end{eqnarray}
Here the explicit logarithmic dependence of $\ZpOm$ is expressed in terms of $\logums=\log(\mu^2/\mMS^2)$, as in eq.~\eqref{eq:sigmass_formula}.  
The explicit expression of 3-loop results for the $\MSbar$ quark mass and $\sigma$-mass relation constant $Z_{\bar{\sigma}}$ is presented in appendix~\ref{appendix:Zsigmabar}.
Analogously to $Z_{\sigma}$, the pure QCD part of $Z_{\bar{\sigma}}$ is also in full agreement with the result reported in Ref.~\cite{Chen:2025iul}.

\subsection{Numerical impacts on mass relations}
\label{sec:num_impact}

To illustrate explicitly the numerical impact of the QED corrections presented in the preceding sections on the heavy-quark mass relations, we give below explicit numbers for various quark masses evaluated at our chosen benchmark settings. 
These numbers are intended to give a quick idea on the size and hence potential impact of the QED effects on the universal mass conversion factors, rather than to constitute a dedicated phenomenological mass extraction. 
Specifically, we consider the pole-to-$\MSbar$ and pole-to-$\sigma$ mass conversions for the $t$-quark, as well as the $\MSbar$-to-pole $\MSbar$-to-$\sigma$ mass conversions for the $b$- and $c$-quarks, evaluated by means of the three-loop QCD+QED conversion relations derived in the preceding subsections.

Let us first consider the case of $t$-quark.
Taking as input the current PDG-average value of the $t$-quark mass $m_t^{\mathrm{os}} = 172.60 \pm 0.27$~GeV~\cite{ParticleDataGroup:2026} as the pole mass (setting aside the dispute over this theoretical interpretation), we obtain the following values for the $\MSbar$ mass and $\sigma$-mass of the $t$-quark:
\begin{align} \label{eq:mtMSmtMS}
   \mMS_t(\mMS_t) = 162.78(26)~\text{GeV}\,,
   \qquad
   \SigmaMass^t = 158.70(25)~\text{GeV}\,,
\end{align}
using the perturbative relations~\eqref{eq:Zff_exp} and~\eqref{eq:Zsigma_exp} with $\as(\mu)$ and $\aew(\mu)$ at the scale $m_t^{\mathrm{os}}$.
Here $\mMS_t(\mMS_t)$ denotes the scale-invariant $\MSbar$ mass, namely $\mMS_t(\mu)$ evaluated at its own scale $\mu = \mMS_t$, obtained iteratively from the pole-to-$\MSbar$ relation with the running couplings $\alpha_s^{(6)}(\mMS_t) = 0.1084$ and $\alpha_e(\mMS_t)=1/127.78$; 
the scale- and scheme-invariant $\sigma$-mass is evaluated at $\mu = m_t^{\mathrm{os}}$ with $\alpha_s^{(6)}(m_t^{\mathrm{os}}) = 0.1076$ and $\alpha_e(m_t^{\mathrm{os}})=1/127.77$. 
For comparison, when the QED effects are turned off --- namely only the QCD effects are incorporated up to 3-loop --- the same procedure yields $\mMS_t(\mMS_t) = 162.95$~GeV and $\SigmaMass^t = 158.95$~GeV, respectively. 
The differences from the central numbers in \eqref{eq:mtMSmtMS} are about $-0.17$ and $-0.25$~GeV, respectively, comparable in magnitude to the input-induced uncertainties on the extracted $\mMS_t(\mMS_t)$ and $\SigmaMass^t$ (quoted inside the brackets)\footnote{The quoted uncertainties reflect the propagation of the uncertainty of the input $m_t^{\mathrm{os}}$ through the conversion factors, which is well-approximated by the multiplicative conversions; perturbative truncation uncertainties and the uncertainty associated with the input $\alpha_s$ value are not included.} as well as to the most stringent HL-LHC projection, namely the direct-measurement target of about $0.17~\GeV$~\cite{Schwienhorst:2022yqu}, while lying below the $0.6~\GeV$ uncertainty projected for an ATLAS $t\bar t+1$-jet pole-mass determination~\cite{ATLAS:2025rva}; these QED effects are hence not quite negligible at the current input precision and certainly not given more and more precision inputs in the near future. 
Concerning the mixed terms therein, the two-loop $\mathcal O(\alpha_s\alpha_e)$ contributions amount to about $+17.4~\mathrm{MeV}$ for $\mMS_t(\mMS_t)$ and $+36~\mathrm{MeV}$ for $\SigmaMass^t$, whereas the three-loop mixed contributions are about $+3.8$ and $+0.22~\mathrm{MeV}$, respectively.
These mixed shifts are well below the present direct top-mass uncertainty of about $0.27~\GeV$ and also below the HL-LHC projections quoted above.
On the other hand, these mixed contributions in mass conversions are comparable to the experimental and theory targets of future $e^+e^-\to t\bar t$ threshold studies; a recent FCC-ee analysis quotes an experimental precision of $6.8~\MeV$ and a missing-higher-order theory impact of about $35~\MeV$ for the top mass in the potential-subtracted scheme~\cite{Defranchis:2025eet}.
Comparable threshold-scan sensitivities are anticipated at the other proposed lepton colliders, with statistical precision of about $10$--$20~\MeV$ at the ILC and CLIC and total uncertainties of $40$--$75~\MeV$ dominated by the theory input~\cite{Horiguchi:2013wra,CLICdp:2018esa,Schwienhorst:2022yqu}, while a dedicated CEPC study quotes a statistical precision of $9~\MeV$~\cite{Li:2022iav}.
We emphasize that this is a comparison of magnitudes rather than a prediction for the shift of an actual threshold-mass fit, which would require the corresponding process-dependent corrections.

In the case of the masses of the $b$- and $c$-quarks, we take as the inputs their high-precision $\MSbar$ masses fitted by PDG, $\overline{m}_b(\mu=\overline{m}_b) = 4.186 \pm 0.006$~GeV and $\overline{m}_c(\mu=\overline{m}_c) = 1.2729\pm 0.0045$~GeV~\cite{ParticleDataGroup:2026}. 
Using the three-loop mixed QCD+QED $\MSbar$-to-pole mass conversion relation~\eqref{eq:Zff_exp}, we obtain the following values of the pole mass of the $b$- and $c$-quark:
\begin{align} \label{eq:mOSBC}
    \mOS^{b} = 4.9266(71)~\text{GeV}\,,
    \qquad
    \mOS^{c} = 1.9341(68)~\text{GeV}\,.
\end{align}
Here the $\MSbar$-to-pole mass conversions are evaluated at the scales $\mu = \overline{m}_b$ and $\mu = \overline{m}_c$, respectively, with the 5-flavor coupling $\alpha_s^{(5)}(\overline{m}_b) = 0.2242$ and $\alpha_e(\mMS_b)=1/128.88$, as well as the 4-flavor perturbative QCD-coupling value $\alpha_s^{(4)}(\overline{m}_c) = 0.38$\footnote{This value is determined by using a three-loop $\alpha_s$-running from $\alpha_s^{(5)}(m_Z) = 0.1179$ with the decoupling of $b$-quark around its mass scale, which was cross-checked against RunDec\cite{Chetyrkin:2000yt}.} and $\alpha_e(\mMS_c)=1/129.24$.
When only the QCD effects are incorporated up to three loops, the same procedure yields $\mOS^{b} = 4.9255$~GeV and $\mOS^{c} = 1.9328$~GeV, respectively.
The differences from the central numbers in \eqref{eq:mOSBC} are about $+1.1$ and $+1.3~\MeV$, respectively, a factor of five or more below the input-induced uncertaintis on the extracted $\mOS^{b}$ and $\mOS^{c}$.
Compared to the $t$-quark case, the fraction of the QED contributions here is smaller, mainly due to the smaller size of the electric charge of a down-type quark and the larger value of $\alpha_s$ at the much lower scales. 
Moreover, the latest lattice-QCD determinations of these masses --- which already incorporate QED effects in the simulations --- report an uncertainty of $6.3~\MeV$ ($0.15\%$) for $b$ quark and $3.4~\MeV$ for $c$ quark, and anticipate a further reduction of their uncertainties by almost a factor of two in the near term~\cite{Colquhoun:2025pnh,Hatton:2021syc}; the QED-induced shifts found here, most notably for the $c$-quark, then become comparable to the projected precision.
On the other hand, using the three-loop mixed QCD+QED $\MSbar$-to-$\sigma$ mass conversion relation in eq.~\eqref{eq:Zsigmabar_exp}, the corresponding $\sigma$-masses as
\begin{align} \label{eq:msigmaBC}
    \SigmaMass^{b} = 3.9686(57)~\text{GeV}\,,
    \qquad
    \SigmaMass^{c} = 1.1682(41)~\text{GeV}\,.
\end{align}
Here the $\MSbar$-to-$\sigma$ conversions are evaluated at the scales $\mu = \overline{m}_b$ and $\mu = \overline{m}_c$ with the same input $\as(\mu)$ and $\aew(\mu)$ values as in the $\MSbar$-to-pole conversions above.
When only the QCD effects are incorporated up to three loops, the same procedure yields $\SigmaMass^{b} = 3.9691$~GeV and $\SigmaMass^{c} = 1.1688$~GeV, respectively.
The differences from the central numbers in \eqref{eq:msigmaBC} are about $-0.5$ and $-0.6~\MeV$, respectively, roughly one order of magnitude below the input-induced uncertaintis on the extracted $\SigmaMass^{b}$ and $\SigmaMass^{c}$. 
This is in part due to the fact that the $\MSbar$-to-$\sigma$ relation converges markedly better than the $\MSbar$-to-pole series underlying~\eqref{eq:mOSBC}, which is free of the issue of the leading infrared renormalon.\\

Beyond the mass conversions quantified above, the renormalization constants obtained in this work enter a broader range of precision observables as universal ultraviolet building blocks.
On the one hand, the on-shell and $\MSbar$ mass renormalization constants are needed for deriving the relations between the pole mass and the short-distance and/or threhold masses --- including the trace-anomaly-subtracted $\sigma$-mass considered above --- used in various precision determinations, for example top-mass extractions from $e^+e^-\to t\bar t$ threshold studies, where the mixed pole-to-$\MSbar$ and pole-to-$\sigma$ relations obtained here are ingredients for extending such mass-scheme conversion chains consistently in QCD+QED~\cite{Defranchis:2025eet,Chen:2025iul}. 
Refs.~\cite{Colquhoun:2025pnh,Hatton:2021syc} reported recently a remarkably accurate result for the $\MSbar$ $b$-quark mass determined using Lattice methods that has an uncertainty less than $0.2\%$, where the incorporation of QED effects became already necessary for quark-mass determination at this precision. 
On the another hand, the on-shell wave-function renormalization constant is a mandatory ingredient of any UV-renormalized amplitude with external on-shell heavy quarks, and therefore enters the mixed QCD+QED (and electroweak) corrections to processes, which we briefly illustrate with three representative examples.
First, the decay $H\to b\bar b$, observed at the LHC with a signal strength currently measured at the $\mathcal O(10\%)$ level~\cite{ATLAS:2018kot,CMS:2018nsn,ATLAS:2024yzu}, is projected to be measured at the few-percent level at the HL-LHC~\cite{Cepeda:2019klc} and probed at the (sub-)percent level at future $e^+e^-$ Higgs factories~\cite{deBlas:2019rxi,ILC:2013jhg,FCC:2018evy,CEPCStudyGroup:2018ghi}; its inclusive decay width with quark-mass effects was determined to $\mathcal O(\alpha_s^3)$ in QCD and beyond~\cite{Chetyrkin:1995pd,Davies:2017xsp,Mondini:2019gid,Wang:2024ilc} and to $\mathcal O(\alpha\alpha_s)$ in mixed QCD-EW~\cite{Mihaila:2015lwa}, and promoting these predictions to complete mixed QCD+QED (and electroweak) accuracy requires the on-shell mass and wave-function counterterms computed here.
Second, for the top-quark decay $t\to bW$, the $W$-helicity fractions and the top width are currently determined at the LHC at the percent level, respectively~\cite{CMS:2020ezf,ATLAS:2022rms,ParticleDataGroup:2026}, with markedly improved precision anticipated at the HL-LHC~\cite{Azzi:2017iwa,ATLAS:2018kci,Collaboration:2927676,ATLAS:2025rva, Azzi:2019yne,Schwienhorst:2022yqu} and at future lepton colliders~\cite{ILC:2013jhg,FCC:2018evy,CEPCStudyGroup:2018ghi,CLICdp:2018esa};
the QCD corrections to the inclusive width, the $W$-helicity fractions, and semi-inclusive differential distributions have recently been pushed to N$^3$LO~\cite{Chen:2023osm,Chen:2026gin,Chen:2026jwl}, extending the earlier $\mathcal O(\alpha_s^2)$ results~\cite{Chen:2022wit,Czarnecki:2010gb,Gao:2012ja,Brucherseifer:2013iv}, an accuracy whose mixed QCD+QED and electroweak counterparts again build upon the renormalization constants obtained in this work.
Third, the cross sections and forward-backward asymmetries of $e^+e^-\to b\bar b$ and $t\bar t$ have been measured at the percent level at LEP/SLC~\cite{ALEPH:2005ab,ALEPH:2009ogm} (with the $b$-quark asymmetry recently re-analysed with higher precision using modern QCD tools~\cite{dEnterria:2020cgt}) and are targeted at the per-mille level or better at future lepton colliders~\cite{Amjad:2013tlv,Amjad:2015mma,CLICdp:2018esa,Cobal:2025shq,Rohrig:2025bea}; theoretical predictions are known to NNLO in QCD with full quark-mass effects~\cite{Chen:2016zbz,Bernreuther:2016ccf,Gao:2014nva,Gao:2014eea}, with the inclusive cross section now available at N$^3$LO~\cite{Chen:2022vzo} together with the requisite three-loop massive form factors~\cite{Henn:2016kjz,Lee:2018rgs,Blumlein:2019oas,Fael:2022miw,Fael:2022rgm};
the mixed on-shell counterterms provided here are prerequisites for the corresponding QCD+QED and electroweak refinements, whose impact on such non-inclusive observables may exceed a naive estimate based on the size of the couplings. 
In all these applications, the results of this work are universal ultraviolet-renormalization building blocks,
providing the previously missing mixed three-loop QCD+QED ingredients for a complete on-shell renormalization at this perturbative order.

\section{Conclusion}
\label{sec:Conclusion}

In this work we have presented a complete calculation of the mixed corrections in QCD+QED defined with one massive quark and $n_l$ massless quarks to the on-shell quark mass and wave-function renormalization constants up to three-loop order, including all contributions of $\mathcal{O}(\alpha_s^m \alpha^n)$ with $m+n=3$. 
The calculation is performed in dimensional regularization and expressed in terms of renormalized gauge couplings, with the mixed $\MSbar$ renormalization of $\alpha_s$ and $\alpha$ consistently taken into account. 
We note that the structure of the electric charge dependence in our mixed results covers the full charge assignments for all six quark flavors in the Standard Model, up to the three-loop order in question.

Owing to the infrared finiteness of the on-shell quark mass, we extract from the on-shell mass renormalization constant the result for the $\MSbar$ mass renormalization constant and hence the quark-mass anomalous dimension in the presence of both QCD and QED interactions. 
Accordingly, the mixed QCD and QED corrections to the matching relation between the heavy-quark pole mass and the $\MSbar$ mass are derived up to three-loop order. 
Furthermore, we have derived the explicit conversion formulae between the pole mass and the trace-anomaly subtracted $\sigma$-mass of a heavy quark up to three loops, incorporating both QCD and QED effects.
We have carried out explicit numerical analysis at a few chosen benchmark points, for a quick idea of the impact, showing that the QED contributions lower the extracted top-quark $\mMS_t(\mMS_t)$ and $\SigmaMass^t$ masses from the given input pole mass by about $-0.2$~GeV, comparable to the uncertainty induced by the input pole-mass uncertainty, whereas the mixed terms are at or below the few-MeV level not only for the top, but also for the bottom and charm quarks.  
While not limiting for present mass determinations, our analytic results present the complete three-loop calculation for the on-shell quark renormalization in mixed QCD+QED, within the approximation where all quarks except for one are treated as massless, needed for future per-mille-level precision studies.

\acknowledgments

We thank Long-Bin Chen for comments on the manuscript, and Xin Guan for helpful discussions regarding the numerical evaluation of master integrals through asymptotic expansions.
The work was supported in part by grants from the Department of Science and Technology of Shandong province tsqn202312052 and 2024HWYQ-005.
The work of M.N.~has been supported by the European Research Council (ERC) under the European Union's Horizon 2020 research and innovation program grant agreement 101019620 (ERC Advanced Grant TOPUP). 
Z.L.~is supported by the National Natural Science Foundation of China under grants  No.~12475085.
The authors gratefully acknowledge the valuable discussions and insights provided by the members of the China Collaboration of Precision Testing and New Physics. 

\appendix
\section{Results for $Z_{\as}$ and $Z_{\aew}$ up to two-loops in QCD+QED}
\label{appendix:Alpha}

For the sake of future reference, we document below the explicit expressions for the $\MSbar$ coupling renormalization constants $Z_{\as}$ and $Z_{\aew}$ defined in eq.~\eqref{eq:aR_def} up to two-loops in mixed QCD+QED, 
\begin{align}\label{eq:aR_expr}
Z_{\as} = {}& 1
- \as\Bigg[\big(11 C_A - 2\,(n_h+n_l)\big)\frac{1}{3 \eps}\Bigg]
+ \as \aew\Bigg[ \big(\eQ^2 n_h+\eq^2 \nq+\eqA^2 \nqA\big)\frac{1}{\eps} \Bigg]
\nn &
+ \as^2\Bigg[
\big(11 C_A-2\,(n_h+n_l)\big)^2\frac{1}{9 \eps^2}
+ \Big(-\frac{17}{3}C_A^2+\frac{5}{3}C_A\,(n_h+n_l)+C_F\,(n_h+n_l)\Big)\frac{1}{\eps}
\Bigg]
\, ,
\nn
Z_{\aew} = {}& 1
+ \aew\, \Bigg[\,\Big(\eQ^2 n_h+\eq^2 \nq+\eqA^2 \nqA\Big) N_c \frac{4}{3 \eps} \Bigg] 
+ \as\aew \Bigg[\, \Big(\eQ^2 n_h+\eq^2 \nq+\eqA^2 \nqA\Big) N_c C_F \frac{2}{\eps} \Bigg] 
\nn &
+ \aew^2\Bigg[
\big(\eQ^2 n_h+\eq^2 \nq+\eqA^2 \nqA\big)^2 N_c^2 \frac{16}{9 \eps^2}
+ \big(\eQ^4 n_h+\eq^4 \nq+\eqA^4 \nqA\big)N_c \frac{2}{\eps}
\Bigg]
\,. 
\end{align}
The expressions for the anomalous dimensions $\beta_s$ and $\beta_e$ in~eq.~\eqref{eq:TASmass_formula} can be directly read-off, respectively, from the coefficients of the simple $1/\epsilon$ poles of $Z_{\as}$ and $Z_{\aew}$ in ~\eqref{eq:aR_expr},  up to a factor $-l$ at $l$-loop order.

\section{Results for $\ZmOS$ at three-loops}
\label{appendix:ZmOS}

For the sake of readability, we restate here the perturbative expansion of the on-shell quark mass renormalization constant $\ZmOS$, as presented in eq.~\eqref{eq:ZmOS_exp}, up to three loops in mixed QCD+QED:
\begin{align}\label{eq:ZmOS_exp2}
\ZmOS = 1 +
\sum_{l=1}^{3}
\sum^{i+j=l}_{i=0}
\as^i\, \aew^j\, Z_{m,l}^{(i,j)}\,.
\end{align}

At three-loops, it is convenient to further decompose the pure QCD contribution $Z_{m,3}^{(3,0)}$ defined in eq.~\eqref{eq:ZmOS_exp2} into components corresponding to the different color structures:
\begin{align}
Z_{m,3}^{(3,0)} = {}&
C_F^3\, Z_{m,\sss{F^3}}^{(3,0)}
+ C_A C_F^2\, Z_{m,\sss{AF^2}}^{(3,0)}
+ C_A^2 C_F\, Z_{m,\sss{A^2F}}^{(3,0)}
+ C_F^2 n_h\, Z_{m,\sss{F^2h}}^{(3,0)}
\nn 
&
+ C_F^2 n_l\, Z_{m,\sss{F^2l}}^{(3,0)}
+ C_A C_F n_h\, Z_{m,\sss{AFh}}^{(3,0)}
+ C_A C_F n_l\, Z_{m,\sss{AFl}}^{(3,0)}
\nn 
&
+ C_F n_h^2\, Z_{m,\sss{Fh^2}}^{(3,0)}
+ C_F n_l^2\, Z_{m,\sss{Fl^2}}^{(3,0)}
+ C_F n_h n_l \, Z_{m,\sss{Fhl}}^{(3,0)}\,,
\end{align}
where,
\begin{align}
Z_{m,\sss{F^3}}^{(3,0)} = {}&
 - \frac{9}{2 \eps^{3}}
- \frac{1}{\eps^{2}}
\Big(
 \frac{63}{4}
+ \frac{27}{2} \logu
\Big)
+ \frac{1}{\eps}
\Big(
- \frac{457}{8}
+ \frac{111 }{8}\pi^{2}
- 24 \pi^{2}\log2
+ 36 \zeta_3
- \frac{189}{4} \logu
\nn
&
- \frac{81}{4} \logu^{2}
\Big)
- \frac{14225}{48}
- \frac{6037}{48} \pi^{2}
- \frac{146}{15} \pi^{4}
+ 320 \pi^{2}\log2
+ 80 \pi^{2}\log^{2}2
- 8 \log^{4}2
\nn 
&
+ \frac{153}{2} \zeta_3
- 4 \pi^{2}\zeta_3
+ 40 \zeta_5
- 192\Li_4\!\big(\tfrac{1}{2}\big)
-
\Big(
 \frac{1371}{8}
- \frac{333}{8} \pi^{2}
+ 72 \pi^{2}\log2
- 108 \zeta_3
\Big) \logu
\nn
&
- \frac{567}{8} \logu^{2}
- \frac{81}{4} \logu^{3}
\,,
\nn
Z_{m,\sss{AF^2}}^{(3,0)} = {}&
- \frac{33}{2 \eps^{3}}
+ \frac{1}{\eps^{2}}
\Big(
\frac{49}{12}
- \frac{33}{2} \logu
\Big)
+ \frac{1}{\eps}
\Big(
\frac{3311}{24}
- \frac{43}{8} \pi^{2}
+ 12 \pi^{2}\log2
- 18 \zeta_3
+ \frac{379}{4} \logu
\nn
&
+ \frac{33}{4} \logu^{2}
\Big)
+ 
\frac{100247}{144}
+ \frac{6545}{144} \pi^{2}
+ \frac{1867}{135} \pi^{4}
+ \frac{400}{9} \pi^{2}\log2
- \frac{712}{9} \pi^{2}\log^{2}2
\nn 
&
- \frac{140}{9} \log^{4}2
- \frac{3995}{6} \zeta_3
- 76 \pi^{2}\zeta_3
+ 180 \zeta_5
- \frac{1120}{3}\Li_4\!\big(\tfrac{1}{2}\big)
\nn
&
+ 
\Big(
\frac{14311}{24}
- \frac{1135}{24} \pi^{2}
+ \frac{284}{3} \pi^{2}\log2
- 142 \zeta_3
\Big)\logu
+ \frac{1797}{8} \logu^{2}
+ \frac{121}{4} \logu^{3}
\,,
\nn
Z_{m,\sss{A^2F}}^{(3,0)}= {}&
 -\frac{121}{9 \eps^{3}}
+ \frac{1679}{54 \eps^{2}}
- \frac{11413}{324 \eps}
- \frac{1322545}{1944}
- \frac{1955}{54} \pi^2
- \frac{179}{54} \pi^4
- \frac{920}{9} \pi^2 \log2
\nn 
&
+ \frac{176}{9} \pi^2 \log^2 2
+ \frac{88}{9} \log^4 2
+ \frac{2686}{9} \zeta_3
+ 51 \pi^2 \zeta_3
- 130 \zeta_5
+ \frac{704}{3}\Li_4\!\big(\tfrac{1}{2}\big)
\nn
&
+ 
\Big(
- \frac{13243}{27}
+ \frac{88}{9} \pi^2
- \frac{88}{3} \pi^2 \log2
+ 44 \zeta_3
\Big)\logu
- \frac{2341}{18} \logu^2
- \frac{121}{9} \logu^3 \,,
\nn
Z_{m,\sss{F^2h}}^{(3,0)}= {}&
\frac{3}{\eps^{3}}
+ \frac{1}{\eps^{2}}
\Big(
- \frac{5}{6}
+ 3 \logu
\Big)
+ \frac{1}{\eps}
\Big(
- \frac{281}{12}
+ \frac{17}{4} \pi^{2}
- 8 \zeta_3
- \frac{23}{2} \logu
- \frac{3}{2} \logu^{2}
\Big)
- \frac{5257}{72}
\nn
&
- \frac{1327}{216} \pi^{2}
+ \frac{182}{135} \pi^{4}
+ \frac{40}{9} \pi^{2}\log2
- \frac{32}{9} \pi^{2}\log^{2}2
+ \frac{32}{9} \log^{4}2
+ \frac{37}{3} \zeta_3
+ \frac{256}{3}\Li_4\!\big(\tfrac{1}{2}\big)
\nn
&
+ 
\Big(
- \frac{1145}{12}
+ \frac{221}{12} \pi^{2}
- \frac{32}{3} \pi^{2}\log2
- 8 \zeta_3
\Big)\logu
- \frac{117}{4} \logu^{2}
- \frac{11}{2} \logu^{3}
\,,
\nn
Z_{m,\sss{F^2l}}^{(3,0)}= {}&
\frac{3}{\eps^{3}}
+ \frac{1}{\eps^{2}}
\Big(
- \frac{5}{6}
+ 3 \logu
\Big)
+ \frac{1}{\eps}
\Big(
- \frac{65}{12}
- \frac{7}{4} \pi^{2}
- 8 \zeta_3
- \frac{23}{2} \logu
- \frac{3}{2} \logu^{2}
\Big)
+ 
\frac{575}{72}
\nn
&
+ \frac{1091}{72} \pi^{2}
- \frac{238}{135} \pi^{4}
- \frac{352}{9} \pi^{2}\log2
+ \frac{64}{9} \pi^{2}\log^{2}2
+ \frac{32}{9} \log^{4}2
+ \frac{145}{3} \zeta_3
+ \frac{256}{3}\Li_4\!\big(\tfrac{1}{2}\big)
\nn
&
+ 
\Big(
- \frac{497}{12}
+ \frac{5}{12} \pi^{2}
- \frac{32}{3} \pi^{2}\log2
- 8 \zeta_3
\Big)\logu
- \frac{117}{4} \logu^{2}
- \frac{11}{2} \logu^{3}
\,,
\nn
Z_{m,\sss{AFh}}^{(3,0)}= {}&
 \frac{44}{9 \eps^3}
- \frac{242}{27 \eps^2}
+ \frac{1}{\eps}\Big( \frac{278}{81} + 8 \zeta_3 \Big)
+ 
\frac{144959}{486}
- \frac{898}{9} \pi^2
- \frac{172}{135} \pi^4
+ \frac{1024}{9} \pi^2 \log2
\nn &
+ \frac{16}{9} \pi^2 \log^2 2
- \frac{16}{9} \log^4 2
- \frac{218}{9} \zeta_3
+ 4 \pi^2 \zeta_3
- 20 \zeta_5
- \frac{128}{3}\Li_4\!\big(\tfrac{1}{2}\big)
\nn &
+ \Big(
\frac{4664}{27}
- \frac{104}{9} \pi^2
+ \frac{16}{3} \pi^2 \log2
+ 16 \zeta_3
\Big)\logu 
+ \frac{373}{9} \logu^2
+ \frac{44}{9} \logu^3
\,,
\nn
Z_{m,\sss{AFl}}^{(3,0)}= {}&
\frac{44}{9 \eps^3}
- \frac{242}{27 \eps^2}
+ \frac{1}{\eps}\Big( \frac{278}{81} + 8 \zeta_3 \Big)
+ 
\frac{70763}{486}
+ \frac{350}{27} \pi^2
+ \frac{38}{135} \pi^4
+ \frac{176}{9} \pi^2 \log2
\nn &
- \frac{32}{9} \pi^2 \log^2 2
- \frac{16}{9} \log^4 2
+ \frac{178}{9} \zeta_3
- \frac{128}{3}\Li_4\!\big(\tfrac{1}{2}\big)
\nn &
+  \Big(
\frac{3476}{27}
+ \frac{28}{9} \pi^2
+ \frac{16}{3} \pi^2 \log2
+ 16 \zeta_3
\Big)\logu
+ \frac{373}{9} \logu^2
+ \frac{44}{9} \logu^3
\,,
\nn
Z_{m,\sss{Fh^2}}^{(3,0)}= {}&
 - \frac{4}{9 \eps^3}
+ \frac{10}{27 \eps^2}
+ \frac{35}{81 \eps}
+ 
- \frac{9481}{486}
+ \frac{64}{135} \pi^2
+ \frac{88}{9} \zeta_3
+  \Big(
- \frac{394}{27}
+ \frac{16}{9} \pi^2
\Big)\logu
\nn &
- \frac{26}{9} \logu^2
- \frac{4}{9} \logu^3
\,,
\nn
Z_{m,\sss{Fl^2}}^{(3,0)}= {}&
- \frac{4}{9 \eps^3}
+ \frac{10}{27 \eps^2}
+ \frac{35}{81 \eps}
- \frac{2353}{486}
- \frac{52}{27} \pi^2
- \frac{56}{9} \zeta_3
+  \Big(
- \frac{178}{27}
- \frac{8}{9} \pi^2
\Big)\logu
\nn &
- \frac{26}{9} \logu^2
- \frac{4}{9} \logu^3
\,,
\nn
Z_{m,\sss{Fhl}}^{(3,0)}= {}&
 - \frac{8}{9 \eps^3}
+ \frac{20}{27 \eps^2}
+ \frac{70}{81 \eps}
- \frac{5917}{243}
+ \frac{52}{27} \pi^2
+ \frac{32}{9} \zeta_3
+  \Big(
- \frac{572}{27}
+ \frac{8}{9} \pi^2
\Big)\logu
\nn &
- \frac{52}{9} \logu^2
- \frac{8}{9} \logu^3
\,.
\end{align}
The mixed QCD+QED corrections at three-loops are given by
\begin{align}
Z_{m,3}^{(2,1)} = {}&
\eQ^2 C_F^2 \Bigg[
- \frac{27}{2 \eps^3}
+ \frac{1}{\eps^2}\Big(
- \frac{189}{4}
- \frac{81}{2} \logu
\Big)
+ \frac{1}{\eps}\Big(
- \frac{1371}{8}
+ \frac{333}{8} \pi^2
- 72 \pi^2 \log2
\nn &
+ 108 \zeta_3
- \frac{567}{4} \logu
- \frac{243}{4} \logu^2
\Big)
- \frac{14225}{16}
- \frac{6037}{16} \pi^2
- \frac{146}{5} \pi^4
+ 960 \pi^2 \log2
\nn &
+ 240 \pi^2 \log^2 2
- 24 \log^4 2
+ \frac{459}{2} \zeta_3
- 12 \pi^2 \zeta_3
+ 120 \zeta_5
- 576\Li_4\!\big(\tfrac{1}{2}\big)
\nn &
+ \Big(
- \frac{4113}{8}
+ \frac{999}{8} \pi^2
- 216 \pi^2 \log2
+ 324 \zeta_3
\Big)\logu 
- \frac{1701}{8} \logu^2
- \frac{243}{4} \logu^3
\,\Bigg]
\nn &
+ \eQ^2  C_A C_F \Bigg[
- \frac{33}{2 \eps^3}
+ \frac{1}{\eps^2}\Big(
\frac{49}{12}
- \frac{33}{2} \logu
\Big)
+ \frac{1}{\eps}\Big(
\frac{3311}{24}
- \frac{43}{8} \pi^2
+ 12 \pi^2 \log2
\nn &
- 18 \zeta_3
+ \frac{379}{4} \logu
+ \frac{33}{4} \logu^2
\Big)
+ 
\frac{100247}{144}
+ \frac{6545}{144} \pi^2
+ \frac{1867}{135} \pi^4
+ \frac{400}{9} \pi^2 \log2
\nn &
- \frac{712}{9} \pi^2 \log^2 2
- \frac{140}{9} \log^4 2
- \frac{3995}{6} \zeta_3
- 76 \pi^2 \zeta_3
+ 180 \zeta_5
- \frac{1120}{3}\Li_4\!\big(\tfrac{1}{2}\big)
\nn &
+ \Big(
\frac{14311}{24}
- \frac{1135}{24} \pi^2
+ \frac{284}{3} \pi^2 \log2
- 142 \zeta_3
\Big)\logu 
+ \frac{1797}{8} \logu^2
+ \frac{121}{4} \logu^3
\,\Bigg]
\nn &
+ \eQ^2 n_h  C_F \Bigg[
\frac{3}{\eps^3}
+ \frac{1}{\eps^2}\Big(
-\frac{5}{6}
+ 3 \logu
\Big)
+ \frac{1}{\eps}\Big(
-\frac{281}{12}
+ \frac{17}{4} \pi^2
- 8 \zeta_3
- \frac{23}{2} \logu
- \frac{3}{2} \logu^2
\Big)
\nn &
-\frac{5257}{72}
- \frac{1327}{216} \pi^2
+ \frac{182}{135} \pi^4
+ \frac{40}{9} \pi^2 \log2
- \frac{32}{9} \pi^2 \log^2 2
+ \frac{32}{9} \log^4 2
+ \frac{37}{3} \zeta_3
\nn &
+ \frac{256}{3}\Li_4\!\big(\tfrac{1}{2}\big)
+\Big(
-\frac{1145}{12}
+ \frac{221}{12} \pi^2
- \frac{32}{3} \pi^2 \log2
- 8 \zeta_3
\Big) \logu 
- \frac{117}{4} \logu^2
- \frac{11}{2} \logu^3
\,\Bigg]
\nn &
+ \eQ^2 n_l  C_F\Bigg[\,
\frac{3}{\eps^3}
+ \frac{1}{\eps^2}\Big(
\frac{7}{6}
+ 3 \logu
\Big)
+ \frac{1}{\eps}\Big(
-\frac{155}{12}
- \frac{7}{4} \pi^2
- \frac{23}{2} \logu
- \frac{3}{2} \logu^2
\Big)
\nn &
-\frac{4387}{72}
+ \frac{947}{72} \pi^2
- \frac{44}{27} \pi^4
- \frac{352}{9} \pi^2 \log2
+ \frac{64}{9} \pi^2 \log^2 2
+ \frac{32}{9} \log^4 2
+ \frac{277}{3} \zeta_3
\nn &
+ \frac{256}{3}\Li_4\!\big(\tfrac{1}{2}\big)
+\Big(
-\frac{863}{12}
+ \frac{5}{12} \pi^2
- \frac{32}{3} \pi^2 \log2
+ 16 \zeta_3
\Big) \logu 
- \frac{129}{4} \logu^2
- \frac{11}{2} \logu^3
\,\Bigg]
\nn &
+ (\eq^2 \nq+\eqA^2 \nqA) C_F\Bigg[
-\frac{2}{\eps^2}
+ \frac{1}{\eps}\Big(
\frac{15}{2}
- 8 \zeta_3
\Big)
+ 
\frac{827}{12}
+ 2 \pi^2
- \frac{2}{15} \pi^4
- 44 \zeta_3
\nn &
+  \Big(
\frac{61}{2}
- 24 \zeta_3
\Big)\logu
+ 3 \logu^2
\Bigg]\,,
\end{align}
and
\begin{align}
Z_{m,3}^{(1,2)} = {}&
 \eQ^4 n_h N_c C_F   \Bigg[\,
\frac{6}{\eps^3}
- \frac{1}{\eps^2}\Big(\frac{5}{3} - 6 \logu \Big)
- \frac{1}{\eps}\Big( \frac{281}{6}
- \frac{17}{2} \pi^2
+ 16 \zeta_3
+ 23 \logu
+ 3 \logu^2 \Big)
\nn &
- \frac{5257}{36}
- \frac{1327}{108} \pi^2
+ \frac{364}{135} \pi^4
+ \frac{80}{9} \pi^2 \log2
- \frac{64}{9} \pi^2 \log^2 2
+ \frac{64}{9} \log^4 2
+ \frac{74}{3} \zeta_3
\nn &
+ \frac{512}{3}\Li_4\!\big(\tfrac{1}{2}\big)
+ \Big(
- \frac{1145}{6}
+ \frac{221}{6} \pi^2
- \frac{64}{3} \pi^2 \log2
- 16 \zeta_3
\Big)\logu 
- \frac{117}{2} \logu^2
- 11 \logu^3
\,\Bigg] 
\nn &
+ \eQ^2 (\eq^2 \nq + \eqA^2 \nqA )N_c C_F  \Bigg[\,
\frac{6}{\eps^3}
- \frac{1}{\eps^2}\Big(\frac{5}{3} - 6 \logu \Big)
-\frac{1}{\eps}\Big( 
 \frac{65}{6}
+ \frac{7}{2} \pi^2
+ 16 \zeta_3
+ 23 \logu
+ 3 \logu^2
\Big)
\nn &
+ \frac{575}{36}
+ \frac{1091}{36} \pi^2
- \frac{476}{135} \pi^4
- \frac{704}{9} \pi^2 \log2
+ \frac{128}{9} \pi^2 \log^2 2
+ \frac{64}{9} \log^4 2
+ \frac{290}{3} \zeta_3
\nn &
+ \frac{512}{3}\Li_4\!\big(\tfrac{1}{2}\big)
+ \Big(
- \frac{497}{6}
+ \frac{5}{6} \pi^2
- \frac{64}{3} \pi^2 \log2
- 16 \zeta_3
\Big)\logu 
- \frac{117}{2} \logu^2
- 11 \logu^3
\,\Bigg]
\nn &
+ \eQ^4 C_F\Bigg[
-\frac{27}{2 \eps^3}
+ \frac{1}{\eps^2}\Big(
-\frac{189}{4}
- \frac{81}{2} \logu
\Big)
+ \frac{1}{\eps}\Big(
-\frac{1371}{8}
+ \frac{333}{8} \pi^2
- 72 \pi^2 \log2
\nn &
+ 108 \zeta_3
- \frac{567}{4} \logu
- \frac{243}{4} \logu^2
\Big)
-\frac{14225}{16}
- \frac{6037}{16} \pi^2
- \frac{146}{5} \pi^4
+ 960 \pi^2 \log2
\nn &
+ 240 \pi^2 \log^2 2
- 24 \log^4 2
+ \frac{459}{2} \zeta_3
- 12 \pi^2 \zeta_3
+ 120 \zeta_5
- 576\Li_4\!\big(\tfrac{1}{2}\big)
\nn &
+ \Big(
-\frac{4113}{8}
+ \frac{999}{8} \pi^2
- 216 \pi^2 \log2
+ 324 \zeta_3
\Big)\logu 
- \frac{1701}{8} \logu^2
- \frac{243}{4} \logu^3
\,\Bigg]
\,.
\end{align}
For the pure three-loop QED contribution, quartic structures in the flavor multiplicities, namely $n_h^2$, $n_h\nq$, $n_h\nqA$, $\nq^2$, $\nqA^2$, and $\nq\nqA$, emerge. 
Taking into account the corresponding combinations of electric charges, it is therefore convenient to further decompose $Z_{m,3}^{(0,3)}$, defined in eq.~\eqref{eq:ZmOS_exp2}, according to the distinct quark-flavor structures as follows:
\begin{align}
Z_{m,3}^{(0,3)} = {}&
\eQ^6  n_h^2 N_c^2 Z^{(0,3)}_{m,h^2}
+ \eQ^4 \eq^2 n_h \nq N_c^2  Z^{(0,3)}_{m,hq}
+ \eQ^4 \eqA^2 n_h\nqA N_c^2 Z^{(0,3)}_{m,h\qA}
+ \eQ^2 \eq^4 \nq^2 N_c^2 Z^{(0,3)}_{m,q^2}
\nn
&
+ \eQ^2 \eqA^4 \nqA^2 N_c^2  Z^{(0,3)}_{m,{\qA}^2}
+ \eQ^2 \eq^2  \eqA^2 \nq\nqA N_c^2 Z^{(0,3)}_{m,q\qA}
+ \eQ^2 \eq^4 \nq N_c  Z^{(0,3)}_{m,\eq^4}
+ \eQ^2 \eqA^4 \nqA N_c Z^{(0,3)}_{m,\eqA^4}
\nn 
&
+ \eQ^4 \eq^2 \nq N_c  Z^{(0,3)}_{m,\eq^2}
+ \eQ^4 \eqA^2 \nqA N_c Z^{(0,3)}_{m,\eqA^2}
+ \eQ^6 n_h N_c Z^{(0,3)}_{m,h} 
+ \eQ^6 Z^{(0,3)}_{m,Q}\,,
\end{align}
where,
\begin{align}
 Z^{(0,3)}_{m,h^2} = {}&
 -\frac{16}{9 \eps^3}
+ \frac{40}{27 \eps^2}
+ \frac{140}{81 \eps}
-\frac{18962}{243}
+ \frac{256}{135} \pi^2
+ \frac{352}{9} \zeta_3
+ \Big(
-\frac{1576}{27}
+ \frac{64}{9} \pi^2
\Big)\logu
\nn &
- \frac{104}{9} \logu^2
- \frac{16}{9} \logu^3 \,,
\nn
Z^{(0,3)}_{m,hq} = {} &
 -\frac{32}{9 \eps^3}
+ \frac{80}{27 \eps^2}
+ \frac{280}{81 \eps}
-\frac{23668}{243}
+ \frac{208}{27} \pi^2
+ \frac{128}{9} \zeta_3
+ \Big(
-\frac{2288}{27}
+ \frac{32}{9} \pi^2
\Big)\logu
\nn &
- \frac{208}{9} \logu^2
- \frac{32}{9} \logu^3 
\,, \nn
Z^{(0,3)}_{m,h\qA} = {} &
Z^{(0,3)}_{m,hq} 
\,, \nn 
Z^{(0,3)}_{m,q^2} = {}
& -\frac{16}{9 \eps^3} 
+ \frac{40}{27 \eps^2} 
+ \frac{140}{81 \eps} 
-\frac{4706}{243} 
- \frac{208}{27} \pi^2 
- \frac{224}{9} \zeta_3
+ \Big(-\frac{712}{27}
- \frac{32}{9} \pi^2 \Big)\logu
\nn &
- \frac{104}{9} \logu^2
- \frac{16}{9} \logu^3
\,, \nn
Z^{(0,3)}_{m,{\qA}^2} = {}& Z^{(0,3)}_{m,q^2} 
\,, \nn
Z^{(0,3)}_{m,q\qA} = {}
& -\frac{32}{9 \eps^3} 
+ \frac{80}{27 \eps^2} 
+ \frac{280}{81 \eps} 
-\frac{9412}{243} 
- \frac{416}{27} \pi^2 
- \frac{448}{9} \zeta_3
+ \Big(-\frac{1424}{27} 
- \frac{64}{9} \pi^2 \Big)\logu
\nn &
- \frac{208}{9} \logu^2
- \frac{32}{9} \logu^3
\,, \nn
Z^{(0,3)}_{m,\eq^4}  = {}
& -\frac{4}{\eps^2} 
+ \frac{1}{\eps} \big(15 - 16  \zeta_3 \big)
+ 
\frac{827}{6} + 4 \pi^2 - \frac{4}{15} \pi^4 - 88  \zeta_3
+ (61 - 48  \zeta_3)\logu + 6  \logu^2
\,, \nn
 Z^{(0,3)}_{m,\eqA^4} = {}&
 Z^{(0,3)}_{m,\eq^4} 
 \,, \nn
Z^{(0,3)}_{m,\eq^2} = {}
& \frac{6}{\eps^3} 
+ \frac{1}{\eps^2} 
\Big(\frac{7}{3} + 6  \logu\Big)
+ \frac{1}{\eps} \Big(-\frac{155}{6} - \frac{7}{2} \pi^2 - 23  \logu - 3  \logu^2
\Big)
-\frac{4387}{36} 
+ \frac{947}{36} \pi^2 
\nn &
- \frac{88}{27} \pi^4 
- \frac{704}{9} \pi^2 \log2 
+ \frac{128}{9} \pi^2 \log^2 2 
+ \frac{64 }{9} \log^4 2 
+ \frac{554 }{3} \zeta_3 
+ \frac{512}{3} \Li_4\!\big(\frac{1}{2}\big) 
\nn & 
+ \Big( -\frac{863}{6} + \frac{5}{6} \pi^2 - \frac{64}{3} \pi^2 \log2 + 32  \zeta_3 \Big) \logu 
- \frac{129}{2}  \logu^2 - 11  \logu^3
\,, \nn
Z^{(0,3)}_{m,\eqA^2} = {}&
Z^{(0,3)}_{m,\eq^2} 
\,, \nn
Z^{(0,3)}_{m,h} = {}
& \frac{6}{\eps^3} 
+ \frac{1}{\eps^2} \Big(-\frac{5}{3} + 6  \logu \Big)
- \frac{1}{\eps} \Big(
\frac{281}{6} 
- \frac{17}{2} \pi^2 
+ 16  \zeta_3 
+ 23  \logu 
+ 3  \logu^2 \Big)
-\frac{5257}{36} 
- \frac{1327}{108} \pi^2 
\nn & 
+ \frac{364}{135} \pi^4 
+ \frac{80}{9} \pi^2 \log2 
- \frac{64}{9} \pi^2 \log^2 2 
+ \frac{64}{9} \log^4 2
+ \frac{74}{3} \zeta_3
+ \frac{512}{3} \Li_4\!\big(\frac{1}{2}\big) 
\nn & 
+ \Big( -\frac{1145}{6} + \frac{221}{6} \pi^2 - \frac{64}{3} \pi^2 \log2 - 16  \zeta_3 \Big) \logu 
- \frac{117}{2}  \logu^2 - 11  \logu^3
\,, \nn
Z^{(0,3)}_{m,Q} = {}
& -\frac{9}{2 \eps^3} 
+ \frac{1}{\eps^2} \Big(
-\frac{63}{4} 
- \frac{27}{2}  \logu
\Big)
+ \frac{1}{\eps} \Big(
 -\frac{457}{8} 
 + \frac{111}{8} \pi^2 
 - 24 \pi^2 \log2 
 + 36  \zeta_3 
 - \frac{189}{4}  \logu 
 \nn &
 - \frac{81}{4}  \logu^2
 \Big)
-\frac{14225}{48} 
- \frac{6037}{48} \pi^2 
- \frac{146}{15} \pi^4 
+ 320 \pi^2 \log2 
+ 80 \pi^2 \log^2 2 
- 8 \log^4 2 
\nn &
+ \frac{153}{2}  \zeta_3 
- 4 \pi^2  \zeta_3 
+ 40  \zeta_5 
- 192 \Li_4\!\big(\frac{1}{2}\big) 
- \Big( \frac{1371}{8} 
- \frac{333}{8} \pi^2 
+ 72 \pi^2 \log2 
- 108  \zeta_3 \Big) \logu 
\nn &
- \frac{567}{8}  \logu^2 
- \frac{81}{4}  \logu^3
\,.
\end{align}

\section{Results for $\ZQOS$ at three-loops}
\label{appendix:ZQOS}

For the sake of readability, we restate here the perturbative expansion of the on-shell heavy-quark wave-function renormalization constant $\ZQOS$, as presented in eq.~\eqref{eq:ZQOS_exp}, up to three loops in mixed QCD+QED:
\begin{align}\label{eq:ZQOS_exp2}
\ZQOS = 1 +
\sum_{l=1}^{3}
\sum^{i+j=l}_{i=0}
\as^i\, \aew^j\, Z_{Q,l}^{(i,j)}\,.
\end{align}

At three-loops, it is convenient to further decompose the pure QCD contribution $Z_{Q,3}^{(3,0)}$ defined in eq.~\eqref{eq:ZQOS_exp2} into components corresponding to the different color structures:
\begin{align}
Z_{Q,3}^{(3,0)} = {}&
C_F^3\, Z_{Q,\sss{F^3}}^{(3,0)}
+ C_A C_F^2\, Z_{Q,\sss{AF^2}}^{(3,0)}
+ C_F^2 n_h\, Z_{Q,\sss{F^2h}}^{(3,0)}
+ C_F^2 n_l\, Z_{Q,\sss{F^2l}}^{(3,0)}
\nn 
&
+ C_A^2 C_F\, Z_{Q,\sss{A^2F}}^{(3,0)}
+ C_A C_F n_h\, Z_{Q,\sss{AFh}}^{(3,0)}
+ C_A C_F n_l\, Z_{Q,\sss{AFl}}^{(3,0)}
\nn 
&
+ C_F n_h^2\, Z_{Q,\sss{Fh^2}}^{(3,0)}
+ C_F n_l^2\, Z_{Q,\sss{Fl^2}}^{(3,0)}
+ C_F n_h n_l \, Z_{Q,\sss{Fhl}}^{(3,0)}\,,
\end{align}
where,
\begin{align}
Z_{Q,\sss{F^3}}^{(3,0)} = {}&
-\frac{9}{2\eps^3}
- \frac{1}{\eps^2}\Big(
  \frac{81}{4}
  + \frac{27}{2}\logu
\Big)
- \frac{1}{\eps}\Big(
  \frac{1039}{8}
  - \frac{303}{8}\pi^2
  + 48\pi^2\log2
  - 72\zeta_3
  + \frac{243}{4}\logu
  + \frac{81}{4}\logu^2
\Big)
\nn &
-\frac{10823}{48}
- \frac{58321}{144}\pi^2
- \frac{328}{15}\pi^4
+ \frac{2740}{3}\pi^2\log2
+ 192\pi^2\log^2 2
- \frac{80}{3}\log^4 2
- \frac{739}{2}\zeta_3
\nn &
+ 8\pi^2\zeta_3
- 20 \zeta_5
- 640\Li_4\!\big(\tfrac{1}{2}\big)
+ \Big(
  -\frac{3117}{8}
  + \frac{909}{8}\pi^2
  - 144\pi^2\log2
  + 216 \zeta_3
\Big)\logu
\nn &
- \frac{729}{8}\logu^2
- \frac{81}{4}\logu^3
\,, \nn 
Z_{Q,\sss{AF^2}}^{(3,0)} = {}&
-\frac{33}{2\eps^3}
+ \frac{1}{\eps^2}\Big(
  \frac{95}{12}
  - \frac{33}{2}\logu
\Big)
+ \frac{1}{\eps}\Big(
  \frac{1787}{8}
  - \frac{131}{8}\pi^2
  + 24\pi^2\log2
  - 40 \zeta_3
  + \frac{469}{4}\logu
  + \frac{33}{4}\logu^2
\Big)
\nn &
+ \frac{136945}{144}
+ \frac{29695}{144}\pi^2
+ \frac{1793}{54}\pi^4
- \frac{1510}{9}\pi^2\log2
- \frac{1880}{9}\pi^2\log^2 2
- \frac{220}{9}\log^4 2
- \frac{6913}{6} \zeta_3
\nn &
- 180\pi^2\zeta_3
+ 580 \zeta_5
- \frac{1760}{3}\Li_4\!\big(\tfrac{1}{2}\big)
+ \Big(
  \frac{25609}{24}
  - \frac{3335}{24}\pi^2
  + \frac{568}{3}\pi^2\log2
  - 296 \zeta_3
\Big)\logu
\nn &
+ \frac{2155}{8}\logu^2
+ \frac{121}{4}\logu^3
\,, \nn 
Z_{Q,\sss{A^2F}}^{(3,0)} = {}&
-\frac{121}{9\eps^3}
+ \frac{2009}{54\eps^2}
+ \frac{1}{\eps}\Big(
  -\frac{12793}{324}
  - \frac{8}{135} \pi^4
  - \frac{1}{12} \xi
  + \frac{2}{135} \pi^4\xi
  - \frac{3}{2}\zeta_3
  - \frac{3}{4}\zeta_3\xi
\Big)
- \frac{1654711}{1944}
- \frac{4339}{54}\pi^2
\nn &
- \frac{3419}{360}\pi^4
- \frac{13}{12}\xi
- \frac{1}{4} \pi^2\xi
+ \frac{17}{432} \pi^4\xi
- \frac{1300}{9}\pi^2\log2
+ \frac{508}{9}\pi^2\log^2 2
+ \frac{170}{9}\log^4 2
+ \frac{5857}{9} \zeta_3
\nn &
+ \frac{1016}{9}\pi^2\zeta_3
- \frac{13}{4}\zeta_3\xi
+ \frac{4}{9}\pi^2\zeta_3\xi
- \frac{1184}{3} \zeta_5
+ \frac{7}{6}\zeta_5\xi
+ \frac{1360}{3}\Li_4\!\big(\tfrac{1}{2}\big)
+ \Big(
  -\frac{36977}{54}
  + \frac{110}{3}\pi^2
  \nn &
  - \frac{8}{45}\pi^4
  - \frac{1}{4} \xi
  + \frac{2}{45} \pi^4\xi
  - \frac{176}{3} \pi^2 \log 2
  + \frac{167}{2} \zeta_3
  - \frac{9}{4} \zeta_3\xi
\Big)\logu
- \frac{2671}{18}\logu^2
- \frac{121}{9}\logu^3
\,, \nn
Z_{Q,\sss{F^2h}}^{(3,0)} = {}&
-\frac{1}{\eps^2}\Big(
\frac{7}{6}
+6 \logu
\Big)
- \frac{1}{\eps}\Big(
\frac{707}{12}
- \frac{15}{2} \pi^2
+ \frac{29}{2} \logu
+15 \logu^2
\Big)
-\frac{76897}{216}
- \frac{11551}{648} \pi^2
\nn &
+ \frac{62}{45} \pi^4
+ \frac{112}{9}\pi^2 \log2
-16\pi^2 \log^2 2
+16\log^4 2
+ \frac{1763}{9} \zeta_3
+384 \Li_4\!\big(\tfrac{1}{2}\big)
\nn &
+ \Big(
-\frac{2891}{12}
+ \frac{233}{6} \pi^2
- \frac{64}{3}\pi^2 \log2
+32 \zeta_3
\Big)\logu
- \frac{143}{4}\logu^2
-19\logu^3
\,, \nn 
Z_{Q,\sss{F^2l}}^{(3,0)} = {}&
\frac{3}{\eps^3}
- \frac{1}{\eps^2}\Big(
\frac{19}{6}
- 3 \logu
\Big)
- \frac{1}{\eps}\Big(
\frac{235}{12}
+ \frac{7}{4} \pi^2
+ 8 \zeta_3
+ \frac{41}{2} \logu
+ \frac{3}{2} \logu^2
\Big)
-\frac{3083}{72}
+ \frac{2845}{72} \pi^2
\nn &
- \frac{458}{135} \pi^4
- \frac{752}{9} \pi^2 \log2
+ \frac{64}{9} \log^4 2
+ \frac{128}{9} \pi^2 \log^2 2
+ \frac{473}{3} \zeta_3
+ \frac{512}{3} \Li_4\!\big(\tfrac{1}{2}\big)
\nn &
+ \Big(
-\frac{1475}{12}
+ \frac{133}{12} \pi^2
- \frac{64}{3} \pi^2 \log2
+ 8 \zeta_3
\Big)\logu
- \frac{179}{4} \logu^2
- \frac{11}{2} \logu^3
\,, \nn 
Z_{Q,\sss{AFh}}^{(3,0)} = {}&
\frac{1}{6 \eps^3}(1-\xi)
+ \frac{1}{\eps^2}\Big(
-\frac{28}{9}
+ \frac{1}{2} \xi
+ \Big(
-\frac{41}{6}
- \frac{1}{2} \xi
\Big)\logu
\Big)
+ \frac{1}{\eps}\Big(
\frac{52}{27}
- \frac{41}{72} \pi^2
- \frac{35}{18} \xi
- \frac{1}{24}\pi^2\xi
\nn &
+ \Big(
\frac{166}{9}
+ \frac{3}{2} \xi
\Big)\logu
+ \Big(
-\frac{35}{12}
- \frac{3}{4} \xi
\Big)\logu^2
\Big)
+ 
\frac{49901}{81}
- \frac{36019}{162} \pi^2
- \frac{68}{45} \pi^4
+ \frac{407}{54} \xi
\nn &
+ \frac{1}{8} \pi^2\xi
+ \frac{2560}{9} \pi^2 \log2
- \frac{32}{3} \log^4 2
+ \frac{96}{9} \pi^2 \log^2 2
- 256\Li_4\!\big(\tfrac{1}{2}\big)
- 154 \zeta_3
+ \frac{22}{3} \pi^2\zeta_3
\nn &
- \frac{7}{6} \zeta_3\xi
- 30 \zeta_5
+ \Big(
\frac{8282}{27}
- \frac{641}{24} \pi^2
- \frac{35}{6} \xi
- \frac{1}{8} \pi^2\xi
+ \frac{32}{3} \pi^2 \log2
- 16 \zeta_3
\Big)\logu
\nn &
+ \Big(
70
+ \frac{9}{4} \xi
\Big)\logu^2
+ \Big(
\frac{247}{36}
- \frac{3}{4} \xi
\Big)\logu^3
\,, \nn 
Z_{Q,\sss{AFl}}^{(3,0)} = {}&
\frac{44}{9 \eps^3}
- \frac{338}{27 \eps^2}
+ \frac{1}{\eps}\Big(
\frac{626}{81}
+ 8 \zeta_3
\Big)
+ \Big(
\frac{111791}{486}
+ \frac{26}{3} \pi^2
+ \frac{76}{135} \pi^4
+ \frac{376}{9}\pi^2 \log2
\nn &
- \frac{64}{9} \pi^2 \log^2 2
- \frac{32}{9} \log^4 2
- \frac{140}{9} \zeta_3
- \frac{256}{3} \Li_4\!\big(\tfrac{1}{2}\big)
+ \Big(
\frac{5408}{27}
- \frac{16}{9} \pi^2
+ \frac{32}{3} \pi^2 \log2
\nn &
+ 8 \zeta_3
\Big) \logu
+ \frac{469}{9} \logu^2
+ \frac{44}{9} \logu^3
\Big)
\,, \nn 
Z_{Q,\sss{Fh^2}}^{(3,0)} = {}&
\frac{2}{9 \eps^2}
- \frac{1}{\eps}\Big(
\frac{5}{27}
+ \frac{4}{3}\logu^2
\Big)
-\frac{8425}{162}
+ \frac{32}{45} \pi^2
+ \frac{112}{3} \zeta_3
+ \Big(
-\frac{962}{27}
+ \frac{10}{3} \pi^2
\Big)\logu
\nn &
- \frac{22}{9}\logu^2
- \frac{8}{3}\logu^3
\,, \nn 
Z_{Q,\sss{Fl^2}}^{(3,0)} = {}&
-\frac{4}{9 \eps^3}
+ \frac{22}{27 \eps^2}
+ \frac{5}{81 \eps}
-\frac{5767}{486}
- \frac{76}{27} \pi^2
- \frac{56}{9} \zeta_3
+ \Big(
-\frac{334}{27}
- \frac{8}{9} \pi^2
\Big)\logu
\nn &
- \frac{38}{9}\logu^2
- \frac{4}{9}\logu^3
\,, \nn 
Z_{Q,\sss{Fhl}}^{(3,0)} = {}& 
\frac{1}{\eps^2}\Big(
\frac{4}{9}
+ \frac{4}{3}\logu
\Big)
+ \frac{1}{\eps}\Big(
-\frac{10}{27}
+ \frac{1}{9} \pi^2
- \frac{16}{9}\logu
+ \frac{2}{3}\logu^2
\Big)
-\frac{4721}{81}
+ \frac{152}{27} \pi^2
- \frac{4}{9} \zeta_3
\nn &
+ \Big(
-\frac{1316}{27}
+ \frac{25}{9} \pi^2
\Big)\logu
- \frac{28}{3}\logu^2
- \frac{10}{9}\logu^3
\,.
\end{align}
The mixed QCD+QED corrections at three-loops are given by
\begin{align}
Z_{Q,3}^{(2,1)} = {}&
\eQ^2 C_F^2 
\Bigg[
-\frac{27}{2\eps^3}
+ \frac{1}{\eps^2}\Big(
-\frac{243}{4}
- \frac{81}{2}\logu
\Big)
+ \frac{1}{\eps}\Big(
-\frac{3117}{8}
+ \frac{909}{8} \pi^2
- 144 \pi^2\log2
+ 216 \zeta_3
\nn &
- \frac{729}{4}\logu
- \frac{243}{4}\logu^2
\Big)
-\frac{10823}{16}
- \frac{58321}{48} \pi^2
- \frac{328}{5} \pi^4
+ 2740\pi^2\log2
+ 576\pi^2\log^2 2
\nn &
- 80\log^4 2
- \frac{2217}{2} \zeta_3
+ 24\pi^2\zeta_3
- 60 \zeta_5
- 1920\Li_4\!\big(\tfrac{1}{2}\big)
+ \Big(
-\frac{9351}{8}
+ \frac{2727}{8} \pi^2
\nn &
- 432\pi^2\log2
+ 648 \zeta_3
\Big)\logu
- \frac{2187}{8}\logu^2
- \frac{243}{4}\logu^3
\Bigg]
\nn &
+ \eQ^2  C_A C_F 
\Bigg[
-\frac{33}{2\eps^3}
+ \frac{1}{\eps^2}\Big(
\frac{95}{12}
- \frac{33}{2}\logu
\Big)
+ \frac{1}{\eps}\Big(
\frac{1787}{8}
- \frac{131}{8} \pi^2
+ 24\pi^2\log2
- 40 \zeta_3
\nn &
+ \frac{469}{4}\logu
+ \frac{33}{4}\logu^2
\Big)
+ 
\frac{136945}{144}
+ \frac{29695}{144} \pi^2
+ \frac{1793}{54} \pi^4
- \frac{1510}{9}\pi^2\log2
- \frac{1880}{9}\pi^2\log^2 2
\nn &
- \frac{220}{9}\log^4 2
- \frac{6913}{6} \zeta_3
- 180\pi^2\zeta_3
+ 580 \zeta_5
- \frac{1760}{3}\Li_4\!\big(\tfrac{1}{2}\big)
+ \Big(
\frac{25609}{24}
- \frac{3335}{24} \pi^2
\nn &
+ \frac{568}{3}\pi^2\log2
- 296 \zeta_3
\Big)\logu
+ \frac{2155}{8}\logu^2
+ \frac{121}{4}\logu^3
\Bigg]
\nn &
+ \eQ^2 n_h  C_F \Bigg[
\frac{1}{\eps^2}\Big(
-\frac{7}{6}
- 6\logu
\Big)
+ \frac{1}{\eps}\Big(
-\frac{707}{12}
+ \frac{15}{2} \pi^2
- \frac{29}{2}\logu
- 15\logu^2
\Big)
-\frac{76897}{216}
\nn &
- \frac{11551}{648} \pi^2
+ \frac{62}{45} \pi^4
+ \frac{112}{9}\pi^2\log2
- 16\pi^2\log^2 2
+ 16\log^4 2
+ \frac{1763}{9} \zeta_3
+ 384\Li_4\!\big(\tfrac{1}{2}\big)
\nn &
+ \Big(
-\frac{2891}{12}
+ \frac{233}{6} \pi^2
- \frac{64}{3} \pi^2\log2
+ 32 \zeta_3
\Big)\logu
- \frac{143}{4}\logu^2
- 19\logu^3
\Bigg]
\nn &
+ \eQ^2 n_l  C_F \Bigg[
\frac{3}{\eps^3}
+ \frac{1}{\eps^2}\Big(
-\frac{7}{6}
+ 3\logu
\Big)
+ \frac{1}{\eps}\Big(
-\frac{349}{12}
- \frac{7}{4} \pi^2
- \frac{41}{2}\logu
- \frac{3}{2}\logu^2
\Big)
\nn &
+ \Big(
-\frac{9701}{72}
+ \frac{2701}{72} \pi^2
- \frac{88}{27} \pi^4
- \frac{752}{9}\pi^2\log2
+ \frac{64}{9}\log^4 2
+ \frac{128}{9}\pi^2\log 2
+ \frac{653}{3} \zeta_3
\nn &
+ \frac{512}{3}\Li_4\!\big(\tfrac{1}{2}\big)
+ \Big(
-\frac{1913}{12}
+ \frac{133}{12} \pi^2
- \frac{64}{3}\pi^2\log2
+ 32 \zeta_3
\Big)\logu
- \frac{191}{4}\logu^2
- \frac{11}{2}\logu^3
\Big)
\Bigg]
\nn &
+ (\eq^2 \nq+\eqA^2 \nqA) C_F
\Bigg[ -\frac{2}{\eps^2}
+ \frac{1}{\eps}\Big(
\frac{19}{2}
- 8 \zeta_3
\Big)
+ 
\frac{1103}{12}
+ 2\pi^2
- \frac{2}{15} \pi^4
- 60 \zeta_3
\nn &
+ \Big(
\frac{73}{2}
- 24 \zeta_3
\Big)\logu
+ 3\logu^2
 \Bigg]\,,
\end{align}
and
\begin{align}
Z_{Q,3}^{(1,2)} = {}&
\eQ^4 n_h N_c C_F
\Bigg[
-\frac{1}{\eps^2}\Big(
\frac{7}{3}+12\logu
\Big)
- \frac{1}{\eps}\Big(
\frac{707}{6}
-15\pi^2
+29\logu
+30\logu^2
\Big)
-\frac{76897}{108}
-\frac{11551}{324} \pi^2
\nn &
+\frac{124}{45} \pi^4
+\frac{224}{9}\pi^2\log2
-32\pi^2\log^2 2
+32\log^4 2
+\frac{3526}{9} \zeta_3
+768 \Li_4\!\big(\tfrac{1}{2}\big)
\nn &
+ \Big(
-\frac{2891}{6}
+\frac{233}{3} \pi^2
-\frac{128}{3}\pi^2\log2
+64 \zeta_3
\Big)\logu
-\frac{143}{2}\logu^2
-38\logu^3
\Bigg]
\nn &
+ \eQ^2 (\eq^2 \nq + \eqA^2 \nqA ) N_c C_F
\Bigg[
\frac{6}{\eps^3}
+ \frac{1}{\eps^2}\Big(
-\frac{19}{3}+6\logu
\Big)
+ \frac{1}{\eps}\Big(
-\frac{235}{6}
-\frac{7}{2} \pi^2
-16 \zeta_3
\nn &
-41\logu
-3\logu^2
\Big)
+ \Big(
-\frac{3083}{36}
+\frac{2845}{36} \pi^2
-\frac{916}{135} \pi^4
-\frac{1504}{9}\pi^2\log2
+\frac{128}{9}\log^4 2
\nn &
+\frac{256}{9}\pi^2\log^2 2
+\frac{946}{3} \zeta_3
+\frac{1024}{3}\Li_4\!\big(\tfrac{1}{2}\big)
+ \Big(
-\frac{1475}{6}
+\frac{133}{6} \pi^2
-\frac{128}{3}\pi^2\log2
\nn &
+16 \zeta_3
\Big)\logu
-\frac{179}{2}\logu^2
-11\logu^3
\Big)
\Bigg]
\nn &
+\eQ^4  C_F
\Bigg[
-\frac{27}{2 \eps^3}
+ \frac{1}{\eps^2}\Big(
-\frac{243}{4}
-\frac{81}{2}\logu
\Big)
+ \frac{1}{\eps}\Big(
-\frac{3117}{8}
+\frac{909}{8} \pi^2
-144\pi^2\log2
\nn &
+216 \zeta_3
-\frac{729}{4}\logu
-\frac{243}{4}\logu^2
\Big)
+ \Big(
-\frac{10823}{16}
-\frac{58321}{48} \pi^2
-\frac{328}{5} \pi^4
+2740\pi^2\log2
\nn &
+576\pi^2\log^2 2
-80\log^4 2
-\frac{2217}{2} \zeta_3
+24\pi^2\zeta_3
-60 \zeta_5
-1920\Li_4\!\big(\tfrac{1}{2}\big)
\nn &
+ \Big(
-\frac{9351}{8}
+\frac{2727}{8} \pi^2
-432\pi^2\log2
+648 \zeta_3
\Big)\logu
-\frac{2187}{8}\logu^2
-\frac{243}{4}\logu^3
\Big)
\Bigg]\,.
\end{align}
For the pure three-loop QED contribution, it is convenient to decompose $Z_{Q,3}^{(0,3)}$, defined in eq.~\eqref{eq:ZQOS_exp2}, according to the distinct electric-charge combinations and quark-flavor structures as follows:
\begin{align}
Z_{Q,3}^{(0,3)} = {}&
\eQ^6  n_h^2 N_c^2 Z^{(0,3)}_{Q,h^2}
+ \eQ^4 \eq^2 n_h \nq N_c^2  Z^{(0,3)}_{Q,hq}
+ \eQ^4 \eqA^2 n_h\nqA N_c^2 Z^{(0,3)}_{Q,h\qA}
+ \eQ^2 \eq^4 \nq^2 N_c^2 Z^{(0,3)}_{Q,q^2}
\nn
&
+ \eQ^2 \eqA^4 \nqA^2 N_c^2  Z^{(0,3)}_{Q,{\qA}^2}
+ \eQ^2 \eq^2  \eqA^2 \nq\nqA N_c^2 Z^{(0,3)}_{Q,q\qA}
+ \eQ^2 \eq^4 \nq N_c  Z^{(0,3)}_{Q,\eq^4}
+ \eQ^2 \eqA^4 \nqA N_c Z^{(0,3)}_{Q,\eqA^4}
\nn 
&
+ \eQ^4 \eq^2 \nq N_c  Z^{(0,3)}_{Q,\eq^2}
+ \eQ^4 \eqA^2 \nqA N_c Z^{(0,3)}_{Q,\eqA^2}
+ \eQ^6 n_h N_c Z^{(0,3)}_{Q,h} 
+ \eQ^6 Z^{(0,3)}_{Q,Q}\,,
\end{align}
where,
\begin{align}
 Z^{(0,3)}_{Q,h^2} = {}&
 \frac{8}{9 \eps^2}
+ \frac{1}{\eps}\Big(
-\frac{20}{27}
-\frac{16}{3}\logu^2
\Big)
-\frac{16850}{81}
+\frac{128}{45} \pi^2
+\frac{448}{3} \zeta_3
+ \Big(
-\frac{3848}{27}
+\frac{40}{3} \pi^2
\Big)\logu
\nn &
-\frac{88}{9}\logu^2
-\frac{32}{3}\logu^3
\,, \nn 
Z^{(0,3)}_{Q,hq} = {} &
\frac{1}{\eps^2}\Big(
\frac{16}{9}
+\frac{16}{3}\logu
\Big)
+ \frac{1}{\eps}\Big(
-\frac{40}{27}
+\frac{4}{9} \pi^2
-\frac{64}{9}\logu
+\frac{8}{3}\logu^2
\Big)
+ \Big(
-\frac{18884}{81}
+\frac{608}{27} \pi^2
-\frac{16}{9} \zeta_3
\nn &
+\Big(
-\frac{5264}{27}
+\frac{100}{9} \pi^2
\Big)\logu
-\frac{112}{3}\logu^2
-\frac{40}{9}\logu^3
\Big)
\,, \nn
Z^{(0,3)}_{Q,h\qA} = {} &
Z^{(0,3)}_{Q,hq} 
\,, \nn 
Z^{(0,3)}_{Q,q^2} = {} &
-\frac{16}{9\eps^3}
+\frac{88}{27\eps^2}
+\frac{20}{81\eps}
-\frac{11534}{243}
-\frac{304}{27} \pi^2
-\frac{224}{9} \zeta_3
-\Big(
\frac{1336}{27}
+\frac{32}{9} \pi^2
\Big)\logu
-\frac{152}{9}\logu^2
-\frac{16}{9}\logu^3
\,, \nn
Z^{(0,3)}_{Q,{\qA}^2} = {}
& Z^{(0,3)}_{Q,q^2} 
\,, \nn
Z^{(0,3)}_{Q,q\qA} = {}& 
-\frac{32}{9\eps^3}
+\frac{176}{27\eps^2}
+\frac{40}{81\eps}
-\frac{23068}{243}
-\frac{608}{27} \pi^2
-\frac{448}{9} \zeta_3
-\Big(
\frac{2672}{27}
+\frac{64}{9} \pi^2
\Big)\logu
-\frac{304}{9}\logu^2
-\frac{32}{9}\logu^3
\,, \nn
Z^{(0,3)}_{Q,\eq^4}  = {}& 
-\frac{4}{\eps^2}
+\frac{1}{\eps}\Big(19-16 \zeta_3\Big)
+
\frac{1103}{6}
+4\pi^2
-\frac{4}{15} \pi^4
-120 \zeta_3
+\big(73-48 \zeta_3\big)\logu
+6\logu^2
\,, \nn
 Z^{(0,3)}_{Q,\eqA^4} = {}&
 Z^{(0,3)}_{Q,\eq^4} 
 \,, \nn
Z^{(0,3)}_{Q,\eq^2} = {}& 
\frac{6}{\eps^3}
+\frac{1}{\eps^2}\Big(-\frac{7}{3}+6\logu\Big)
+\frac{1}{\eps}\Big(
-\frac{349}{6}
-\frac{7}{2} \pi^2
-41\logu
-3\logu^2
\Big)
\nn &
-\frac{9701}{36}
+\frac{2701}{36} \pi^2
-\frac{176}{27} \pi^4
-\frac{1504}{9}\pi^2\log2
+\frac{128}{9}\log^4 2
+\frac{256}{9}\pi^2\log^2 2
\nn &
+\frac{1306}{3} \zeta_3
+\frac{1024}{3}\Li_4\!\big(\tfrac{1}{2}\big)
+\Big(
-\frac{1913}{6}
+\frac{133}{6} \pi^2
-\frac{128}{3}\pi^2\log2
+64 \zeta_3
\Big)\logu
\nn &
-\frac{191}{2}\logu^2
-11\logu^3
\,, \nn
 Z^{(0,3)}_{Q,\eqA^2} = {}&
 Z^{(0,3)}_{Q,\eq^2} 
 \,, \nn
Z^{(0,3)}_{Q,h} = {}& 
\frac{1}{\eps^2}\Big(
-\frac{7}{3}
-12\logu
\Big)
+ \frac{1}{\eps}\Big(
-\frac{707}{6}
+15\pi^2
-29\logu
-30\logu^2
\Big)
-\frac{76897}{108}
-\frac{11551}{324} \pi^2
\nn &
+\frac{124}{45} \pi^4
+\frac{224}{9}\pi^2\log2
-32\pi^2\log^2 2
+32\log^4 2
+\frac{3526}{9} \zeta_3
+768 \Li_4\!\big(\tfrac{1}{2}\big)
\nn &
+ \Big(
-\frac{2891}{6}
+\frac{233}{3} \pi^2
-\frac{128}{3}\pi^2\log2
+64 \zeta_3
\Big)\logu
-\frac{143}{2}\logu^2
-38\logu^3
\,, \nn
Z^{(0,3)}_{Q,Q} = {}& 
-\frac{9}{2\eps^3}
+ \frac{1}{\eps^2}\Big(
-\frac{81}{4}
-\frac{27}{2}\logu
\Big)
+ \frac{1}{\eps}\Big(
-\frac{1039}{8}
+\frac{303}{8} \pi^2
-48\pi^2\log2
+72 \zeta_3
-\frac{243}{4}\logu
\nn &
-\frac{81}{4}\logu^2
\Big)
-\frac{10823}{48}
-\frac{58321}{144} \pi^2
-\frac{328}{15} \pi^4
+\frac{2740}{3}\pi^2\log2
+192\pi^2\log^2 2
-\frac{80}{3}\log^4 2
\nn &
-\frac{739}{2} \zeta_3
+8\pi^2\zeta_3
-20 \zeta_5
-640 \Li_4\!\big(\tfrac{1}{2}\big)
- \Big(
\frac{3117}{8}
-\frac{909}{8} \pi^2
+144\pi^2\log2
-216 \zeta_3
\Big)\logu
\nn &
-\frac{729}{8}\logu^2
-\frac{81}{4}\logu^3
\,.
\end{align}
The covariant-gauge fixing parameter $\xi$ appearing in the above expressions is defined through the gluon propagator $\frac{i}{k^2} \left(-g^{\mu\nu} + \xi \,\frac{k^{\mu} k^{\mu}}{k^2} \right)$, and the same gauge-fixing parameter is adopted for the photon propagator. 
We note that the gauge-parameter  $\xi$-dependence in field renormalization constants in dimensional regularization first arises at three-loops in QCD~\cite{Melnikov:2000zc}. In contrast, the pure three-loop QED contributions in dimensional regularization are $\xi$-independent. 
The $\xi$-(in)dependence of $\ZQOS$ in QED in dimensional and other regularization schemes were discussed extensively in refs.~\cite{Landau:1955gai,Johnson:1959zz,Fukuda:1978jy,Broadhurst:1991fy,Melnikov:2000zc}.

\section{Results for $\ZqOS$ up to three-loops}
\label{appendix:ZqOS}

For completeness, we provide the perturbative result for the on-shell wave-function renormalization constant $\ZqOS$ for a massless quark with charge $\eq$, up to three loops in mixed QCD+QED: 
\begin{align}
\ZqOS ={}&
1 
+
\as^2 \Bigg\{
C_F n_h
\Bigg[
\frac{1}{2\eps}
-\frac{5}{12}
+\logu
+\eps\Big(
\frac{89}{72}
+\frac{1}{12}\pi^2
-\frac{5}{6}\logu
+\logu^2
\Big)
\Bigg]
\Bigg\}
\nn &
+ \aew^2\Bigg\{
\eq^2\eQ^2 N_c n_h
\Bigg[
\frac{1}{\eps}
-\frac{5}{6}
+2\logu
+\eps\Big(
\frac{89}{36}
+\frac{1}{6}\pi^2
-\frac{5}{3}\logu
+2\logu^2
\Big)
\Bigg]
\Bigg\}
\nn &
+\as^3\Bigg\{
C_F^2 n_h
\Bigg[
\frac{1}{3\eps^2}
+\frac{1}{\eps}\Big(
-\frac{1}{2}
+\logu
\Big)
+
\frac{443}{36}
+\frac{1}{12}\pi^2
+4 \zeta_3
-\frac{3}{2}\logu
+\frac{3}{2}\logu^2
\Bigg]
\nn &
+ C_A C_F n_h
\Bigg[
\frac{1}{6\eps^3}(1-\xi)
+\frac{1}{\eps^2}\Big(
-\frac{28}{9}
+\frac{1}{2}\xi
+\frac{1}{2}(1-\xi)\logu
\Big)
\nn
&+\frac{1}{\eps}\Big(
\frac{196}{27}
+\frac{1}{24}\pi^2
-\frac{35}{18}\xi
-\frac{1}{24}\pi^2\xi
+\Big(-\frac{17}{3}+\frac{3}{2}\xi\Big)\logu
+\frac{3}{4}(1-\xi)\logu^2
\Big)
\nn
&
-\frac{1204}{81}
-\frac{17}{36}\pi^2
+\frac{407}{54}\xi
+\frac{1}{8}\pi^2\xi
-\frac{17}{6} \zeta_3
-\frac{7}{6}\xi \zeta_3
+\Big(
\frac{337}{18}
+\frac{1}{8}\pi^2(1-\xi)
-\frac{35}{6}\xi
\Big)\logu
\nn &
+\Big(
-\frac{29}{6}
+\frac{9}{4}\xi
\Big)\logu^2
+\frac{3}{4}(1-\xi)\logu^3
\Bigg]
+C_F n_h^2
\Bigg[
\frac{2}{9\eps^2}
-\frac{5}{27\eps}
-\frac{35}{162}
-\frac{2}{3}\logu^2
\Bigg]
\nn &
+ C_F n_h n_l
\Bigg[
\frac{4}{9\eps^2}
+\frac{1}{\eps}\Big(
-\frac{10}{27}
+\frac{2}{3}\logu
\Big)
+
-\frac{35}{81}
+\frac{1}{18}\pi^2
-\frac{5}{9}\logu
+\frac{1}{3}\logu^2
\Big)
\Bigg]
\Bigg\}
\nn &
+ \as^2\aew\Bigg\{
\eq^2 n_h C_F 
\Bigg[
\frac{1}{3\eps^2}
+\frac{1}{\eps}\Big(
-\frac{3}{2}
+\logu
\Big)
+
\frac{173}{36}
+\frac{1}{12}\pi^2
-\frac{9}{2}\logu
+\frac{3}{2}\logu^2
\Bigg]
\nn &
+\eQ^2 n_h C_F 
\Bigg[
\frac{1}{\eps}
+\Big(
\frac{15}{2}
+4 \zeta_3
+3\logu
\Big)
\Bigg]
\Bigg\}
\nn &
+\as\aew^2\Bigg\{
\eq^2\eQ^2 n_h N_c C_F 
\Bigg[
\frac{2}{3\eps^2}
+\frac{1}{\eps}\Big(
-1
+2\logu
\Big)
+
\frac{443}{18}
+\frac{1}{6}\pi^2
-3\logu
+3\logu^2
+8 \zeta_3
\Bigg]
\Bigg\}
\nn &
+\aew^3 \Bigg\{
\eq^2\eQ^4 n_h^2 N_c^2 
\Bigg[
\frac{8}{9\eps^2}
-\frac{20}{27\eps}
-\frac{70}{81}
-\frac{8}{3}\logu^2
\Bigg]
+ \eq^2  \big(\eQ^2\eq^2 n_h\nq +\eQ^2 \eqA^2  n_h \nqA \big) N_c^2
\Bigg[
\frac{16}{9\eps^2}
\nn &
+\frac{1}{\eps}\Big(
-\frac{40}{27}
+\frac{8}{3}\logu
\Big)
-\frac{140}{81}
+\frac{2}{9}\pi^2
-\frac{20}{9}\logu
+\frac{4}{3}\logu^2
\Bigg]
\nn &
+ \eq^4\eQ^2 n_h N_c 
\Bigg[
\frac{2}{3\eps^2}
+\frac{1}{\eps}\Big(
-3
+2\logu
\Big)
+
\frac{173}{18}
+\frac{1}{6}\pi^2
-9\logu
+3\logu^2
\Bigg]
\nn &
+ \eq^2\eQ^4 N_c n_h
\Bigg[
\frac{2}{\eps}
+
15
+6\logu
+8 \zeta_3
\Bigg]
\Bigg\} \,.
\end{align}
The pure QCD part in the above result agrees with the expression given in ref.~\cite{Chen:2025iul}. 
Similary to the case of $\ZQOS$, the gauge-parameter $\xi$-dependence of $\ZqOS$ in dimensional regularization also first arises at three-loops in QCD, whereas the pure three-loop QED contributions are $\xi$-independent in dimensional regularization~\cite{Broadhurst:1991fy,Melnikov:2000zc}.

Note that the on-shell wave-function renormalization constant $\ZqOS$ of a massless quark receives non-vanishing contributions only in the presence of a massive closed quark loop. 
Consequently, the pure QED corrections at two-loop order and the mixed QCD–QED contributions at three-loop order contain only terms linear in the heavy-flavor multiplicity $n_h$.
Compared to its massive counterpart $\ZQOS$ in eq.~\eqref{eq:ZQOS_exp}, the quartic flavor structures in the three-loop order pure QED contribution to $\ZqOS$, formulated for two distinct massless quark flavors with electric charges $e_q$ and $e_{\qA}$ and multiplicities $n_q$ and $n_{\qA}$, are considerably simpler, namely $n_h^2, n_h\nq$, and $n_h\nqA$.

\section{Results for $\ZmMSb$ up to three-loops} 
\label{appendix:ZmMS}

The $\MSbar$ quark mass renormalization constant $\ZmMSb$, defined in eq.~\eqref{eq:mMS_def}, up to three loops in mixed QCD+QED can be conveniently cast into the following form:
\begin{align}\label{eq:ZmMS_exp}
\ZmMSb = 1 +
\sum_{l=1}^{3}
\sum^{i+j=l}_{i=0}
\as^i\, \aew^j\,Z_{\mMS,l}^{(i,j)}\,.
\end{align}
The expression for the quark-mass anomalous dimension can be directly read-off from the coefficient of the simple $1/\epsilon$ pole (up to a factor $-l$) of $\ZmMSb$ in eq.~\eqref{eq:ZmMS_exp}.

In the following, we provide the explicit analytic expressions for the coefficients in the perturbative expansion~\eqref{eq:ZmMS_exp} of the $\MSbar$ mass renormalization constant $\ZmMSb$.
At one loop, the QCD and QED contributions to the $\MSbar$ quark mass renormalization constant read
\begin{align}
Z_{\mMS,1}^{(1,0)} = 
- C_F\frac{3}{\eps}\,,
\qquad
Z_{\mMS,1}^{(0,1)} = 
- \eQ^2\frac{3}{\eps}\,.
\end{align}
The two-loop QCD+QED $\MSbar$ quark mass renormalization constants are given by
\begin{align}
Z_{\mMS,2}^{(2,0)} = {}&
C_A C_F\Big(
  \frac{11}{2 \eps^2}
  - \frac{97}{12 \eps}
\Big)
+ C_F^2\Big(
  \frac{9}{2 \eps^2}
  - \frac{3}{4 \eps}
\Big)
+ C_F(n_h+n_l)\Big(
  -\frac{1}{\eps^2}
  + \frac{5}{6 \eps}
\Big)
\,, \nn
Z_{\mMS,2}^{(1,1)} ={} &
\eQ^2 C_F \Big(
  \frac{9}{\eps^2}
  - \frac{3}{2 \eps}
\Big)
 \,, \nn
Z_{\mMS,2}^{(0,2)} ={} &
 (\eQ^4 n_h + \eQ^2 \eq^2 \nq + \eQ^2 \eqA^2 \nqA )N_c\Big(
  -\frac{2}{\eps^2}
  + \frac{5}{3 \eps}
\Big)
+ \eQ^4\Big(
  \frac{9}{2 \eps^2}
  - \frac{3}{4 \eps}
\Big)\,.
\end{align}
At three-loop order, it is convenient to further decompose the pure QCD contribution $Z_{\mMS,3}^{(3,0)}$ defined in eq.~\eqref{eq:ZmMS_exp} into components corresponding to the different color structures:
\begin{align}
Z_{\mMS,3}^{(3,0)} = {}&
C_F^3\, Z_{\mMS,\sss{F^3}}^{(3,0)}
+ C_A C_F^2\, Z_{\mMS,\sss{AF^2}}^{(3,0)}
+ C_A^2 C_F\, Z_{\mMS,\sss{A^2F}}^{(3,0)}
+ C_F^2 n_h\, Z_{\mMS,\sss{F^2h}}^{(3,0)}
\nn 
&
+ C_F^2 n_l\, Z_{\mMS,\sss{F^2l}}^{(3,0)}
+ C_A C_F n_h\, Z_{\mMS,\sss{AFh}}^{(3,0)}
+ C_A C_F n_l\, Z_{\mMS,\sss{AFl}}^{(3,0)}
\nn 
&
+ C_F n_h^2\, Z_{\mMS,\sss{Fh^2}}^{(3,0)}
+ C_F n_l^2\, Z_{\mMS,\sss{Fl^2}}^{(3,0)}
+ C_F n_h n_l \, Z_{\mMS,\sss{Fhl}}^{(3,0)}\,,
\end{align}
where,
\begin{align}
Z_{\mMS,\sss{F^3}}^{(3,0)} = {}&
  - \frac{9}{2 \eps^3}
  + \frac{9}{4 \eps^2}
  - \frac{43}{2 \eps}
\,,
&
Z_{\mMS,\sss{AF^2}}^{(3,0)} = {}&
  - \frac{33}{2 \eps^3}
  + \frac{313}{12 \eps^2}
  + \frac{43}{4 \eps}
\,,
\nn
Z_{\mMS,\sss{A^2F}}^{(3,0)}= {}&
  - \frac{121}{9 \eps^3}
  + \frac{1679}{54 \eps^2}
  - \frac{11413}{324 \eps}
 \,,
&Z_{\mMS,\sss{F^2h}}^{(3,0)} =  {}&  
  \frac{3}{\eps^3}
  - \frac{29}{6 \eps^2}
  + \frac{1}{\eps}\Big(\frac{23}{3} - 8 \zeta_3\Big)
\,,
\nn
Z_{\mMS,\sss{F^2l}}^{(3,0)}= {}&
Z_{\mMS,\sss{F^2h}}^{(3,0)}
\,,
&Z_{\mMS,\sss{AFh}}^{(3,0)}= {}&
  \frac{44}{9 \eps^3}
  - \frac{242}{27 \eps^2}
  + \frac{1}{\eps}\Big(\frac{278}{81} + 8 \zeta_3\Big)
\,,
\nn
Z_{\mMS,\sss{AFl}}^{(3,0)}= {}&
Z_{\mMS,\sss{AFh}}^{(3,0)}
\,,
&Z_{\mMS,\sss{Fh^2}}^{(3,0)}= {}&
  - \frac{4}{9 \eps^3}
  + \frac{10}{27 \eps^2}
 + \frac{35}{81 \eps}
\,,
\nn
Z_{\mMS,\sss{Fl^2}}^{(3,0)}= {}&
Z_{\mMS,\sss{Fh^2}}^{(3,0)}
\,,
&Z_{\mMS,\sss{Fhl}}^{(3,0)}= {}&
  - \frac{8}{9 \eps^3}
  + \frac{20}{27 \eps^2}
  + \frac{70}{81 \eps}
\,.
\end{align}
The mixed QCD+QED corrections at three-loops are given by
\begin{align}
Z_{\mMS,3}^{(2,1)} = {}&
\eQ^{2} C_F^2\Bigg[
  - \frac{27}{2 \eps^{3}}
  + \frac{27}{4 \eps^{2}}
  - \frac{129}{2 \eps}
\Bigg]
+ \eQ^{2}\, C_A C_F\Bigg[
  - \frac{33}{2 \eps^{3}}
  + \frac{313}{12 \eps^{2}}
  + \frac{43}{4 \eps}
\Bigg]
\nn &
+\eQ^{2}\, n_h C_F\Bigg[
  \frac{3}{\eps^{3}}
  - \frac{29}{6 \eps^{2}}
  + \frac{1}{\eps}\Big(\frac{23}{3} - 8 \zeta_3\Big)
\Bigg]
+ \eQ^{2} n_l C_F \Bigg[
  \frac{3}{\eps^{3}}
  - \frac{17}{6 \eps^{2}}
  + \frac{1}{6 \eps}
\Bigg]
\nn &
+ (\eq^{2} \nq + \eqA^{2} \nqA) C_F \Bigg[
  - \frac{2}{\eps^{2}}
  + \frac{1}{\eps}\Big(\frac{15}{2} - 8 \zeta_3\Big)
\Bigg]
\,,
\end{align}
and
\begin{align}
Z_{\mMS,3}^{(1,2)} = {}&
(\eQ^{4}+\eQ^{2} \eq^{2}+\eQ^{2} \eqA^{2}) N_c\Bigg[
  \frac{6}{\eps^{3}}
  - \frac{29}{3 \eps^{2}}
  + \frac{1}{\eps}\Big(\frac{46}{3} - 16 \zeta_3\Big)
\Bigg]
+ \eQ^{4}\Bigg[
  - \frac{27}{2 \eps^{3}}
  + \frac{27}{4 \eps^{2}}
  - \frac{129}{2 \eps}
\Bigg]
\,.
\end{align}
For the pure three-loop QED contribution, it is convenient to organize $Z_{\mMS,3}^{(0,3)}$, defined in eq.~\eqref{eq:ZmMS_exp}, into separate components according to the different electric-charge combinations and quark-flavor structures as follows:
\begin{align}
Z_{\mMS,3}^{(0,3)} = {}&
\eQ^6  n_h^2 N_c^2 Z^{(0,3)}_{\mMS,h^2}
+ \eQ^4 \eq^2 n_h \nq N_c^2  Z^{(0,3)}_{\mMS,hq}
+ \eQ^4 \eqA^2 n_h\nqA N_c^2 Z^{(0,3)}_{\mMS,h\qA}
+ \eQ^2 \eq^4 \nq^2 N_c^2 Z^{(0,3)}_{\mMS,q^2}
\nn
&
+ \eQ^2 \eqA^4 \nqA^2 N_c^2  Z^{(0,3)}_{\mMS,{\qA}^2}
+ \eQ^2 \eq^2  \eqA^2 \nq\nqA N_c^2 Z^{(0,3)}_{\mMS,q\qA}
+ \eQ^2 \eq^4 \nq N_c  Z^{(0,3)}_{\mMS,\eq^4}
+ \eQ^2 \eqA^4 \nqA N_c Z^{(0,3)}_{\mMS,\eqA^4}
\nn 
&
+ \eQ^4 \eq^2 \nq N_c  Z^{(0,3)}_{\mMS,\eq^2}
+ \eQ^4 \eqA^2 \nqA N_c Z^{(0,3)}_{\mMS,\eqA^2}
+ \eQ^6 n_h N_c Z^{(0,3)}_{\mMS,h} 
+ \eQ^6 Z^{(0,3)}_{\mMS,Q}\,,
\end{align}
where,
\begin{align}
 Z^{(0,3)}_{\mMS,h^2} = {}&
  - \frac{16}{9 \eps^{3}}
  + \frac{40}{27 \eps^{2}}
  + \frac{140}{81 \eps}
 \,,
&Z^{(0,3)}_{\mMS,hq} = {} &
  - \frac{32}{9 \eps^{3}}
  + \frac{80}{27 \eps^{2}}
  + \frac{280}{81 \eps}
\,, \nn
Z^{(0,3)}_{\mMS,h\qA} = {} &
Z^{(0,3)}_{\mMS,hq} 
\,, 
& Z^{(0,3)}_{\mMS,q^2} = {}& 
Z^{(0,3)}_{\mMS,h^2}
\,, \nn
Z^{(0,3)}_{\mMS,{\qA}^2} = {}& Z^{(0,3)}_{\mMS,h^2}
\,, 
&Z^{(0,3)}_{\mMS,q\qA} = {}& 
Z^{(0,3)}_{\mMS,hq} 
\,,  \nn
Z^{(0,3)}_{\mMS,\eq^4}  = {}&
    - \frac{4}{\eps^{2}}
    + \frac{1}{\eps}( 15 - 16 \zeta_3 )
\,,
 &Z^{(0,3)}_{\mMS,\eqA^4} = {}&
 Z^{(0,3)}_{\mMS,\eq^4} 
 \,, \nn
Z^{(0,3)}_{\mMS,\eq^2} = {}& 
    \frac{6}{\eps^{3}}
    - \frac{17}{3 \eps^{2}}
    + \frac{1}{3 \eps}
\,, 
&Z^{(0,3)}_{\mMS,\eqA^2} = {}& 
  Z^{(0,3)}_{\mMS,\eq^2}
\,, \nn
Z^{(0,3)}_{\mMS,h} = {}& 
  \frac{6}{\eps^{3}}
  - \frac{29}{3 \eps^{2}}
  + \frac{1}{\eps}\Big(\frac{46}{3} - 16 \zeta_3\Big)
\,, 
&Z^{(0,3)}_{\mMS,Q} = {}&
  - \frac{9}{2 \eps^{3}}
  + \frac{9}{4 \eps^{2}}
  - \frac{43}{2 \eps}
\,.
\end{align}

\section{Results for $Z_{\bar{\sigma}}$ up to three-loops}
\label{appendix:Zsigmabar}

Expanding the $\MSbar$ and $\sigma$ mass conversion factor $Z_{\bar{\sigma}}$ defined in eq.~\eqref{eq:TASmass_over_MSbar}, through total degree three in $\as$ and $\aew$, with the coupling normalization of eq.~\eqref{eq:aR_def}, gives
\begin{align}\label{eq:Zsigmabar_exp}
Z_{\bar{\sigma}} = 1 +
\sum_{l=1}^{3}
\sum^{i+j=l}_{i=0}
\as^i\, \aew^j\, z_{\bar{\sigma},l}^{(i,j)}\,.
\end{align}

In the following, we provide the explicit analytic expressions for the coefficients in the perturbative expansion~\eqref{eq:Zsigmabar_exp} of the mass conversion factor $Z_{\bar{\sigma}}$.
The one-loop QCD and QED contributions read
\begin{align} 
z_{\bar{\sigma},1}^{(1,0)} = C_F (-2+3 \logums) \qquad z_{\bar{\sigma},1}^{(0,1)} = \eQ^2 (-2+3 \logums)
\end{align}
The two-loop  mixed QCD+QED contributions to the mass relation between $m_{\sigma}$ and $\mMS$ are given by
\begin{align}
z_{\bar{\sigma},2}^{(2,0)} = {}&  C_F n_l \Big(\frac{11}{4}-\frac{2}{3}\pi^2-\frac{1}{3}\logums-\logums^2\Big)+C_F n_h \Big(-\frac{13}{4} + \frac{4}{3}\pi^2 - \frac{1}{3}\logums - \logums^2\Big)
\nn & + C_A C_F \Big(-\frac{123}{8} - \frac{4}{3}\pi^2 + 4 \pi^2 \log2 - 6 \zeta_3 + \frac{53}{6}\logums + \frac{11}{2}\logums^2\Big)
\nn & + C_F^2 \Big(\frac{1}{8} + 5 \pi^2 - 8 \pi^2 \log2 + 12 \zeta_3 - \frac{45}{2}\logums + \frac{9}{2}\logums^2\Big)
\nn
z_{\bar{\sigma},2}^{(1,1)} = {}&  C_F \eQ^2 \Big(\frac{1}{4} + 10 \pi^2 - 16 \pi^2 \log2 + 24 \zeta_3 - 45 \logums + 9 \logums^2\Big)
\nn
z_{\bar{\sigma},2}^{(0,2)} = {}&  \eQ^2 N_c \Big(\eq^2 \nq+\eqA^2 \nqA\Big) \Big(\frac{11}{2} - \frac{4}{3}\pi^2 - \frac{2}{3}\logums - 2 \logums^2\Big)
\nn & 
+ \eQ^4 N_c n_h \Big(-\frac{13}{2} + \frac{8}{3}\pi^2 - \frac{2}{3}\logums - 2 \logums^2\Big)
\nn & + \eQ^4 \Big(\frac{1}{8} + 5 \pi^2 - 8 \pi^2 \log2 + 12 \zeta_3 - \frac{45}{2}\logums + \frac{9}{2}\logums^2\Big)
\end{align}

At three-loops, it is convenient to further decompose the pure QCD contribution $z_{\bar{\sigma},3}^{(3,0)}$ defined in eq.~\eqref{eq:Zsigmabar_exp} into components corresponding to the different color structures:
\begin{align}
z_{\bar{\sigma},3}^{(3,0)} = {}&
C_F^3\, z_{\bar{\sigma},\sss{F^3}}^{(3,0)}
+ C_A C_F^2\, z_{\bar{\sigma},\sss{AF^2}}^{(3,0)}
+ C_A^2 C_F\, z_{\bar{\sigma},\sss{A^2F}}^{(3,0)}
+ C_F^2 n_h\, z_{\bar{\sigma},\sss{F^2h}}^{(3,0)}
\nn 
&
+ C_F^2 n_l\, z_{\bar{\sigma},\sss{F^2l}}^{(3,0)}
+ C_A C_F n_h\, z_{\bar{\sigma},\sss{AFh}}^{(3,0)}
+ C_A C_F n_l\, z_{\bar{\sigma},\sss{AFl}}^{(3,0)}
\nn 
&
+ C_F n_h^2\, z_{\bar{\sigma},\sss{Fh^2}}^{(3,0)}
+ C_F n_l^2\, z_{\bar{\sigma},\sss{Fl^2}}^{(3,0)}
+ C_F n_h n_l \, z_{\bar{\sigma},\sss{Fhl}}^{(3,0)}\,,
\end{align}
where,
\begin{align}
z_{\bar{\sigma},\sss{F^3}}^{(3,0)} = {}&
\frac{2087}{12} + \frac{553}{3} \pi^2 + \frac{4}{3} \pi^4 - 432 \pi^2 \log2 - 32 \pi^2 \log^2 2 + 32 \log^4 2 + 276 \zeta_3 + 4 \pi^2 \zeta_3
\nn &
- 40 \zeta_5 + 768 \Li_4\!\big(\tfrac{1}{2}\big) + \Big( \frac{1503}{8} + 15 \pi^2 - 24 \pi^2 \log2 + 36 \zeta_3 \Big)\logums - \frac{117}{2} \logums^2 + \frac{9}{2} \logums^3
\,, \nn
z_{\bar{\sigma},\sss{AF^2}}^{(3,0)} = {}&
\frac{1031}{72} - \frac{1121}{9} \pi^2 - \frac{260}{27} \pi^4 + \frac{1160}{9} \pi^2 \log2 + \frac{496}{9} \pi^2 \log^2 2 + \frac{32}{9} \log^4 2 + \frac{1090}{3} \zeta_3
\nn &
+ 76 \pi^2 \zeta_3 - 180 \zeta_5 + \frac{256}{3} \Li_4\!\big(\tfrac{1}{2}\big) + \Big( - \frac{6235}{24} + \frac{98}{3} \pi^2 - \frac{140}{3} \pi^2 \log2 + 70 \zeta_3 \Big)\logums
\nn &
- 78 \logums^2 + \frac{33}{2} \logums^3
\,, \nn
z_{\bar{\sigma},\sss{A^2F}}^{(3,0)} = {}&
- \frac{584447}{1944} + \frac{3011}{54} \pi^2 + \frac{179}{54} \pi^4 + \frac{392}{9} \pi^2 \log2 - \frac{176}{9} \pi^2 \log^2 2 - \frac{88}{9} \log^4 2 - \frac{1894}{9} \zeta_3
\nn &
- 51 \pi^2 \zeta_3 + 130 \zeta_5 - \frac{704}{3} \Li_4\!\big(\tfrac{1}{2}\big) + \Big( - \frac{803}{27} - \frac{88}{9} \pi^2 + \frac{88}{3} \pi^2 \log2 - 44 \zeta_3 \Big)\logums
\nn &
+ \frac{889}{18} \logums^2 + \frac{121}{9} \logums^3
\,, \nn
z_{\bar{\sigma},\sss{F^2h}}^{(3,0)} = {}&
\frac{49}{18} + \frac{896}{27} \pi^2 - \frac{182}{135} \pi^4 - \frac{448}{9} \pi^2 \log2 + \frac{32}{9} \pi^2 \log^2 2 - \frac{32}{9} \log^4 2 + \frac{164}{3} \zeta_3
\nn &
- \frac{256}{3} \Li_4\!\big(\tfrac{1}{2}\big) + \Big( - \frac{163}{12} - \frac{8}{3} \pi^2 + \frac{32}{3} \pi^2 \log2 + 8 \zeta_3 \Big)\logums + 15 \logums^2 - 3 \logums^3
\,, \nn
z_{\bar{\sigma},\sss{F^2l}}^{(3,0)} = {}&
- \frac{599}{18} - \frac{64}{9} \pi^2 + \frac{238}{135} \pi^4 + \frac{160}{9} \pi^2 \log2 - \frac{64}{9} \pi^2 \log^2 2 - \frac{32}{9} \log^4 2 - \frac{268}{3} \zeta_3
\nn &
- \frac{256}{3} \Li_4\!\big(\tfrac{1}{2}\big) + \Big( \frac{53}{12} - \frac{26}{3} \pi^2 + \frac{32}{3} \pi^2 \log2 + 8 \zeta_3 \Big)\logums + 15 \logums^2 - 3 \logums^3
\,, \nn
z_{\bar{\sigma},\sss{AFh}}^{(3,0)} = {}&
\frac{22945}{486} + \frac{230}{3} \pi^2 + \frac{172}{135} \pi^4 - \frac{928}{9} \pi^2 \log2 - \frac{16}{9} \pi^2 \log^2 2 + \frac{16}{9} \log^4 2 + \frac{506}{9} \zeta_3
\nn &
- 4 \pi^2 \zeta_3 + 20 \zeta_5 + \frac{128}{3} \Li_4\!\big(\tfrac{1}{2}\big) + \Big( - \frac{188}{27} + \frac{104}{9} \pi^2 - \frac{16}{3} \pi^2 \log2 - 16 \zeta_3 \Big)\logums
\nn &
- \frac{109}{9} \logums^2 - \frac{44}{9} \logums^3
\,, \nn
z_{\bar{\sigma},\sss{AFl}}^{(3,0)} = {}&
\frac{54373}{486} - \frac{182}{27} \pi^2 - \frac{38}{135} \pi^4 - \frac{80}{9} \pi^2 \log2 + \frac{32}{9} \pi^2 \log^2 2 + \frac{16}{9} \log^4 2 + \frac{110}{9} \zeta_3
\nn &
+ \frac{128}{3} \Li_4\!\big(\tfrac{1}{2}\big) + \Big( \frac{1000}{27} - \frac{28}{9} \pi^2 - \frac{16}{3} \pi^2 \log2 - 16 \zeta_3 \Big)\logums - \frac{109}{9} \logums^2 - \frac{44}{9} \logums^3
\,, \nn
z_{\bar{\sigma},\sss{Fh^2}}^{(3,0)} = {}&
- \frac{4703}{486} + \frac{416}{135} \pi^2 - \frac{88}{9} \zeta_3 + \Big( \frac{82}{27} - \frac{16}{9} \pi^2 \Big)\logums + \frac{2}{9} \logums^2 + \frac{4}{9} \logums^3
\,, \nn
z_{\bar{\sigma},\sss{Fl^2}}^{(3,0)} = {}&
- \frac{4055}{486} + \frac{4}{27} \pi^2 + \frac{56}{9} \zeta_3 + \Big( - \frac{134}{27} + \frac{8}{9} \pi^2 \Big)\logums + \frac{2}{9} \logums^2 + \frac{4}{9} \logums^3
\,, \nn
z_{\bar{\sigma},\sss{Fhl}}^{(3,0)} = {}&
- \frac{4379}{243} - \frac{4}{27} \pi^2 - \frac{32}{9} \zeta_3 + \Big( - \frac{52}{27} - \frac{8}{9} \pi^2 \Big)\logums + \frac{4}{9} \logums^2 + \frac{8}{9} \logums^3
\,.
\end{align}
The mixed QCD+QED corrections at three-loops are given by
\begin{align}
z_{\bar{\sigma},3}^{(2,1)} = {}&
\eQ^2 C_F^2 \Bigg[ \frac{2087}{4} + 553 \pi^2 + 4 \pi^4 - 1296 \pi^2 \log2 - 96 \pi^2 \log^2 2 + 96 \log^4 2 + 828 \zeta_3
\nn &
+ 12 \pi^2 \zeta_3 - 120 \zeta_5 + 2304 \Li_4\!\big(\tfrac{1}{2}\big) + \Big( \frac{4509}{8} + 45 \pi^2 - 72 \pi^2 \log2 + 108 \zeta_3 \Big)\logums
\nn &
- \frac{351}{2} \logums^2 + \frac{27}{2} \logums^3 \Bigg] + \eQ^2 C_A C_F \Bigg[ \frac{1031}{72} - \frac{1121}{9} \pi^2 - \frac{260}{27} \pi^4 + \frac{1160}{9} \pi^2 \log2
\nn &
+ \frac{496}{9} \pi^2 \log^2 2 + \frac{32}{9} \log^4 2 + \frac{1090}{3} \zeta_3 + 76 \pi^2 \zeta_3 - 180 \zeta_5 + \frac{256}{3} \Li_4\!\big(\tfrac{1}{2}\big)
\nn &
+ \Big( - \frac{6235}{24} + \frac{98}{3} \pi^2 - \frac{140}{3} \pi^2 \log2 + 70 \zeta_3 \Big)\logums - 78 \logums^2 + \frac{33}{2} \logums^3 \Bigg]
\nn &
+ \eQ^2 n_h C_F \Bigg[ \frac{49}{18}
+ \frac{896}{27} \pi^2 - \frac{182}{135} \pi^4 - \frac{448}{9} \pi^2 \log2 + \frac{32}{9} \pi^2 \log^2 2 - \frac{32}{9} \log^4 2 
+ \frac{164}{3} \zeta_3
\nn &
 - \frac{256}{3} \Li_4\!\big(\tfrac{1}{2}\big)
+ \Big( - \frac{163}{12} - \frac{8}{3} \pi^2 + \frac{32}{3} \pi^2 \log2 + 8 \zeta_3 \Big)\logums + 15 \logums^2 - 3 \logums^3 \Bigg] 
\nn &
+ \eQ^2 n_l C_F \Bigg[ - \frac{913}{36}
- \frac{46}{9} \pi^2 + \frac{44}{27} \pi^4 + \frac{160}{9} \pi^2 \log2 - \frac{64}{9} \pi^2 \log^2 2 - \frac{32}{9} \log^4 2 
- \frac{256}{3} \zeta_3
\nn &
 - \frac{256}{3} \Li_4\!\big(\tfrac{1}{2}\big)
+ \Big( \frac{275}{12} - \frac{26}{3} \pi^2 + \frac{32}{3} \pi^2 \log2 - 16 \zeta_3 \Big)\logums + 18 \logums^2 - 3 \logums^3 \Bigg]
\nn &
+ (\eq^2\nq+\eqA^2\nqA)C_F \Bigg[ - \frac{95}{12} - 2 \pi^2 + \frac{2}{15} \pi^4 - 4 \zeta_3 + \Big( - \frac{37}{2} + 24 \zeta_3 \Big)\logums - 3 \logums^2 \Bigg]
\,.
\end{align}
and
\begin{align}
z_{\bar{\sigma},3}^{(1,2)} = {}&
\eQ^4 n_h N_c C_F \Bigg[ \frac{49}{9} + \frac{1792}{27} \pi^2 - \frac{364}{135} \pi^4 - \frac{896}{9} \pi^2 \log2 + \frac{64}{9} \pi^2 \log^2 2 - \frac{64}{9} \log^4 2
+ \frac{328}{3} \zeta_3
\nn &
 - \frac{512}{3} \Li_4\!\big(\tfrac{1}{2}\big) + \Big( - \frac{163}{6} - \frac{16}{3} \pi^2 + \frac{64}{3} \pi^2 \log2 + 16 \zeta_3 \Big)\logums + 30 \logums^2 - 6 \logums^3 \Bigg]
\nn &
+ \eQ^2(\eq^2\nq+\eqA^2\nqA) N_c C_F \Bigg[ - \frac{599}{9} - \frac{128}{9} \pi^2 + \frac{476}{135} \pi^4 + \frac{320}{9} \pi^2 \log2 - \frac{128}{9} \pi^2 \log^2 2
\nn &
- \frac{64}{9} \log^4 2 - \frac{536}{3} \zeta_3 - \frac{512}{3} \Li_4\!\big(\tfrac{1}{2}\big) + \Big( \frac{53}{6} - \frac{52}{3} \pi^2 + \frac{64}{3} \pi^2 \log2 + 16 \zeta_3 \Big)\logums + 30 \logums^2
\nn &
- 6 \logums^3 \Bigg] + \eQ^4 C_F \Bigg[ \frac{2087}{4} + 553 \pi^2 + 4 \pi^4 - 1296 \pi^2 \log2 - 96 \pi^2 \log^2 2 + 96 \log^4 2
\nn &
+ 828 \zeta_3 + 12 \pi^2 \zeta_3 - 120 \zeta_5 + 2304 \Li_4\!\big(\tfrac{1}{2}\big)
+ \Big( \frac{4509}{8} + 45 \pi^2 - 72 \pi^2 \log2 + 108 \zeta_3 \Big)\logums 
\nn &
- \frac{351}{2} \logums^2 + \frac{27}{2} \logums^3 \Bigg]
\,.
\end{align}
For the pure three-loop QED contribution, it is convenient to decompose $z_{\bar{\sigma},3}^{(0,3)}$, defined in eq.~\eqref{eq:Zsigmabar_exp}, according to the distinct electric-charge combinations and quark-flavor structures as follows:
\begin{align}
z_{\bar{\sigma},3}^{(0,3)} = {}&
\eQ^6 n_h^2 N_c^2\, z_{\bar{\sigma},\sss{h^2}}^{(0,3)}
+ \eQ^4 \eq^2 n_h \nq N_c^2\, z_{\bar{\sigma},\sss{hq}}^{(0,3)}
+ \eQ^4 \eqA^2 n_h \nqA N_c^2\, z_{\bar{\sigma},\sss{h\qA}}^{(0,3)}
+ \eQ^2 \eq^4 \nq^2 N_c^2\, z_{\bar{\sigma},\sss{q^2}}^{(0,3)}
\nn 
&
+ \eQ^2 \eqA^4 \nqA^2 N_c^2\, z_{\bar{\sigma},\sss{{\qA}^2}}^{(0,3)}
+ \eQ^2 \eq^2 \eqA^2 \nq \nqA N_c^2\, z_{\bar{\sigma},\sss{q\qA}}^{(0,3)}
+ \eQ^2 \eq^4 \nq N_c\, z_{\bar{\sigma},\sss{\eq^4}}^{(0,3)}
+ \eQ^2 \eqA^4 \nqA N_c\, z_{\bar{\sigma},\sss{\eqA^4}}^{(0,3)}
\nn 
&
+ \eQ^4 \eq^2 \nq N_c\, z_{\bar{\sigma},\sss{\eq^2}}^{(0,3)}
+ \eQ^4 \eqA^2 \nqA N_c\, z_{\bar{\sigma},\sss{\eqA^2}}^{(0,3)}
+ \eQ^6 n_h N_c\, z_{\bar{\sigma},\sss{h}}^{(0,3)}
+ \eQ^6\, z_{\bar{\sigma},\sss{Q}}^{(0,3)}\,,
\end{align}
where,
\begin{align}
z_{\bar{\sigma},\sss{h^2}}^{(0,3)} = {}&
- \frac{9406}{243} + \frac{1664}{135} \pi^2 - \frac{352}{9} \zeta_3 + \Big( \frac{328}{27} - \frac{64}{9} \pi^2 \Big)\logums + \frac{8}{9} \logums^2 + \frac{16}{9} \logums^3
\,, \nn
z_{\bar{\sigma},\sss{hq}}^{(0,3)} = {}&
- \frac{17516}{243} - \frac{16}{27} \pi^2 - \frac{128}{9} \zeta_3 + \Big( - \frac{208}{27} - \frac{32}{9} \pi^2 \Big)\logums + \frac{16}{9} \logums^2 + \frac{32}{9} \logums^3
\,, \nn
z_{\bar{\sigma},\sss{h\qA}}^{(0,3)} = {}&
z_{\bar{\sigma},\sss{hq}}^{(0,3)}
\,, \nn
z_{\bar{\sigma},\sss{q^2}}^{(0,3)} = {}&
- \frac{8110}{243} + \frac{16}{27} \pi^2 + \frac{224}{9} \zeta_3 + \Big( - \frac{536}{27} + \frac{32}{9} \pi^2 \Big)\logums + \frac{8}{9} \logums^2 + \frac{16}{9} \logums^3
\,, \nn
z_{\bar{\sigma},\sss{{\qA}^2}}^{(0,3)} = {}&
z_{\bar{\sigma},\sss{q^2}}^{(0,3)}
\,, \nn
z_{\bar{\sigma},\sss{q\qA}}^{(0,3)} = {}&
- \frac{16220}{243} + \frac{32}{27} \pi^2 + \frac{448}{9} \zeta_3 + \Big( - \frac{1072}{27} + \frac{64}{9} \pi^2 \Big)\logums + \frac{16}{9} \logums^2 + \frac{32}{9} \logums^3
\,, \nn
z_{\bar{\sigma},\sss{\eq^4}}^{(0,3)} = {}&
- \frac{95}{6} - 4 \pi^2 + \frac{4}{15} \pi^4 - 8 \zeta_3 + \Big( - 37 + 48 \zeta_3 \Big)\logums - 6 \logums^2
\,, \nn
z_{\bar{\sigma},\sss{\eqA^4}}^{(0,3)} = {}&
z_{\bar{\sigma},\sss{\eq^4}}^{(0,3)}
\,, \nn
z_{\bar{\sigma},\sss{\eq^2}}^{(0,3)} = {}&
- \frac{913}{18} - \frac{92}{9} \pi^2 + \frac{88}{27} \pi^4 + \frac{320}{9} \pi^2 \log2 - \frac{128}{9} \pi^2 \log^2 2 - \frac{64}{9} \log^4 2 - \frac{512}{3} \zeta_3
\nn &
- \frac{512}{3} \Li_4\!\big(\tfrac{1}{2}\big) + \Big( \frac{275}{6} - \frac{52}{3} \pi^2 + \frac{64}{3} \pi^2 \log2 - 32 \zeta_3 \Big)\logums + 36 \logums^2 - 6 \logums^3
\,, \nn
z_{\bar{\sigma},\sss{\eqA^2}}^{(0,3)} = {}&
z_{\bar{\sigma},\sss{\eq^2}}^{(0,3)}
\,, \nn
z_{\bar{\sigma},\sss{h}}^{(0,3)} = {}&
\frac{49}{9} + \frac{1792}{27} \pi^2 - \frac{364}{135} \pi^4 - \frac{896}{9} \pi^2 \log2 + \frac{64}{9} \pi^2 \log^2 2 - \frac{64}{9} \log^4 2 + \frac{328}{3} \zeta_3
\nn &
- \frac{512}{3} \Li_4\!\big(\tfrac{1}{2}\big) + \Big( - \frac{163}{6} - \frac{16}{3} \pi^2 + \frac{64}{3} \pi^2 \log2 + 16 \zeta_3 \Big)\logums + 30 \logums^2 - 6 \logums^3
\,, \nn
z_{\bar{\sigma},\sss{Q}}^{(0,3)} = {}&
\frac{2087}{12} + \frac{553}{3} \pi^2 + \frac{4}{3} \pi^4 - 432 \pi^2 \log2 - 32 \pi^2 \log^2 2 + 32 \log^4 2 + 276 \zeta_3 + 4 \pi^2 \zeta_3
\nn &
- 40 \zeta_5 + 768 \Li_4\!\big(\tfrac{1}{2}\big) + \Big( \frac{1503}{8} + 15 \pi^2 - 24 \pi^2 \log2 + 36 \zeta_3 \Big)\logums - \frac{117}{2} \logums^2 + \frac{9}{2} \logums^3
\,.
\end{align}

\bibliographystyle{JHEP}
\bibliography{references}

\end{document}